\documentclass[conference]{IEEEtran}
\usepackage[T1]{fontenc}
\usepackage{amsmath,amssymb,amsfonts,bm,mathtools}
\usepackage{amsthm,graphicx,cite}
\usepackage{xcolor,tikz}

\newcommand{\NL}{N_{\mathrm{Leg}}}
\newcommand{\NT}{N_T}
\newcommand{\CN}{\mathcal{CN}}
\newcommand{\Id}{\mathbf I}

\newcommand{\ind}{\mathbf 1}

\newtheorem{proposition}{Proposition}
\newtheorem{theorem}{Theorem}

\newtheorem{remark}{Remark}

\usepackage[unicode,hidelinks]{hyperref}
\providecommand{\symdef}[1]{\hypertarget{def:#1}{}}
\providecommand{\symref}[2]{\hyperlink{def:#1}{#2}}
\renewcommand{\baselinestretch}{0.95}

\title{ {Maurer-Inspired Two-Way Physical-Layer}\\
 {Authentication Against Multi-Antenna Impersonation Attacks}}
\author{
\IEEEauthorblockN{ {Pin-Hsun Lin\IEEEauthorrefmark{1},
Linda Senigagliesi\IEEEauthorrefmark{2}, and
Eduard A. Jorswieck\IEEEauthorrefmark{1}}}
\IEEEauthorblockA{ {\IEEEauthorrefmark{1}Institute for Communications Technology,
Technische Universit\"at Braunschweig, Germany}\\
 {\IEEEauthorrefmark{2}ETIS UMR 8051, CYU, ENSEA, CNRS, Cergy, France}\\
 {\IEEEauthorrefmark{1}\{p.lin,\,e.jorswieck\}@tu-braunschweig.de\quad
\IEEEauthorrefmark{2}linda.senigagliesi@ensea.fr}}
}
\begin{document}
\pagecolor{white}\color{black}
\maketitle

\begin{abstract}
In this work, we investigate Maurer-inspired two-way physical-layer authentication.
Alice seeks authentication by Bob in the presence of a multi-antenna
impersonator, Trudy. Trudy may replace the challenge to Alice, the
response to Bob, or both. To create an information advantage, Bob adds
artificial noise (AN) to his challenge, and Alice adds spatial AN while forwarding the received mixture. Bob predicts the response from trusted preparation, whereas Trudy observes only a noisy version of the challenge, and Alice's AN interferes with Trudy's preparation reception.
Impersonation thus requires matching Bob's fresh randomness beyond
the channel consistency checked by public-pilot one-way authentication.
For quasi-static Gaussian multiple-input multiple-output channels
with time-division duplex (TDD) reciprocity, we derive
missed-detection bounds for attacks that replace the authentication
challenge, the response, or both. These bounds assume attack-free
initial training and preparation and account for estimation error
and AN leakage. With equal antenna numbers at all nodes, we show scenarios where the proposed scheme has
lower worst-case missed detection than specified one-way tests with/without AN.
\end{abstract}


\section{Introduction}
\label{SEC_introduction}
Physical-layer authentication (PLA) compares current channel
information with trusted observations \cite{Xiao2008,Baracca2012,Senigagliesi2021}.
In conventional one-way PLA, Bob estimates Alice's channel from public
pilots and checks consistency with trusted preparation. This relies on
Trudy being unable to reproduce Alice's channel, however, spatial processing can
weaken this protection \cite{SpatialPLASurvey2026}. An attacker knowing
Bob's beamformer can falsify angle-of-arrival features \cite{Srinivasan2024}.
Multi-antenna impersonation attacks can then succeed when Alice's spatial signature
lies in the attacker's achievable span, in the far field \cite{Pham2026},
near field \cite{NearField2026}, and certain reconfigurable intelligent surface (RIS)-assisted channels
\cite{Senigagliesi2025}. 
Challenge-response PLA requires answering a fresh challenge
\cite{Jorswieck2015}. Existing schemes use shared secrets
\cite{Shan2013,Crosara2026,Zhang2026} or secret configurations of
controlled hardware, including RISs \cite{Guglielmi2026} and
amplify-and-forward relays \cite{Rowhani2025}. Artificial noise (AN)
instead commonly protects confidentiality \cite{GoelNegi2008,Lin2013},
channel training \cite{Huang2013}, or a key-derived authentication tag
\cite{AN_TIFS2021}.

Maurer showed how interaction and unequal noisy observations can
support security without an initially shared key \cite{Maurer1993}.
Inspired by this idea, we propose the PLA protocol: let Bob add fresh AN to his challenge and
retain his exact samples. Alice forwards the received mixture and
adds spatial AN in the estimated return-channel nullspace
\cite{GoelNegi2008}. Bob estimates Alice's response from trusted preparation
and his recorded samples, whereas Trudy must infer the challenge
through noise. Thus, an impersonation attack requires reproducing fresh randomness
as well as channel information. Alice's AN additionally obscures
preparation interceptions. The test needs no pre-shared
authentication key. Higher-layer identity
verification is still assumed during setup, and no secret key is distilled.

The main contributions of this paper are listed as follows:
\begin{itemize}
\item We propose a two-way protocol using Bob's recorded AN as fresh authentication
evidence and Alice's spatial AN to obscure preparation observations.
\item We derive the worst-case missed-detection probability bounds covering replacement of the
authentication challenge, response, or both. We assume attack-free
initial training and preparation and remain valid after disclosing
the channels and Bob's preparation records to Trudy. 
\item A channel-strength condition and numerical comparisons establish
lower worst-case missed detection probability than a specified one-way tests with and
without AN, under matched energy constraint and a common false-alarm probability limit.
\end{itemize}

The proper complex Gaussian distribution $\CN(\boldsymbol\mu,\mathbf K)$
has mean $\boldsymbol\mu$ and covariance $\mathbf K$.
The Gamma distribution $\Gamma(k,1)$ has shape $k>0$, unit scale,
and cumulative distribution function (CDF) $G_k$, while $\Gamma(k)$
denotes the Gamma function. The F distribution $F_{a,b}$ has
real degrees of freedom $a,b>0$, with quantile function $F_{a,b}^{-1}$.
Superscripts $\mathsf T,H$ denote transpose and conjugate transpose.
The norms $\|\cdot\|,\|\cdot\|_F$ are Euclidean and Frobenius.
The symbols $\Id_k$, $\ind\{\cdot\}$, and $\preceq$ denote the
$k\times k$ identity matrix, indicator function, and positive semi-definite order.

\section{System Model and Preliminaries}
\label{SEC_system_model}
Alice requests authentication and Bob decides whether to accept her.
Meanwhile, Trudy listens to the wireless exchange and may replace Alice's
response ($T\to B$), Bob's challenge ($T\to A$), or both.
Each legitimate node has $p:=\NL$ antennas. Alice and Bob each
use all $p$ antennas for transmission and activate $N_R<p$
antennas for reception. Under the Gaussian channel model,
the return-channel estimate has rank $N_R$ almost surely,
leaving $d:=p-N_R$ nullspace directions for Alice's spatial AN. Trudy has $q:=\NT$ antennas, and $1\leq N_R<p$.
A session contains initial spatial training, a returned preparation exchange, and one authentication challenge and response. Alice's request starts the session. 

\subsection{Channels, Timing, and Authentication Objective}
\label{SUBSEC_channels}
We assume TDD reciprocity of the effective baseband channels. The quasi-static channel is
$\mathbf H\in\mathbb C^{p\times p}$ from Alice to Bob. Its entries
are independent $\CN(0,S)$, with $S>0$. The reverse full channel is
$\mathbf H^{\mathsf T}$. Let
$\mathbf S=[\Id_{N_R}\ \mathbf0]\in\mathbb R^{N_R\times p}$
select the active receive antennas. The active forward and return channels $\mathbf H_A,\mathbf H_B\in\mathbb C^{N_R\times p}$
have per-entry variance $S$ and satisfy
\begin{equation}
\mathbf H_A=\mathbf S\mathbf H^{\mathsf T},\qquad
\mathbf H_B=\mathbf S\mathbf H.
\label{EQ_TDD_channels}
\end{equation}
They are dependent and constant throughout training, preparation,
and authentication.
Independent sessions use independent $\mathbf H$.
Let $\mathbf G_U\in\mathbb C^{q\times p}$ be the channel from node
$U\in\{A,B\}$ to Trudy. Its entries are independent $\CN(0,S_T)$,
where $S_T>0$, and $\mathbf G_A,\mathbf G_B$ are mutually independent and
independent of the legitimate channels. They also stay constant within a session. Reciprocity gives Trudy's channel to node $U$'s active antennas as
$\mathbf K_U=\mathbf S\mathbf G_U^{\mathsf T}
\in\mathbb C^{N_R\times q}$. 
Legitimate and Trudy receiver noises have independent entries
$\CN(0,\sigma^2)$ and $\CN(0,\sigma_T^2)$, respectively, with positive
variances. Noises are independent across blocks and of all channels.

We assume that initial training and preparation are attack-free.
During the subsequent authentication exchange, Trudy may replace
the challenge, the response, or both.

Bob's decision is $I_B\in\{0,1\}$, with one for acceptance.
Let $\mathcal H_0$ denote an honest exchange. Let $\mathfrak A$ be
the class of attack strategies $\pi$ specifying challenge replacement,
response replacement, or both, and the forged signals subject to
the timing restrictions below. 
The false-alarm probability $P_{\rm FA}$
is honest rejection. The missed detection probability $P_{\rm MD}(\pi)$
is Bob's acceptance of an authentication exchange with at least one
replaced block, and $P_2^\star$ is its
worst case over all three replacement cases:
\begin{equation}
\begin{aligned}
P_{\rm FA}&=\Pr\{I_B=0\mid\mathcal H_0\},\\
P_{\rm MD}(\pi)&=\Pr_\pi\{I_B=1\},\qquad
P_2^\star=\sup_{\pi\in\mathfrak A}P_{\rm MD}(\pi).
\end{aligned}
\label{EQ_objective}
\end{equation}
We seek an achievable protocol with small $P_2^\star$ subject to
$P_{\rm FA}\leq\alpha$, where $\alpha\in(0,1)$. The guarantee concerns
the proposed detector and does not assert optimal detector design.

\subsection{Noisy Spatial Design and Trudy's Information}
\label{SUBSEC_spatial}
Alice and Bob use the same public training matrix with orthogonal
rows of equal energy, $\mathbf P_{\rm tr}\in\mathbb C^{p\times p}$,
to choose Alice's transmit directions and Bob's receive beamformer.
They transmit $\mathbf P_{\rm tr}$ in separate slots, first
$A\to B$, then $B\to A$. With training energy $E_{\rm seed}>0$
per antenna, the matrix satisfies
\begin{equation}
\mathbf P_{\rm tr}\mathbf P_{\rm tr}^H=E_{\rm seed}\Id_p.
\label{EQ_training_pilots}
\end{equation}
This initial training is identical in all comparisons.
Bob's received block is
$\mathbf Y_{\rm tr}=\mathbf H_B\mathbf P_{\rm tr}+\mathbf Z_{\rm tr}$,
where $\mathbf Y_{\rm tr},\mathbf Z_{\rm tr}\in\mathbb C^{N_R\times p}$,
where the receiver-noise matrix $\mathbf Z_{\rm tr}$ has independent
proper complex Gaussian entries, each is distributed as $\CN(0,\sigma^2)$.
Bob's minimum mean-square error channel estimate is
$\widehat{\mathbf H}_B\in\mathbb C^{N_R\times p}$. Its error matrix
$\boldsymbol\Delta_B\in\mathbb C^{N_R\times p}$ has per-entry variance
$\tau\in(0,S)$:
\begin{equation}
\begin{aligned}
\mathbf H_B&=\widehat{\mathbf H}_B+\boldsymbol\Delta_B,\\
\widehat{\mathbf H}_B&=\frac{S\mathbf Y_{\rm tr}\mathbf P_{\rm tr}^H}
{SE_{\rm seed}+\sigma^2},\qquad
\tau=\frac{S\sigma^2}{SE_{\rm seed}+\sigma^2}.
\end{aligned}
\label{EQ_training}
\end{equation}
The estimate and error are independent, with independent zero-mean
proper Gaussian entries of variances $S-\tau$ and $\tau$, respectively.
For Alice-to-Bob transmission, let $s>0$ be the largest singular
value of $\widehat{\mathbf H}_B$. Its corresponding unit-norm
left and right singular vectors are Bob's receive beamformer
$\mathbf u_B\in\mathbb C^{N_R}$ and Alice's transmit beam
$\mathbf f_A\in\mathbb C^p$, respectively.

Since $\widehat{\mathbf H}_B$ has full row rank almost surely,
its nullspace has dimension $d=p-N_R$. Alice's spatial AN
directions are the columns of
$\mathbf B\in\mathbb C^{p\times d}$, an orthonormal basis of
this nullspace. These choices satisfy
\begin{equation}
\begin{aligned}
\widehat{\mathbf H}_B\mathbf f_A&=s\mathbf u_B,
&\widehat{\mathbf H}_B^H\mathbf u_B&=s\mathbf f_A,\\
\widehat{\mathbf H}_B\mathbf B&=\mathbf0,
&\mathbf f_A^H\mathbf B&=\mathbf0.
\end{aligned}
\label{EQ_spatial}
\end{equation}
The directions $\mathbf f_A,\mathbf B,\mathbf u_B$ are public
and known to Trudy. Bob sends Alice $\mathbf f_A,\mathbf B$
through authenticated public feedback. He does not disclose
the singular values or his received training records.

For Bob-to-Alice transmission, Alice uses a fixed public unit
receive beamformer $\mathbf u_A\in\mathbb C^{N_R}$, giving
$\mathbf h:=\mathbf u_A^H\mathbf H_A\in\mathbb C^{1\times p}$.
Bob's public-pilot beam is
$\mathbf f_B=\mathbf e_p=(0,\ldots,0,1)^{\mathsf T}\in\mathbb C^p$,
whose only nonzero entry is a one in position $p$.
Thus, Bob transmits the public pilot only through his $p$th antenna.
By \eqref{EQ_TDD_channels} and $p>N_R$, $\mathbf h\mathbf f_B$
depends on row $p$ of $\mathbf H$, outside Bob's active receive
set, and is therefore independent of $\mathbf H_B$ and its training.
For Alice-to-Bob transmission, the useful gain $g\in\mathbb C$
and AN leakage row $\boldsymbol\ell\in\mathbb C^{1\times d}$
satisfy, by \eqref{EQ_training} and \eqref{EQ_spatial},
\begin{equation}
\begin{aligned}
g&:=\mathbf u_B^H\mathbf H_B\mathbf f_A
=s+\mathbf u_B^H\boldsymbol\Delta_B\mathbf f_A,\\
\boldsymbol\ell&:=\mathbf u_B^H\mathbf H_B\mathbf B=\mathbf u_B^H\boldsymbol\Delta_B\mathbf B.
\end{aligned}
\label{EQ_return_gain}
\end{equation}
The distributions of terms in \eqref{EQ_TDD_channels} and
\eqref{EQ_training} can be derived as:
\begin{equation}
\mathbf h\sim\CN(\mathbf0,S\Id_p),\quad
g\mid s\sim\CN(s,\tau),\quad
\boldsymbol\ell\sim\CN(\mathbf0,\tau\Id_d).
\label{EQ_effective_laws}
\end{equation}

The distributions of $g-s$ and $\boldsymbol\ell$ follow from
\eqref{EQ_training} and the orthonormal directions in \eqref{EQ_spatial}.
The distribution of $\mathbf h=\mathbf u_A^H\mathbf H_A$ follows from
\eqref{EQ_TDD_channels} and the fixed unit vector $\mathbf u_A$.
It is marginal and does not assert independence from return training.
For later analysis, let $\mathbf h_\perp\in\mathbb C^{1\times d}$
contain the last $d$ entries of $\mathbf h$. Equation~\eqref{EQ_TDD_channels}
makes $\mathbf h_\perp$ independent of $\mathbf H_B$ and its training.

For $U\in\{A,B\}$, Trudy's forged block is
$\mathbf X_{T\to U}\in\mathbb C^{q\times n}$, with unrestricted power.
A replacement contains only her signal and receiver noise. A forged
response may use her entire current challenge observation and independent
private randomness. A forged challenge uses only observations available
when its symbols are sent. Neither can depend on Alice's subsequent
response or Bob's decision. Trudy may process her observation of Bob's current challenge
to construct a forged response, but she cannot use Alice's
current response. Under $T\to A$
alone, Alice forwards the forgery with fresh AN. When Trudy replaces the response ($T\to B$), Bob receives
only her forged signal and receiver noise. This applies both
to response-only replacement and to replacement of both blocks.

\section{Proposed Two-Way Authentication Protocol}
\label{SEC_protocol}
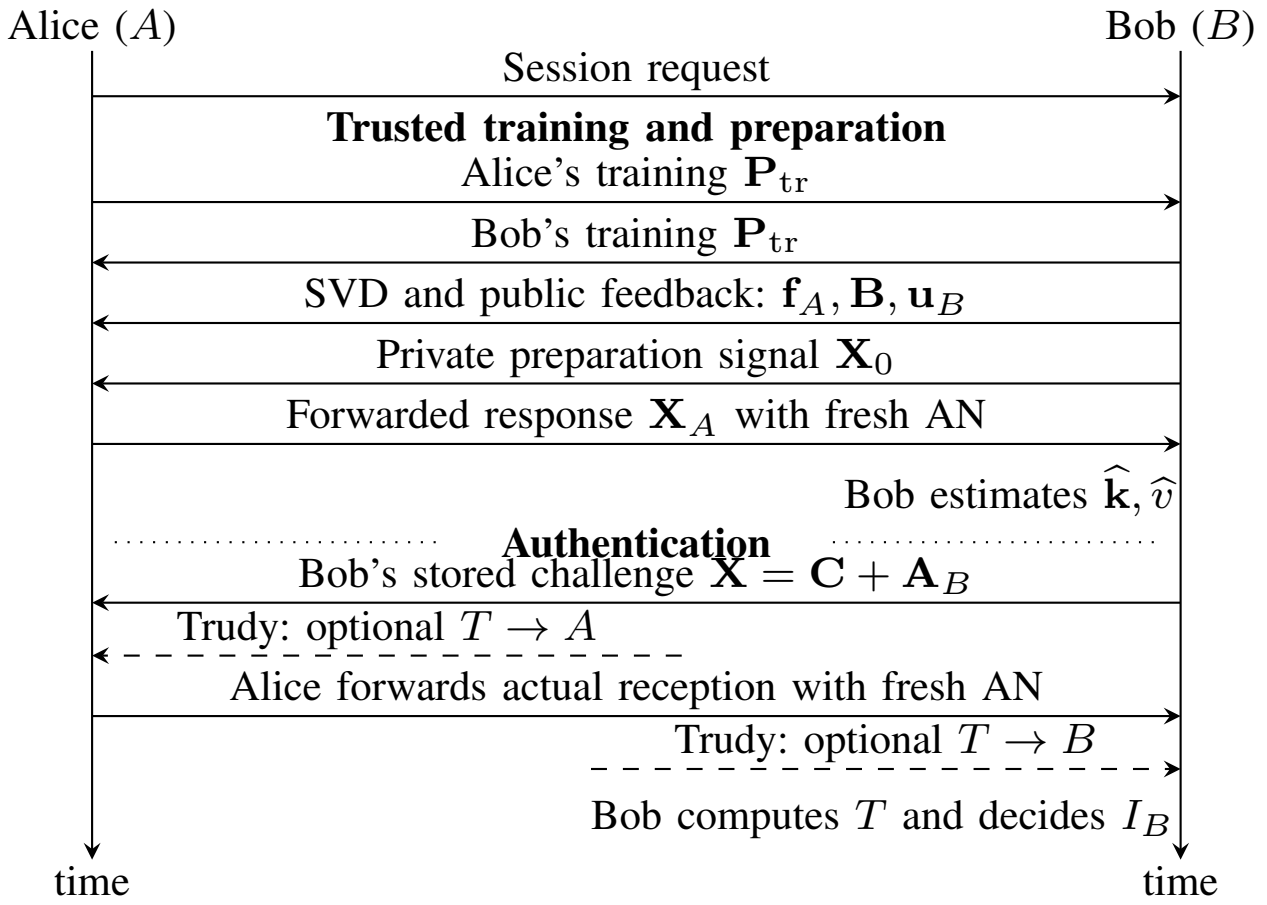
\begin{figure}[!t]
\centering
\begin{tikzpicture}[x=1cm,y=1cm,>=stealth,font=\footnotesize,
every node/.style={align=center,inner sep=1pt}]
\node at (0.3,0.15) {Alice ($A$)};
\node at (7.5,0.15) {Bob ($B$)};
\draw[->] (0.3,0)--(0.3,-5.35) node[below] {time};
\draw[->] (7.5,0)--(7.5,-5.35) node[below] {time};
\draw[->] (0.3,-0.3)--node[above]{Session request}(7.5,-0.3);
\node at (3.9,-0.53) {\textbf{Trusted training and preparation}};
\draw[->] (0.3,-1.0)--node[above]{Alice's training $\mathbf P_{\rm tr}$}(7.5,-1.0);
\draw[->] (7.5,-1.4)--node[above]{Bob's training $\mathbf P_{\rm tr}$}(0.3,-1.4);
\draw[->] (7.5,-1.8)--node[above]{SVD and public feedback: $\mathbf f_A,\mathbf B,\mathbf u_B$}(0.3,-1.8);
\draw[->] (7.5,-2.2)--node[above]{Private preparation signal $\mathbf X_0$}(0.3,-2.2);
\draw[->] (0.3,-2.6)--node[above]{Forwarded response $\mathbf X_A$ with fresh AN}(7.5,-2.6);
\node[anchor=east] at (7.5,-2.91) {Bob estimates $\widehat{\mathbf k},\widehat v$};
\draw[dotted] (0.45,-3.25)--(2.6,-3.25);
\draw[dotted] (5.2,-3.25)--(7.35,-3.25);
\node at (3.9,-3.25) {\textbf{Authentication}};
\draw[->] (7.5,-3.65)--node[above]{Bob's stored challenge $\mathbf X=\mathbf C+\mathbf A_B$}(0.3,-3.65);
\draw[->,dashed] (4.2,-4.0)--node[above]{Trudy: optional $T\to A$}(0.3,-4.0);
\draw[->] (0.3,-4.4)--node[above]{Alice forwards actual reception with fresh AN}(7.5,-4.4);
\draw[->,dashed] (3.6,-4.75)--node[above]{Trudy: optional $T\to B$}(7.5,-4.75);
\node[anchor=east] at (7.5,-5.08) {Bob computes $T$ and decides $I_B$};
\end{tikzpicture}
\caption{Exchange within one coherence interval. Time runs downward.
Training and preparation are assumed attack-free. Dashed arrows mark
Trudy's optional replacements during authentication. Bob alone verifies identity.}
\label{FIG_protocol}
\end{figure}

Trusted preparation supplies Bob's prediction and noise estimates.
The average challenge and response energies per symbol are $P_B,P_A>0$.
The fractions $0<\beta_B<1$ and $0\leq\beta_A<1$ or their energy are allocated to their AN's.
The proposed protocol is summarized in Figure~\ref{FIG_protocol}.

\subsection{Trusted Preparation}
\label{SUBSEC_calibration}
After higher-layer authentication verifies Alice's identity, a \textit{trusted preparation}
provides Bob with estimates of the equivalent channel and noise variance
for checking her subsequent responses. These estimates may be reused
across authentication rounds while the channels remain unchanged,
although the present analysis considers a single round.

Bob constructs and transmits the preparation matrix
$\mathbf X_0\in\mathbb C^{p\times m}$, with $m>p$.
An achievable construction uses a local matrix
$\mathbf W\in\mathbb C^{p\times m}$ with independent $\CN(0,1)$
entries, independent of all channels and receiver noises:
\begin{equation}
\mathbf X_0=\sqrt{E_0}(\mathbf W\mathbf W^H)^{-1/2}\mathbf W,
\qquad \mathbf X_0\mathbf X_0^H=E_0\Id_p,
\label{EQ_private_probe}
\end{equation}
where the preparation energy per transmit antenna is $E_0:=mP_B/p>0$. The matrix inverse square root exists almost surely, which produces a
uniform random orthogonal row frame with total energy $mP_B$. 

Let $L$ denote the block length, equal to $m$ in preparation and
$n$ in authentication. During preparation, $L=m$, and Alice receives
$\mathbf H_A\mathbf X_0+\mathbf Z_A$, where
$\mathbf Z_A\in\mathbb C^{N_R\times L}$ is her receiver noise.
Alice forwards the received preparation signal along $\mathbf f_A$
and adds her spatial AN:
\begin{equation}
\begin{aligned}
\mathbf X_A&=\mathbf f_A\kappa\mathbf u_A^H
(\mathbf H_A\mathbf X_0+\mathbf Z_A)+\mathbf B\mathbf A_A,\,\kappa^2=\frac{(1-\beta_A)P_A}{SP_B+\sigma^2},
\end{aligned}
\label{EQ_response}
\end{equation}
where the scaling
factor $\kappa>0$ sets the honest average response energy and 
$\mathbf A_A\in\mathbb C^{d\times L}$ is AN with independent
$\CN(0,\beta_AP_A/d)$ entries. 
Since $\mathbf X_0$ is independent of $\mathbf H_A$ and satisfies
$\|\mathbf X_0\|_F^2=mP_B$, the scaling factor $\kappa$ in
\eqref{EQ_response} ensures
$\mathbb E_{\mathbf X_A}[\|\mathbf X_A\|_F^2]/m=P_A$.

During preparation, Bob's received signal and receiver noise are
$\mathbf Y_B,\mathbf Z_B\in\mathbb C^{N_R\times m}$, respectively.
He applies the receiver beamformer $\mathbf u_B=\widehat{\mathbf H}_B\mathbf f_A/s$ obtained from the
SVD in \eqref{EQ_spatial}.
Let $\mathbf y,\mathbf e\in\mathbb C^{1\times m}$ denote the
beamformed received signal and total noise, respectively.
Let $\mathbf k\in\mathbb C^{1\times p}$ be the equivalent channel
row and $v_0>0$ the conditional noise variance per entry.
Substituting \eqref{EQ_response} and collecting the signal and noise
terms gives
\begin{align}
\mathbf y&\!=\!\mathbf u_B^H\mathbf Y_B
\!=\!\mathbf u_B^H(\mathbf H_B\mathbf X_A+\mathbf Z_B)\!\overset{\mathrm{(a)}}{=}\!
\kappa(\mathbf u_B^H\mathbf H_B\mathbf f_A)
(\mathbf u_A^H\mathbf H_A)\mathbf X_0\notag\\
&\quad+\kappa(\mathbf u_B^H\mathbf H_B\mathbf f_A)
\mathbf u_A^H\mathbf Z_A+(\mathbf u_B^H\mathbf H_B\mathbf B)\mathbf A_A
+\mathbf u_B^H\mathbf Z_B\notag\\
&\overset{\mathrm{(b)}}{=}\mathbf k\mathbf X_0+\mathbf e,
\qquad \mathbf k:=\kappa g\mathbf h,\notag\\
\mathbf e&:=\kappa g\mathbf u_A^H\mathbf Z_A
+\boldsymbol\ell\mathbf A_A+\mathbf u_B^H\mathbf Z_B,\label{EQ_echo}
\end{align}
where (a) substitutes \eqref{EQ_response} and expands
the products. Step (b) uses
$g=\mathbf u_B^H\mathbf H_B\mathbf f_A$ and
$\boldsymbol\ell=\mathbf u_B^H\mathbf H_B\mathbf B$
from \eqref{EQ_return_gain}, together with
$\mathbf h=\mathbf u_A^H\mathbf H_A$
defined in Section~\ref{SUBSEC_spatial}.
The three terms in $\mathbf e$ are forwarded Alice noise, leaked AN,
and Bob noise. Conditional on the physical channels and initial training,
the entries of $\mathbf e$ are independent $\CN(0,v_0)$, $v_0{=}\kappa^2|g|^2\sigma^2
+\frac{\beta_AP_A}{d}\|\boldsymbol\ell\|^2+\sigma^2$.

To predict Alice's authentication response, Bob estimates $\mathbf k$ and $v_0$
from his recorded $\mathbf X_0$ and preparation observation
$\mathbf y_0:=\mathbf y$ in \eqref{EQ_echo}.
Let $\widehat{\mathbf k}\in\mathbb C^{1\times p}$ denote the
least-squares estimate of $\mathbf k$, and let $\widehat v>0$
denote the residual estimate of $v_0$.
Using $\mathbf X_0\mathbf X_0^H=E_0\Id_p$ and
$\nu:=m-p>0$ complex residual degrees of freedom, Bob computes
\begin{equation}
\widehat{\mathbf k}=\frac{\mathbf y_0\mathbf X_0^H}{E_0},
\qquad
\widehat v=
\frac{\|\mathbf y_0-\widehat{\mathbf k}\mathbf X_0\|^2}{\nu}.
\label{EQ_calibration}
\end{equation}
Conditional on the channels and initial training,
$\widehat{\mathbf k}-\mathbf k$ and $\widehat v$ are independent.
This Gaussian least-squares property gives the false-alarm guarantee
in Proposition~\ref{PROP_FA}.

\subsection{Bob's Authentication Challenge}
\label{SUBSEC_B_transmission}
After preparation, Bob sends a fresh random challenge whose samples
he records for verification. For  block length $n\geq1$, the challenge uses a public pilot row
$\mathbf p^{\mathsf T}\in\mathbb C^{1\times n}$ with
$\|\mathbf p\|^2=n$.
Bob forms $\mathbf C=\sqrt{(1-\beta_B)P_B}\mathbf f_B\mathbf p^{\mathsf T}
\in\mathbb C^{p\times n}$ and an AN matrix $\mathbf A_B\in\mathbb C^{p\times n}$
with independent
$\CN(0,b)$ entries, $b=\beta_BP_B/p$.
Bob transmits and retains
\begin{equation}
\mathbf X=\mathbf C+\mathbf A_B,\qquad
\mathbb E_{\mathbf A_B}[\|\mathbf X\|_F^2]=nP_B.
\label{EQ_challenge}
\end{equation}
Trudy can observe this block before choosing a forged response.
Alice receives $\mathbf Y_{A,1}=\mathbf H_A\mathbf X+\mathbf Z_A\in\mathbb C^{N_R\times n}$ in an honest exchange,
with receiver noise $\mathbf Z_A\in\mathbb C^{N_R\times n}$.
Under challenge replacement ($T\to A$), it instead equals
$\mathbf K_A\mathbf X_{T\to A}+\mathbf Z_A$.

\subsection{Alice's Authentication Response}
\label{SUBSEC_response}
Alice returns the received challenge so that Bob can check it
using his preparation estimates. In \eqref{EQ_response}, she sets
$L=n$ and replaces $\mathbf H_A\mathbf X_0+\mathbf Z_A$ by her actual
reception $\mathbf Y_{A,1}$, including any forged challenge.
She uses fresh spatial AN independent of preparation.
She does not decode or subtract Bob's AN.
Bob then receives $\mathbf H_B\mathbf X_A+\mathbf Z_B$, where
$\mathbf Z_B\in\mathbb C^{N_R\times n}$ is his receiver noise.
Under response replacement ($T\to B$), the signal
$\mathbf H_B\mathbf X_A$ is replaced by
$\mathbf K_B\mathbf X_{T\to B}$.

\subsection{Bob's Verification}
\label{SUBSEC_decision}
Bob compares the received response with his prediction from trusted
preparation to distinguish between
\[
\begin{array}{ll}
\mathcal H_0: & \text{Neither block is replaced},\\
\mathcal H_1: & \text{At least one block is replaced}.
\end{array}
\]
Let $\mathbf y_1\in\mathbb C^{1\times n}$ be Bob's received response
after applying $\mathbf u_B^H$. His prediction residual is
$\mathbf r\in\mathbb C^{1\times n}$:
\begin{equation}
\mathbf r:=\mathbf y_1-\widehat{\mathbf k}\mathbf X,
\qquad \mathbf X=\mathbf C+\mathbf A_B.
\label{EQ_cancellation}
\end{equation}
To account for both preparation-estimation error and current noise,
let $\mathbf e_0\in\mathbb C^{1\times m}$ denote the preparation
noise $\mathbf e$ in \eqref{EQ_echo}, and let
$\mathbf e_1\in\mathbb C^{1\times n}$ contain the same three
noise terms during authentication, using fresh independent
receiver noises and AN. Under $\mathcal H_0$,
$\mathbf r=(\mathbf k-\widehat{\mathbf k})\mathbf X+\mathbf e_1$.
All covariances below are conditional on the channels, initial
training, $\mathbf X_0$, and current $\mathbf X$.
Using \eqref{EQ_echo}, \eqref{EQ_calibration}, and
$\mathbf X_0\mathbf X_0^H=E_0\Id_p$ from \eqref{EQ_private_probe},
we obtain
\begin{align}
\widehat{\mathbf k}-\mathbf k
&=\frac{\mathbf e_0\mathbf X_0^H}{E_0},\quad
\operatorname{Cov}\bigl((\mathbf k-\widehat{\mathbf k})^H\bigr)
=\frac{v_0}{E_0}\Id_p,
\label{EQ_preparation_error_covariance}\\
\operatorname{Cov}(\mathbf r^H)
&=v_0\left(\Id_n+\frac{\mathbf X^H\mathbf X}{E_0}\right)
=:v_0\mathbf V_X.
\label{EQ_residual_covariance}
\end{align}
Let $T\geq0$ denote the residual
statistic and $\eta>0$ its acceptance threshold.
For $\delta\in(0,\alpha)$, let $R_X>0$ be the $(1-\delta)$ quantile
of $\|\mathbf X\|_F^2$. The statistic $T/n$ compares average whitened
residual energy with the preparation noise estimate $\widehat v$.
Then, Bob uses the following detector:
\begin{equation}
\begin{aligned}
I_B&=\ind\{\|\mathbf X\|_F^2\leq R_X,\ T\leq\eta\},\\
T&=\frac{\mathbf r\mathbf V_X^{-1}\mathbf r^H}{\widehat v},
\qquad
\eta=nF_{2n,2\nu}^{-1}(1-\alpha+\delta).
\end{aligned}
\label{EQ_decision}
\end{equation}
Here, $I_B=1$ declares $\mathcal H_0$, and $I_B=0$ declares
$\mathcal H_1$. Bob accepts only when both the residual statistic and the challenge
energy satisfy their thresholds. 
By Proposition~\ref{PROP_FA}, $T/n\sim F_{2n,2\nu}$ under
$\mathcal H_0$, so the threshold $\eta$ gives residual-test
false-alarm probability $\alpha-\delta$.
The challenge-energy condition rejects with probability $\delta$
under $\mathcal H_0$ and ensures
$\mathbf V_X=\Id_n+\frac{\mathbf X^H\mathbf X}{E_0}
\preceq\left(1+\frac{R_X}{E_0}\right)\Id_n
$
on acceptance. This matrix bound is used to upper-bound missed
detection in Theorem~\ref{THM_echo_bound}. The condition
$\|\mathbf X\|_F^2\leq R_X$ is an additional acceptance requirement,
separate from the average transmit-energy constraint in
\eqref{EQ_challenge}.

\section{Security Analysis}
\label{SEC_security}
Given attack-free initial training and preparation, we establish
$P_{\rm FA}\leq\alpha$ for honest authentication and upper-bound
$P_2^\star$ for replacement of the challenge, response, or both.
Comparing this upper bound with a one-way lower bound then identifies
when two-way authentication has lower worst-case missed detection probability.
The proofs of Proposition~\ref{PROP_FA} and
Theorem~\ref{THM_echo_bound} are given in
Appendices~\ref{APP_echo} and~\ref{APP_security}.
Appendix~\ref{APP_oneway} derives the one-way acceptance probability
and the coefficients in Proposition~\ref{PROP_two_way_comparison}.

\subsection{False Alarms with Noisy Cancellation}
We first show that Bob's decision rule in \eqref{EQ_decision}
satisfies the false-alarm constraint $P_{\rm FA}\leq\alpha$
even with noisy preparation and Alice's AN leakage at Bob.

\begin{proposition}
\label{PROP_FA}
For fixed channels and initial training,
under $\mathcal H_0$, the statistic $T$ and thresholds
defined in \eqref{EQ_decision} give
\begin{equation}
T/n\sim F_{2n,2\nu},\qquad P_{\rm FA}\leq\alpha.
\label{EQ_FA_law}
\end{equation}
\end{proposition}

Conditional on the channels and initial training,
$\nu\widehat v/v_0\sim\Gamma(\nu,1)$ and
$\mathbf r\mathbf V_X^{-1}\mathbf r^H/v_0\sim\Gamma(n,1)$
are independent by \eqref{EQ_calibration} and \eqref{EQ_cancellation}.
Their normalized ratio gives $T/n$ in \eqref{EQ_FA_law}.
The union bound applied to \eqref{EQ_decision} gives
$P_{\rm FA}\leq\alpha$.

\subsection{A Bound Including the Intercepted Current Challenge}
Bob knows his transmitted AN, whereas Trudy observes it through noise.
We use Trudy's remaining uncertainty to
upper-bound $P_2^\star$ in \eqref{EQ_objective}.

\begin{theorem}
\label{THM_echo_bound}
Give Trudy all channel matrices and Bob's preparation records
$(\mathbf X_0,\mathbf y_0,\widehat{\mathbf k},\widehat v)$.
Let $c_X:=1+R_X/E_0$, and define the normalized equivalent-channel
strength $\mathcal A:=\kappa^2|g|^2S/v_0$ and its CDF
$F_{\mathcal A}(\theta):=\Pr\{\mathcal A<\theta\}$.
Let $Z_n\sim\Gamma(n,1)$, $Z_\nu\sim\Gamma(\nu,1)$, and
$Z_q\sim\Gamma(q,1)$ be independent auxiliary variables.
For any fixed threshold $0<\theta<S/\sigma^2$, we have
\begin{equation}
\begin{aligned}
P_2^\star\leq\overline P_2
&:=\min\!\biggl\{1,F_{\mathcal A}(\theta)\\
&+\mathbb E_{Z_n,Z_\nu,Z_q}\!\left[G_p\!\left(
\frac{\eta c_X Z_\nu(b^{-1}+S_TZ_q/\sigma_T^2)}
{\nu Z_n(\theta+E_0^{-1})}
\right)\right]\biggr\}.
\end{aligned}
\label{EQ_upper_scalar}
\end{equation}
\end{theorem}
The expectation in \eqref{EQ_upper_scalar}
bounds acceptance for the remaining sessions using Trudy's uncertainty
about Bob's challenge. Choosing $\theta$ tightens $\overline P_2$
without changing Bob's detector.

\begin{remark}
\label{REM_alice_AN}
Theorem~\ref{THM_echo_bound} gives Trudy Bob's preparation records,
so its guarantee does not require Alice's AN to hide those records.
It still accounts for AN leakage and reduced forwarding energy.
For Trudy's actual observations, Proposition~\ref{PROP_information} establishes
an estimation-error lower bound for $\mathbf h_\perp$, defined in Section~\ref{SUBSEC_spatial}, that is
nondecreasing in $\beta_A$ under an additional-disclosure model
(proved in Appendix~\ref{APP_information}). The bound in Theorem~\ref{THM_echo_bound} also holds for
$\beta_A=0$, when Alice transmits no AN.
\end{remark}

\begin{figure*}[!t]

\centering
\includegraphics[width=\textwidth]{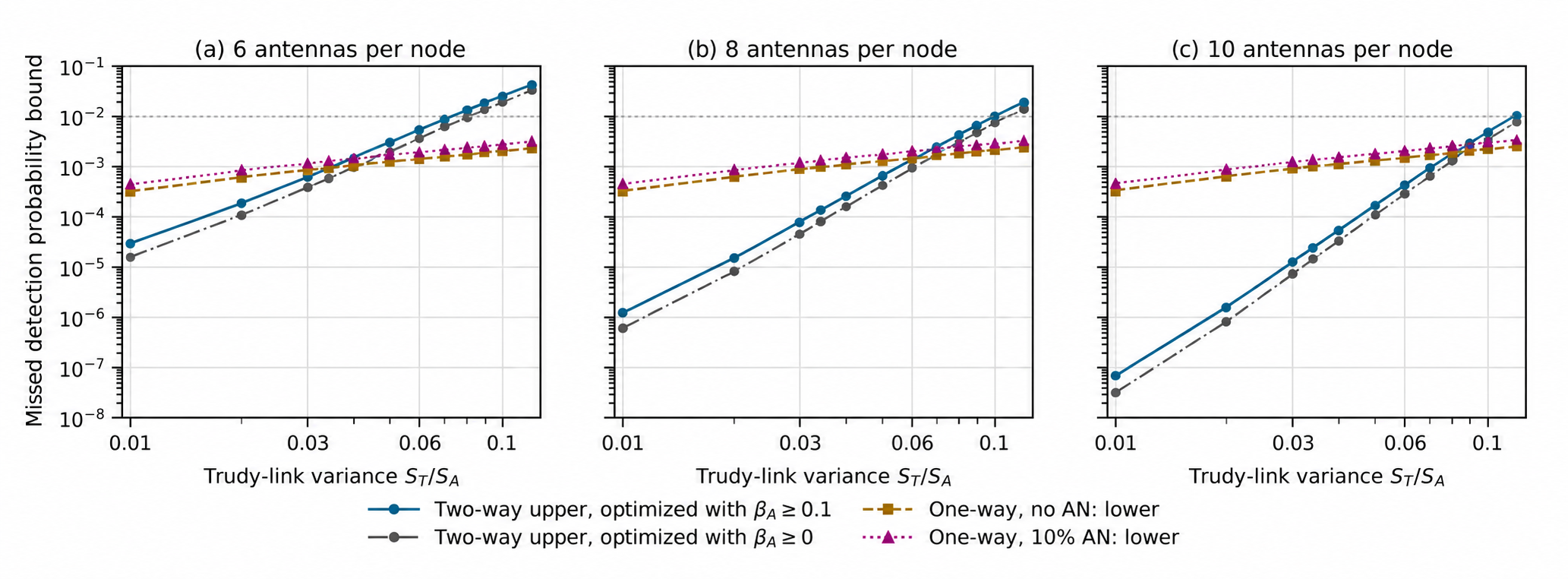}
\caption{Comparison for (a) six, (b) eight, and (c) ten antennas per node. Two-way curves show the upper bound \eqref{EQ_upper_scalar}
after solving \eqref{EQ_AN_allocation}. One-way curves show the exact success probability $P_{\mathrm{MD},1}$ of the feasible attack in
Section~\ref{SUBSEC_oneway}, computed using \eqref{EQ_one_lower}.
The horizontal axes use $S_A=S$. The dotted line marks $0.01$.}
\label{FIG_results}
\end{figure*}

\subsection{Comparison with One-Way Authentication}
\label{SUBSEC_oneway}

We adopt a one-way baseline based on conventional channel-consistency
PLA, which compares a current channel estimate with a trusted reference
using a variance-normalized squared difference
\cite{Xiao2008,Senigagliesi2021}. We specialize this principle to the
quasi-static scalar channel obtained after transmit beamforming and
receive combining. To match the two-way resource constraint, the one-way
scheme uses the same spatial training, $2m$ preparation symbols, and
$2n$ authentication symbols, each with energy $P_1=(P_A+P_B)/2$.
Alice transmits a unit-modulus public pilot along $\mathbf f_A$ and
spatial AN along $\mathbf B$, with AN fraction $\beta\in[0,1)$,
which need not equal $\beta_A$ in the two-way scheme.
The case $\beta=0$ gives the channel-consistency baseline without AN,
while $\beta>0$ adds a spatial-AN extension.
Bob applies $\mathbf u_B^H$, then divides each combined sample by the
known pilot symbol multiplied by $\sqrt{(1-\beta)P_1}$.

Let $\widehat g_0,\widehat g_1\in\mathbb C$ be the preparation and
current mean gain estimates, and let $\widehat v_1>0$ be the 
sample variance of the normalized preparation observations.
Bob uses $\widehat v_1$ to account for noise in the gain difference
$\widehat g_1-\widehat g_0$. Bob accepts if
\begin{equation}
\frac{|\widehat g_1-\widehat g_0|^2}
{\widehat v_1(1/(2n)+1/(2m))}
\leq F_{2,4m-2}^{-1}(1-\alpha).
\label{EQ_one_threshold}
\end{equation}
Under $\mathcal H_0$, the normalized squared mean difference in
\eqref{EQ_one_threshold} follows $F_{2,4m-2}$.
Choosing its $(1-\alpha)$ quantile as the threshold gives
false-alarm probability $\alpha$.

We compare the two-way upper bound $\overline P_2$ in
\eqref{EQ_upper_scalar} with the acceptance probability of a
feasible attack against \eqref{EQ_one_threshold}.
Let $\widehat{\mathbf h}_T\in\mathbb C^q$ be Trudy's minimum
mean-square error estimate of $(\mathbf u_B^H\mathbf K_B)^{\mathsf T}$
from initial training. In the replaced authentication block,
Bob receives Trudy's forged pilot and receiver noise, with no
contribution from Alice's signal.
Trudy uses a constant $a>0$, chosen before the channel realization,
as a guess for Bob's private preparation estimate $\widehat g_0$.
She sends the public pilot with beam
$\sqrt{(1-\beta)P_1}\,a\widehat{\mathbf h}_T^*/
\|\widehat{\mathbf h}_T\|^2$,
giving $\mathbb E_{\widehat g_1\mid\widehat{\mathbf h}_T}[\widehat g_1]=a$.

Let $\rho:=S_T/S>0$ measure Trudy's link strength, and vary $S_T$ alone, keeping all other parameters, including $\beta_A,\beta_B,\beta,a$
and $\theta$, fixed. For the one-way test with fixed AN fraction $\beta$, denote the missed detection probability for the specified
attack by $P_{\mathrm{MD},1}(\rho)$ and its supremum over all admissible attacks by $P_{\mathrm{MD},1}^{\star}(\rho)$. We seek a condition on $\rho$ ensuring that the proposed two-way
test has lower worst-case missed-detection probability than the
one-way test in \eqref{EQ_one_threshold}.

\begin{proposition}
\label{PROP_two_way_comparison}
Assume $n>p$ and fix $R>0$. There exist constants $A_0,A_1,A_2,c>0$, specified explicitly in \eqref{EQ_explicit_quadratic_coefficients}
and \eqref{EQ_explicit_one_way_linear} in Appendix~\ref{APP_comparison_proof}, which depend only on public statistics and on parameters that change neither detector, such that for $0<\rho\leq R$
\begin{equation}
\begin{aligned}
\overline P_2(\rho)&\leq A_0+A_1\rho+A_2\rho^2,\\
P_{\mathrm{MD},1}(\rho)&\geq c\rho.
\end{aligned}
\label{EQ_explicit_envelopes}
\end{equation}
Hence, if $c>A_1+2\sqrt{A_0A_2}$, then
$P_2^\star(\rho)<P_{\mathrm{MD},1}^{\star}(\rho)$ for
$\rho_-<\rho<\rho_+$ and $\rho\leq R$, where
\begin{equation}
\rho_\pm:=\frac{c-A_1\pm\sqrt{(c-A_1)^2-4A_0A_2}}{2A_2}.
\label{EQ_explicit_roots}
\end{equation}
\end{proposition}

\begin{remark}
Proposition~\ref{PROP_two_way_comparison} is a sufficient condition, so its interval is conservative.
Direct evaluation of $\overline P_2$ and $P_{\mathrm{MD},1}$ in Section~\ref{SEC_numerical} establishes the advantage over a wider range. 
\end{remark}

\subsection{Optimizing the AN Energy Allocation}
\label{SUBSEC_AN_allocation}
We choose AN power allocation from public statistics before training,
accounting for Trudy's uncertainty, useful response energy, and AN
leakage. Let $\beta_{A,\min}\in[0,1)$ be Alice's minimum AN fraction.
We then formulate the following optmization problem:
\begin{align}
\underset{\beta_A,\beta_B,\theta}{\operatorname{minimize}}
&\,\, \overline P_2(\beta_A,\beta_B;\theta)\notag\\
\text{subject to}
&\,\, \beta_{A,\min}\!\leq\!\beta_A\!<\!1,\, 0\!<\!\beta_B\!< \!1,\,
0\!<\!\theta\!<\!S/\sigma^2.\label{EQ_AN_allocation}
\end{align}

For fixed $\beta_B$ and $\theta$, increasing $\beta_A$ makes
$F_{\mathcal A}(\theta)$ nondecreasing, while the other term in
\eqref{EQ_upper_scalar} remains unchanged. Therefore,
$\beta_A=\beta_{A,\min}$ minimizes $\overline P_2$, although it
need not minimize the actual missed-detection probability.
A positive $\beta_{A,\min}$ retains Alice's AN for the preparation
concealment discussed in Remark~\ref{REM_alice_AN}.
With $\beta_A=\beta_{A,\min}$, we perform a two-dimensional
numerical search over $(\beta_B,\theta)$ because the remaining
objective is non-convex. Every feasible pair yields a valid
upper bound, even if the search does not find the global minimum.

\section{Numerical Results}
\label{SEC_numerical}
We evaluate the analytical probabilities at the following setting: $p=q\in\{6,8,10\}$,
$N_R=2$, $S=P_A=P_B=1$,
$\sigma^2=\sigma_T^2=0.01$, $E_{\rm seed}=0.5$, $m=1024$,
$n=16$, $\alpha=0.05$, and $\delta=0.001$.
For $S_T/S\in[0.01,0.12]$, we consider
$0.01\leq\beta_B\leq0.99$ and
$\beta_A=\beta_{A,\min}\in\{0,0.1\}$. Both tests obey the false-alarm limit $0.05$. The one-way value is
$0.05$, and the two-way value is
$1-(1-\delta)(1-\alpha+\delta)=0.0499$.

For $p=q=6$, $\beta_A=0.1$, $\beta_B=0.99$, $a=2.8$,
$\theta=40$, and $R=0.005$, the coefficient construction gives
$A_0<3\times10^{-6},\, A_1<9.1\times10^{-4},\,
A_2<0.67,\, c>0.0117$
for each $\beta\in\{0,0.1\}$, using the parameters in
Appendix~\ref{APP_example_parameters}.
The difference between the two-way upper bound and the one-way
lower bound in \eqref{EQ_explicit_envelopes} is bounded above by $3\times10^{-6}-0.0108\rho+0.67\rho^2$.
This convex polynomial is negative at both endpoints, so
Proposition~\ref{PROP_two_way_comparison} certifies the entire interval
\begin{equation}
3\times10^{-4}\leq S_T/S\leq5\times10^{-3}.
\label{EQ_explicit_example_range}
\end{equation}
Figure~\ref{FIG_results} shows a two-way advantage at larger
channel-variance ratios $S_T/S$ as $p=q$ increases.
For the one-way attacks, Trudy's fixed guesses for Bob's private
preparation gain estimate $\widehat g_0$ are $a=2.8,3.2,3.5$
for $p=q=6,8,10$, respectively. We have the following three main observations:
1) A strict two-way advantage exists when both schemes use the same total transmit energy and blocklength, $P_{\rm FA}\leq0.05$, and have equal antenna counts at all nodes. At $p=q=6$ and $S_T/S=0.034$, the two-way upper bound $\overline P_2=9.15\times10^{-4}$ for $\beta_A=0.1$ is below the one-way lower bounds $9.54\times10^{-4}$ and $1.27\times10^{-3}$ for $\beta=0$ and
$0.1$, respectively. Thus, $P_2^\star<P_{\mathrm{MD},1}^{\star}$
for both specified one-way tests, with/without AN. Since $\overline P_2$ upper-bounds and $P_{\mathrm{MD},1}$ lower-bounds the respective worst cases, the actual gap is at least the reported one, and is larger than the reported numerical convergence changes
in Appendix~\ref{APP_numerical_checks}. 
2) We also observe that the advantage of the two-way PLA still exists when Bob in the one-way scheme has perfect knowledge of the
current channel and noise variance. The corresponding test and
feasible-attack calculation are specified in Appendix~\ref{APP_perfect_benchmark}. At the same setting, both perfect-CSI lower bounds exceed $\overline P_2$, with the
smaller equal to $9.39\times10^{-4}$. Hence, the proved advantage is not solely attributable to estimation error at the one-way verifier. 
3) In the evaluated settings, increasing $p=q$ from 6 to 10
lowers the two-way upper bound at fixed $S_T/S$ and
establishes an advantage over the specified one-way tests
at larger channel-variance ratios.
The evaluated ratios $S_T/S=0.034,0.060,0.081$ certify an advantage
over both noisy-preparation one-way tests for $p=q=6,8,10$,
respectively.
At $S_T/S=0.01$, the plotted gap grows from about one order of
magnitude at six antennas to more than three at ten.

\section{Conclusion}
We investigated Maurer-inspired two-way physical-layer authentication
against a multi-antenna impersonator. Bob uses trusted preparation and
recorded AN to predict Alice's forwarded response, while Alice's spatial
AN obscures Trudy's preparation observations. For quasi-static Gaussian
multiple-input multiple-output channels with TDD reciprocity, we derived
missed-detection bounds for attacks replacing the authentication
challenge, response, or both. The bounds assume attack-free initial
training and preparation and retain estimation error and AN leakage.
With equal antenna counts at all nodes, these bounds establish lower
worst-case missed detection than the specified one-way tests with and
without AN under the stated channel conditions.

\clearpage
\appendices

\section{Equivalent Channel and Honest Calibration}
\label{APP_echo}
We derive the honest response and use its noise covariance to
calibrate Bob's false-alarm probability in \eqref{EQ_decision}.
\symdef{notation}
For the calculations below, $\Id_k$ is the $k\times k$ identity matrix
and $\ind\{\cdot\}$ is an event indicator. The distribution
$\CN(\boldsymbol\mu,\mathbf K)$ is proper complex Gaussian with mean
$\boldsymbol\mu$ and covariance $\mathbf K$. For a row $\mathbf z$,
we specify the covariance of $\mathbf z^H$.
The distribution $\Gamma(k,1)$ has shape $k>0$ and unit scale,
with CDF
\begin{equation}
G_k(x):=\frac1{\Gamma(k)}\int_0^x v^{k-1}e^{-v}\,\mathrm dv,
\qquad x\geq0,
\label{EQ_appendix_gamma_cdf}
\end{equation}
where $\Gamma(\cdot)$ is Euler's Gamma function.
The F distribution $F_{r,t}$ has $r,t>0$ real degrees of freedom,
and $F_{r,t}^{-1}(u)$ is its $u$-quantile for $0<u<1$.

\subsection{Projection, Energy, and Leakage}
Bob predicts the honest response using the equivalent channel $\mathbf k$
in \eqref{EQ_echo}. To justify that prediction, we first determine
the channel projections in $\mathbf k=\kappa g\mathbf h$, then
derive the response noise and verify Alice's energy budget.
\symdef{block}
For block length $L$, let ${\symref{block}{\mathbf X_{\rm in}}}\in\mathbb C^{p\times L}$
be Bob's honest input: ${\hyperref[EQ_private_probe]{\mathbf X_0}}$ with $L=m$ in preparation,
or ${\hyperref[EQ_challenge]{\mathbf X}}$ with $L=n$ in authentication.
Let ${\symref{block}{\mathbf Z_A}},{\symref{block}{\mathbf Z_B}}\in\mathbb C^{N_R\times L}$ be
Alice's and Bob's receiver-noise blocks.
\begingroup
\allowdisplaybreaks[4]
\begin{proof}[Derivation of \eqref{EQ_effective_laws} and \eqref{EQ_echo}]
\emph{Step 1: derive the channel gains and AN leakage.}
Since Bob chooses ${\hyperref[EQ_spatial]{\mathbf u_B}}$ from the singular value decomposition
of ${\hyperref[EQ_training]{\widehat{\mathbf H}_B}}$ in \eqref{EQ_spatial}, we condition on
${\hyperref[EQ_training]{\widehat{\mathbf H}_B}}$ to fix $s,{\hyperref[EQ_spatial]{\mathbf u_B}},{\hyperref[EQ_spatial]{\mathbf f_A}},{\hyperref[EQ_spatial]{\mathbf B}}$.
The independent error ${\hyperref[EQ_training]{\boldsymbol\Delta_B}}$ retains independent
$\CN(0,{\hyperref[EQ_training]{\tau}})$ entries by \eqref{EQ_training}.
Starting from $g$ in \eqref{EQ_return_gain} and ${\hyperref[EQ_projection_proof2]{\boldsymbol\ell}}$
in \eqref{EQ_return_gain}, we obtain
\begin{align}
&g={\hyperref[EQ_spatial]{\mathbf u_B}}^H{\hyperref[EQ_TDD_channels]{\mathbf H_B}}{\hyperref[EQ_spatial]{\mathbf f_A}}
\overset{\mathrm{(a)}}{=}{\hyperref[EQ_spatial]{\mathbf u_B}}^H
({\hyperref[EQ_training]{\widehat{\mathbf H}_B}}+{\hyperref[EQ_training]{\boldsymbol\Delta_B}}){\hyperref[EQ_spatial]{\mathbf f_A}}
\overset{\mathrm{(b)}}{=}s+{\hyperref[EQ_spatial]{\mathbf u_B}}^H{\hyperref[EQ_training]{\boldsymbol\Delta_B}}{\hyperref[EQ_spatial]{\mathbf f_A}},
\label{EQ_projection_proof1}\\
&{\hyperref[EQ_projection_proof2]{\boldsymbol\ell}}={\hyperref[EQ_spatial]{\mathbf u_B}}^H{\hyperref[EQ_TDD_channels]{\mathbf H_B}}{\hyperref[EQ_spatial]{\mathbf B}}
\overset{\mathrm{(c)}}{=}{\hyperref[EQ_spatial]{\mathbf u_B}}^H
({\hyperref[EQ_training]{\widehat{\mathbf H}_B}}+{\hyperref[EQ_training]{\boldsymbol\Delta_B}}){\hyperref[EQ_spatial]{\mathbf B}}
\overset{\mathrm{(d)}}{=}{\hyperref[EQ_spatial]{\mathbf u_B}}^H{\hyperref[EQ_training]{\boldsymbol\Delta_B}}{\hyperref[EQ_spatial]{\mathbf B}}.
\label{EQ_projection_proof2}
\end{align}
In \eqref{EQ_projection_proof1}--\eqref{EQ_projection_proof2},
(a) and (c) substitute \eqref{EQ_training}.
Step (b) uses ${\hyperref[EQ_training]{\widehat{\mathbf H}_B}}{\hyperref[EQ_spatial]{\mathbf f_A}}=s{\hyperref[EQ_spatial]{\mathbf u_B}}$
and $\|{\hyperref[EQ_spatial]{\mathbf u_B}}\|=1$, whereas (d) uses
${\hyperref[EQ_training]{\widehat{\mathbf H}_B}}{\hyperref[EQ_spatial]{\mathbf B}}=\mathbf0$ from \eqref{EQ_spatial}.
Thus the true channel leaves AN leakage through
${\hyperref[EQ_training]{\boldsymbol\Delta_B}}$ even though the estimated channel nulls it.

\symdef{projectederror}
Both errors depend on the channel-error row after receive combining,
${\symref{projectederror}{\mathbf z}}\in\mathbb C^{1\times p}$.
For $1\leq i\leq N_R$ and $1\leq j\leq p$, let $u_{B,i}$ and
${\hyperref[EQ_training]{\Delta_{B,ij}}}$ denote the corresponding entries of ${\hyperref[EQ_spatial]{\mathbf u_B}}$
and ${\hyperref[EQ_training]{\boldsymbol\Delta_B}}$, and let $z_j$ be entry $j$ of ${\symref{projectederror}{\mathbf z}}$.
To determine its Gaussian distribution, compute its mean and covariance:
\begin{align}
{\symref{projectederror}{\mathbf z}}&:={\hyperref[EQ_spatial]{\mathbf u_B}}^H{\hyperref[EQ_training]{\boldsymbol\Delta_B}},
\label{EQ_combined_error_definition}\\
z_j&=\sum_{i=1}^{N_R}u_{B,i}({\hyperref[EQ_training]{\widehat{\mathbf H}_B}})^*{\hyperref[EQ_training]{\Delta_{B,ij}}},
\label{EQ_combined_error_entry}\\
\mathbb E_{{\hyperref[EQ_training]{\boldsymbol\Delta_B}}\mid{\hyperref[EQ_training]{\widehat{\mathbf H}_B}}}[z_j]
&\overset{\mathrm{(a)}}{=}\sum_{i=1}^{N_R}u_{B,i}^*
\mathbb E_{{\hyperref[EQ_training]{\Delta_{B,ij}}}}[{\hyperref[EQ_training]{\Delta_{B,ij}}}]=0,\notag\\
\mathbb E_{{\hyperref[EQ_training]{\boldsymbol\Delta_B}}\mid{\hyperref[EQ_training]{\widehat{\mathbf H}_B}}}[|z_j|^2]
&\overset{\mathrm{(b)}}{=}{\hyperref[EQ_training]{\tau}}\sum_{i=1}^{N_R}|u_{B,i}|^2
\overset{\mathrm{(c)}}{=}{\hyperref[EQ_training]{\tau}},
\label{EQ_combined_error_variance}\\
{\symref{projectederror}{\mathbf z}}^H\mid{\hyperref[EQ_training]{\widehat{\mathbf H}_B}}
&\overset{\mathrm{(d)}}{\sim}\CN(\mathbf0,{\hyperref[EQ_training]{\tau}}\Id_p),\notag\\
\mathbb E_{{\symref{projectederror}{\mathbf z}}\mid{\hyperref[EQ_training]{\widehat{\mathbf H}_B}}}[{\symref{projectederror}{\mathbf z}}^H{\symref{projectederror}{\mathbf z}}]
&={\hyperref[EQ_training]{\tau}}\Id_p.
\label{EQ_combined_error_law}
\end{align}
In \eqref{EQ_combined_error_definition}--\eqref{EQ_combined_error_law},
(a) uses the independence of ${\hyperref[EQ_training]{\boldsymbol\Delta_B}}$ and
${\hyperref[EQ_training]{\widehat{\mathbf H}_B}}$ and the zero means in \eqref{EQ_training}.
Step (b) uses independent error entries of variance ${\hyperref[EQ_training]{\tau}}$, and
(c) uses $\|{\hyperref[EQ_spatial]{\mathbf u_B}}\|=1$. In (d), each $z_j$ is a linear
combination of proper Gaussian entries, and different $j$ use
independent columns of ${\hyperref[EQ_training]{\boldsymbol\Delta_B}}$.

The relations $g-s={\symref{projectederror}{\mathbf z}}{\hyperref[EQ_spatial]{\mathbf f_A}}$ and
${\hyperref[EQ_projection_proof2]{\boldsymbol\ell}}={\symref{projectederror}{\mathbf z}}{\hyperref[EQ_spatial]{\mathbf B}}$ from
\eqref{EQ_projection_proof1}--\eqref{EQ_combined_error_definition}
now give their variances and mutual dependence in one calculation:
\begin{align}
&\mathbb E_{g\mid{\hyperref[EQ_training]{\widehat{\mathbf H}_B}}}[|g-s|^2]
\overset{\mathrm{(a)}}{=}{\hyperref[EQ_spatial]{\mathbf f_A}}^H
\mathbb E_{{\symref{projectederror}{\mathbf z}}\mid{\hyperref[EQ_training]{\widehat{\mathbf H}_B}}}[{\symref{projectederror}{\mathbf z}}^H{\symref{projectederror}{\mathbf z}}]{\hyperref[EQ_spatial]{\mathbf f_A}}
={\hyperref[EQ_training]{\tau}}{\hyperref[EQ_spatial]{\mathbf f_A}}^H{\hyperref[EQ_spatial]{\mathbf f_A}}={\hyperref[EQ_training]{\tau}},
\label{EQ_projected_gain_variance}\\
&\operatorname{Cov}({\hyperref[EQ_projection_proof2]{\boldsymbol\ell}}^H\mid{\hyperref[EQ_training]{\widehat{\mathbf H}_B}})
\overset{\mathrm{(b)}}{=}{\hyperref[EQ_spatial]{\mathbf B}}^H
\mathbb E_{{\symref{projectederror}{\mathbf z}}\mid{\hyperref[EQ_training]{\widehat{\mathbf H}_B}}}[{\symref{projectederror}{\mathbf z}}^H{\symref{projectederror}{\mathbf z}}]{\hyperref[EQ_spatial]{\mathbf B}}
\notag\\
&\qquad={\hyperref[EQ_training]{\tau}}{\hyperref[EQ_spatial]{\mathbf B}}^H{\hyperref[EQ_spatial]{\mathbf B}}={\hyperref[EQ_training]{\tau}}\Id_d,
\label{EQ_projected_leakage_covariance}\\
&\mathbb E_{g,{\hyperref[EQ_projection_proof2]{\boldsymbol\ell}}\mid{\hyperref[EQ_training]{\widehat{\mathbf H}_B}}}
[(g-s)^*{\hyperref[EQ_projection_proof2]{\boldsymbol\ell}}]
\overset{\mathrm{(c)}}{=}{\hyperref[EQ_spatial]{\mathbf f_A}}^H
\mathbb E_{{\symref{projectederror}{\mathbf z}}\mid{\hyperref[EQ_training]{\widehat{\mathbf H}_B}}}[{\symref{projectederror}{\mathbf z}}^H{\symref{projectederror}{\mathbf z}}]{\hyperref[EQ_spatial]{\mathbf B}}
={\hyperref[EQ_training]{\tau}}{\hyperref[EQ_spatial]{\mathbf f_A}}^H{\hyperref[EQ_spatial]{\mathbf B}}=\mathbf0,
\label{EQ_projection_crosscov}\\
&(g-s)\mid{\hyperref[EQ_training]{\widehat{\mathbf H}_B}}\overset{\mathrm{(d)}}{\sim}\CN(0,{\hyperref[EQ_training]{\tau}}),\quad
{\hyperref[EQ_projection_proof2]{\boldsymbol\ell}}^H\mid{\hyperref[EQ_training]{\widehat{\mathbf H}_B}}
\overset{\mathrm{(e)}}{\sim}\CN(\mathbf0,{\hyperref[EQ_training]{\tau}}\Id_d).
\label{EQ_projected_error_laws}
\end{align}
In \eqref{EQ_projected_gain_variance}--\eqref{EQ_projected_error_laws},
(a)--(c) substitute $g-s={\symref{projectederror}{\mathbf z}}{\hyperref[EQ_spatial]{\mathbf f_A}}$ and
${\hyperref[EQ_projection_proof2]{\boldsymbol\ell}}={\symref{projectederror}{\mathbf z}}{\hyperref[EQ_spatial]{\mathbf B}}$, then use
\eqref{EQ_combined_error_law} and the orthogonality in \eqref{EQ_spatial}.
Steps (d) and (e) use the zero means and Gaussianity of these
linear projections. Their joint proper Gaussian distribution and
zero cross-covariance imply conditional independence.
To obtain the distribution given only $s$ and determine whether
$\boldsymbol\ell$ depends on $(s,g)$, average this conditional
product over estimates with singular value $s=t>0$.
For density arguments $\gamma\in\mathbb C$ and $\mathbf v\in\mathbb C^d$,
\begin{align}
&f_{g,\boldsymbol\ell^H\mid s}(\gamma,\mathbf v\mid t)\notag\\
&\quad\overset{\mathrm{(a)}}{=}
\mathbb E_{\widehat{\mathbf H}_B\mid s=t}
[f_{g,\boldsymbol\ell^H\mid\widehat{\mathbf H}_B}
(\gamma,\mathbf v\mid\widehat{\mathbf H}_B)]\notag\\
&\quad\overset{\mathrm{(b)}}{=}
\mathbb E_{\widehat{\mathbf H}_B\mid s=t}
\left[\frac{e^{-(|\gamma-s|^2+\|\mathbf v\|^2)/\tau}}
{(\pi\tau)^{d+1}}\right]\notag\\
&\quad\overset{\mathrm{(c)}}{=}
\frac{e^{-|\gamma-t|^2/\tau}}{\pi\tau}
\frac{e^{-\|\mathbf v\|^2/\tau}}{(\pi\tau)^d}.
\label{EQ_return_gain_conditional_density}
\end{align}
In \eqref{EQ_return_gain_conditional_density}, (a) uses conditional
total probability and $s=\sigma_{\max}(\widehat{\mathbf H}_B)$.
Step (b) substitutes the independent Gaussian distributions in
\eqref{EQ_projected_error_laws}, and (c) uses $s=t$.
The first factor gives $g\mid s\sim\CN(s,\tau)$.
The second factor does not depend on $t$ or $\gamma$, so
$\boldsymbol\ell$ is independent of $(s,g)$.
Also, the $d$ independent squared magnitudes in
$\|{\hyperref[EQ_projection_proof2]{\boldsymbol\ell}}\|^2/{\hyperref[EQ_training]{\tau}}$ are unit-mean exponentials, so
$\|{\hyperref[EQ_projection_proof2]{\boldsymbol\ell}}\|^2/{\hyperref[EQ_training]{\tau}}\sim\Gamma(d,1)$, with shape $d$ and unit scale.
Finally, the same Gaussian projection calculation applied to
${\hyperref[SUBSEC_spatial]{\mathbf h}}={\hyperref[SUBSEC_spatial]{\mathbf u_A}}^H{\hyperref[EQ_TDD_channels]{\mathbf H_A}}$ in Section~\ref{SUBSEC_spatial}
uses the independent columns in \eqref{EQ_TDD_channels} and
$\|{\hyperref[SUBSEC_spatial]{\mathbf u_A}}\|=1$ to give independent $\CN(0,S)$ entries.
This establishes \eqref{EQ_effective_laws}.
The distribution of ${\hyperref[SUBSEC_spatial]{\mathbf h}}$ is marginal and does not imply
independence from return training.

\emph{Step 2: derive the honest response and total noise.}
Let ${\hyperref[EQ_echo]{\mathbf y}},\mathbf e\in\mathbb C^{1\times L}$ be Bob's
combined response and noise. Substituting ${\symref{block}{\mathbf X_{\rm in}}}$ into
\eqref{EQ_response}, as also used in Section~\ref{SUBSEC_response}, gives
\begin{align}
{\hyperref[EQ_echo]{\mathbf y}}&={\hyperref[EQ_spatial]{\mathbf u_B}}^H{\hyperref[EQ_TDD_channels]{\mathbf H_B}}{\hyperref[EQ_response]{\mathbf X_A}}+{\hyperref[EQ_spatial]{\mathbf u_B}}^H{\symref{block}{\mathbf Z_B}}
\notag\\
&\overset{\mathrm{(a)}}{=}{\hyperref[EQ_spatial]{\mathbf u_B}}^H{\hyperref[EQ_TDD_channels]{\mathbf H_B}}
\bigl[{\hyperref[EQ_spatial]{\mathbf f_A}}{\hyperref[EQ_response]{\kappa}}{\hyperref[SUBSEC_spatial]{\mathbf u_A}}^H
({\hyperref[EQ_TDD_channels]{\mathbf H_A}}{\symref{block}{\mathbf X_{\rm in}}}+{\symref{block}{\mathbf Z_A}})
+{\hyperref[EQ_spatial]{\mathbf B}}{\hyperref[EQ_response]{\mathbf A_A}}\bigr]+{\hyperref[EQ_spatial]{\mathbf u_B}}^H{\symref{block}{\mathbf Z_B}}\notag\\
&\overset{\mathrm{(b)}}{=}{\hyperref[EQ_response]{\kappa}} g{\hyperref[SUBSEC_spatial]{\mathbf h}}{\symref{block}{\mathbf X_{\rm in}}}
+{\hyperref[EQ_response]{\kappa}} g{\hyperref[SUBSEC_spatial]{\mathbf u_A}}^H{\symref{block}{\mathbf Z_A}}
+{\hyperref[EQ_projection_proof2]{\boldsymbol\ell}}{\hyperref[EQ_response]{\mathbf A_A}}+{\hyperref[EQ_spatial]{\mathbf u_B}}^H{\symref{block}{\mathbf Z_B}}.
\label{EQ_echo_proof}
\end{align}
In \eqref{EQ_echo_proof}, (a) substitutes \eqref{EQ_response}
with the appropriate input block. Step (b) uses $g$ in
\eqref{EQ_return_gain}, ${\hyperref[SUBSEC_spatial]{\mathbf h}}$ in Section~\ref{SUBSEC_spatial},
and ${\hyperref[EQ_projection_proof2]{\boldsymbol\ell}}$ in \eqref{EQ_projection_proof2}.
Thus ${\hyperref[EQ_echo]{\mathbf y}}={\hyperref[EQ_echo]{\mathbf k}}{\symref{block}{\mathbf X_{\rm in}}}+\mathbf e$ with
${\hyperref[EQ_echo]{\mathbf k}},\mathbf e$ from \eqref{EQ_echo}.

\symdef{condition}
Let ${\symref{condition}{\mathcal C_0}}$ collect the physical channels, initial training,
and ${\hyperref[EQ_private_probe]{\mathbf X_0}}$, fixing the channel projections but leaving
receiver noises and Alice's AN random. Then
\begin{align}
\operatorname{Cov}(\mathbf e^H\mid{\symref{condition}{\mathcal C_0}})
&\overset{\mathrm{(a)}}{=}
\left({\hyperref[EQ_response]{\kappa}}^2|g|^2{\hyperref[SUBSEC_channels]{\sigma^2}}+
\frac{\beta_AP_A}{d}\|{\hyperref[EQ_projection_proof2]{\boldsymbol\ell}}\|^2+{\hyperref[SUBSEC_channels]{\sigma^2}}\right)\Id_L
\notag\\
&=v_0\Id_L,\qquad
\mathbf e^H\mid{\symref{condition}{\mathcal C_0}}\overset{\mathrm{(b)}}{\sim}\CN(\mathbf0,v_0\Id_L).
\label{EQ_response_noise_covariance}
\end{align}
In \eqref{EQ_response_noise_covariance}, (a) adds the covariances
of the independent forwarded Alice noise, AN leakage, and Bob noise
in \eqref{EQ_echo_proof}, using \eqref{EQ_response} and the unit
norms of ${\hyperref[SUBSEC_spatial]{\mathbf u_A}},{\hyperref[EQ_spatial]{\mathbf u_B}}$. Step (b) uses their proper Gaussian
distributions and independence across symbols.
Thus the same $v_0$ applies to honest preparation and authentication.

\emph{Step 3: verify Alice's average response energy.}
To check the energy budget, average over the honest input and noises:
\begin{align}
\frac{\mathbb E_{{\hyperref[EQ_response]{\mathbf X_A}}}[\|{\hyperref[EQ_response]{\mathbf X_A}}\|_F^2]}{L}
&\overset{\mathrm{(a)}}{=}\frac{{\hyperref[EQ_response]{\kappa}}^2}{L}
\mathbb E_{{\hyperref[SUBSEC_spatial]{\mathbf h}},{\symref{block}{\mathbf X_{\rm in}}},{\symref{block}{\mathbf Z_A}}}
[\|{\hyperref[SUBSEC_spatial]{\mathbf h}}{\symref{block}{\mathbf X_{\rm in}}}+{\hyperref[SUBSEC_spatial]{\mathbf u_A}}^H{\symref{block}{\mathbf Z_A}}\|^2]
\notag\\
&\quad+\frac{\mathbb E_{{\hyperref[EQ_response]{\mathbf A_A}}}[\|{\hyperref[EQ_response]{\mathbf A_A}}\|_F^2]}{L}
\notag\\
&\overset{\mathrm{(b)}}{=}\frac{{\hyperref[EQ_response]{\kappa}}^2}{L}
\left(S\mathbb E_{{\symref{block}{\mathbf X_{\rm in}}}}[\|{\symref{block}{\mathbf X_{\rm in}}}\|_F^2]
+L{\hyperref[SUBSEC_channels]{\sigma^2}}\right)+\beta_AP_A\notag\\
&\overset{\mathrm{(c)}}{=}{\hyperref[EQ_response]{\kappa}}^2(SP_B+{\hyperref[SUBSEC_channels]{\sigma^2}})+\beta_AP_A
\overset{\mathrm{(d)}}{=}P_A.
\label{EQ_energy_proof}
\end{align}
In \eqref{EQ_energy_proof}, (a) uses \eqref{EQ_response},
${\hyperref[EQ_spatial]{\mathbf f_A}}^H{\hyperref[EQ_spatial]{\mathbf B}}=\mathbf0$, $\|{\hyperref[EQ_spatial]{\mathbf f_A}}\|=1$, and
${\hyperref[EQ_spatial]{\mathbf B}}^H{\hyperref[EQ_spatial]{\mathbf B}}=\Id_d$.
Step (b) uses the marginal covariance $S\Id_p$ of ${\hyperref[SUBSEC_spatial]{\mathbf h}}$,
its independence from the honest input, and the independent noise
and AN in \eqref{EQ_response}.
Step (c) uses input energy $mP_B$ in \eqref{EQ_private_probe} or
expected input energy $nP_B$ in \eqref{EQ_challenge}.
Step (d) substitutes ${\hyperref[EQ_response]{\kappa}}^2$ from \eqref{EQ_response}.
This proves the honest response model and energy claim.
The energy equality does not apply to an unrestricted forged challenge.
\end{proof}

\subsection{Proof of Proposition~\ref{PROP_FA}}
\label{APP_FA_proof}
\begin{proof}
To bound $P_{\rm FA}$, calculate the probabilities of
$T>\eta$ and $\|{\hyperref[EQ_challenge]{\mathbf X}}\|_F^2>R_X$ in \eqref{EQ_decision}.
First condition on ${\symref{condition}{\mathcal C_0}}$ defined above.

\emph{Step 1: separate the channel-estimation error from the variance estimate.}
The F-distribution of $T/n$ requires independence of its residual
and ${\hyperref[EQ_calibration]{\widehat v}}$.
\symdef{regression}
Let ${\symref{regression}{\mathbf e_0}}\in\mathbb C^{1\times m}$ be the preparation noise
in \eqref{EQ_echo}, and let
${\symref{regression}{\mathbf P_0}}:={\hyperref[EQ_private_probe]{\mathbf X_0}}^H{\hyperref[EQ_private_probe]{\mathbf X_0}}/E_0\in\mathbb C^{m\times m}$
project onto the preparation row space. By \eqref{EQ_private_probe},
${\symref{regression}{\mathbf P_0}}$ has rank $p$ and $\Id_m-{\symref{regression}{\mathbf P_0}}$ has rank
${\hyperref[EQ_calibration]{\nu}}=m-p>0$.
\symdef{calibrationratio}
Let $W:={\hyperref[EQ_calibration]{\nu}}{\hyperref[EQ_calibration]{\widehat v}}/v_0$ denote the preparation residual energy
normalized by its noise variance. From \eqref{EQ_calibration},
\begin{align}
&{\hyperref[EQ_calibration]{\widehat{\mathbf k}}}-{\hyperref[EQ_echo]{\mathbf k}}
\overset{\mathrm{(a)}}{=}{\symref{regression}{\mathbf e_0}}{\hyperref[EQ_private_probe]{\mathbf X_0}}^H/E_0,\notag\\
&{\hyperref[EQ_calibration]{\nu}}{\hyperref[EQ_calibration]{\widehat v}}\overset{\mathrm{(b)}}{=}
\|{\symref{regression}{\mathbf e_0}}(\Id_m-{\symref{regression}{\mathbf P_0}})\|^2,\notag\\
&\operatorname{Cov}(({\hyperref[EQ_calibration]{\widehat{\mathbf k}}}-{\hyperref[EQ_echo]{\mathbf k}})^H\mid{\symref{condition}{\mathcal C_0}})
\overset{\mathrm{(c)}}{=}\frac{v_0}{E_0^2}{\hyperref[EQ_private_probe]{\mathbf X_0}}{\hyperref[EQ_private_probe]{\mathbf X_0}}^H
=\frac{v_0}{E_0}\Id_p,\notag\\
&({\hyperref[EQ_calibration]{\widehat{\mathbf k}}}-{\hyperref[EQ_echo]{\mathbf k}})^H\mid{\symref{condition}{\mathcal C_0}}
\sim\CN(\mathbf0,(v_0/E_0)\Id_p),\quad
W\mid{\symref{condition}{\mathcal C_0}}\overset{\mathrm{(d)}}{\sim}\Gamma({\hyperref[EQ_calibration]{\nu}},1),\notag\\
&\operatorname{Cov}(({\symref{regression}{\mathbf e_0}}{\symref{regression}{\mathbf P_0}})^H,
({\symref{regression}{\mathbf e_0}}(\Id_m-{\symref{regression}{\mathbf P_0}}))^H\mid{\symref{condition}{\mathcal C_0}})
\notag\\
&\qquad\overset{\mathrm{(e)}}{=}v_0{\symref{regression}{\mathbf P_0}}(\Id_m-{\symref{regression}{\mathbf P_0}})=\mathbf0.
\label{EQ_regression_laws}
\end{align}
In \eqref{EQ_regression_laws}, (a) substitutes \eqref{EQ_echo}
and uses \eqref{EQ_private_probe}. Step (b) also uses
${\hyperref[EQ_private_probe]{\mathbf X_0}}(\Id_m-{\symref{regression}{\mathbf P_0}})=\mathbf0$.
Step (c) transforms the covariance in \eqref{EQ_response_noise_covariance}.
In (d), the residual has ${\hyperref[EQ_calibration]{\nu}}$ independent $\CN(0,v_0)$ coordinates
in an orthonormal basis, so their squared magnitudes divided by
$v_0$ sum to $\Gamma({\hyperref[EQ_calibration]{\nu}},1)$.
Step (e) uses ${\symref{regression}{\mathbf P_0}}^2={\symref{regression}{\mathbf P_0}}$.
Joint proper Gaussianity makes the two projections independent.
Since ${\symref{regression}{\mathbf P_0}}{\hyperref[EQ_private_probe]{\mathbf X_0}}^H={\hyperref[EQ_private_probe]{\mathbf X_0}}^H$, they determine
${\hyperref[EQ_calibration]{\widehat{\mathbf k}}}-{\hyperref[EQ_echo]{\mathbf k}}$ and ${\hyperref[EQ_calibration]{\widehat v}}$ separately,
proving their conditional independence.

\emph{Step 2: derive the distribution of Bob's statistic.}
Under $\mathcal H_0$, let $\mathbf e_1\in\mathbb C^{1\times n}$
denote the fresh authentication noise from \eqref{EQ_echo_proof}.
It is independent of ${\symref{regression}{\mathbf e_0}}$ and ${\hyperref[EQ_challenge]{\mathbf X}}$ given ${\symref{condition}{\mathcal C_0}}$.
To compare with $W$, define the normalized current residual energy
$U:={\hyperref[EQ_cancellation]{\mathbf r}}{\hyperref[EQ_residual_covariance]{\mathbf V_X}}^{-1}{\hyperref[EQ_cancellation]{\mathbf r}}^H/v_0$. Then
\begin{align}
{\hyperref[EQ_cancellation]{\mathbf r}}&\overset{\mathrm{(a)}}{=}
\mathbf e_1-({\hyperref[EQ_calibration]{\widehat{\mathbf k}}}-{\hyperref[EQ_echo]{\mathbf k}}){\hyperref[EQ_challenge]{\mathbf X}},\notag\\
\operatorname{Cov}({\hyperref[EQ_cancellation]{\mathbf r}}^H\mid{\symref{condition}{\mathcal C_0}},{\hyperref[EQ_challenge]{\mathbf X}})
&\overset{\mathrm{(b)}}{=}v_0\Id_n+
\frac{v_0}{E_0}{\hyperref[EQ_challenge]{\mathbf X}}^H{\hyperref[EQ_challenge]{\mathbf X}}=v_0{\hyperref[EQ_residual_covariance]{\mathbf V_X}},\notag\\
U\mid{\symref{condition}{\mathcal C_0}},{\hyperref[EQ_challenge]{\mathbf X}}
&\overset{\mathrm{(c)}}{\sim}\Gamma(n,1),\notag\\
\frac{T}{n}&\overset{\mathrm{(d)}}{=}
\frac{U/n}{W/{\hyperref[EQ_calibration]{\nu}}},\qquad
\left.\frac{T}{n}\right|{\symref{condition}{\mathcal C_0}},{\hyperref[EQ_challenge]{\mathbf X}}
\overset{\mathrm{(e)}}{\sim}F_{2n,2{\hyperref[EQ_calibration]{\nu}}}.
\label{EQ_F_derivation}
\end{align}
In \eqref{EQ_F_derivation}, (a) substitutes the honest response
into \eqref{EQ_cancellation}. Step (b) adds the independent
noise and estimation-error covariances from
\eqref{EQ_response_noise_covariance} and \eqref{EQ_regression_laws}.
In (c), multiplication by ${\hyperref[EQ_residual_covariance]{\mathbf V_X}}^{-1/2}/\sqrt{v_0}$
gives independent $\CN(0,1)$ residual coordinates.
Step (d) substitutes $T$ from \eqref{EQ_decision} and
$W={\hyperref[EQ_calibration]{\nu}}{\hyperref[EQ_calibration]{\widehat v}}/v_0$.
In (e), $U$ depends on $\mathbf e_1$ and ${\symref{regression}{\mathbf e_0}}{\symref{regression}{\mathbf P_0}}$,
whereas $W$ depends on the independent complementary projection
in \eqref{EQ_regression_laws}.
Thus $2U$ and $2W$ are independent chi-square variables with
$2n$ and $2{\hyperref[EQ_calibration]{\nu}}$ degrees of freedom, giving the stated F-distribution.

\emph{Step 3: include rejection by the energy condition.}
To determine $R_X$, use the independent entries
$X_{ij}\sim\CN(C_{ij},b)$ from \eqref{EQ_challenge}, for
$1\leq i\leq p$, $1\leq j\leq n$ and $b>0$.
Here $C_{ij}$ is the corresponding entry of ${\hyperref[EQ_challenge]{\mathbf C}}$.
Write $\chi'^2_k(\lambda)$ for the noncentral chi-square distribution with $k\geq1$ degrees of freedom and noncentrality $\lambda\geq0$. Its CDF at $x\geq0$ is denoted by $\Phi_{k,\lambda}(x)$.
Then
\begin{align}
\frac{2\|{\hyperref[EQ_challenge]{\mathbf X}}\|_F^2}{b}
&=\sum_{i=1}^p\sum_{j=1}^n
\left[\left(\frac{\Re X_{ij}}{\sqrt{b/2}}\right)^2
+\left(\frac{\Im X_{ij}}{\sqrt{b/2}}\right)^2\right]\notag\\
&\overset{\mathrm{(a)}}{\sim}
\chi'^2_{2pn}\!\left(\frac{2\|{\hyperref[EQ_challenge]{\mathbf C}}\|_F^2}{b}\right)
=\chi'^2_{2pn}\!\left(\frac{2n(1-\beta_B)P_B}{b}\right).
\label{EQ_guard_law}
\end{align}
In \eqref{EQ_guard_law}, (a) sums squares of independent real
Gaussian components after dividing by their standard deviation
$\sqrt{b/2}$. Their variances become one and their means remain
in the sum, producing noncentrality $2\|{\hyperref[EQ_challenge]{\mathbf C}}\|_F^2/b$.
The last equality uses ${\hyperref[EQ_challenge]{\mathbf C}}$ in Section~\ref{SUBSEC_B_transmission}.
Thus $R_X$ is $b/2$ times the $(1-\delta)$ quantile of
\eqref{EQ_guard_law}.

The quantiles in \eqref{EQ_decision}, together with
\eqref{EQ_F_derivation} and \eqref{EQ_guard_law}, now give
\begin{align}
\Pr_{\mathcal H_0}\{T>\eta\}
&\overset{\mathrm{(a)}}{=}
\mathbb E_{{\symref{condition}{\mathcal C_0}},{\hyperref[EQ_challenge]{\mathbf X}}}
[\Pr_{\mathcal H_0}\{T>\eta\mid{\symref{condition}{\mathcal C_0}},{\hyperref[EQ_challenge]{\mathbf X}}\}]
\notag\\
&=1-(1-\alpha+\delta)=\alpha-\delta,\notag\\
\Pr_{\mathcal H_0}\{\|{\hyperref[EQ_challenge]{\mathbf X}}\|_F^2>R_X\}
&\overset{\mathrm{(b)}}{=}\delta,\notag\\
P_{\rm FA}&\overset{\mathrm{(c)}}{=}
\Pr_{\mathcal H_0}(\{T>\eta\}\cup\{\|{\hyperref[EQ_challenge]{\mathbf X}}\|_F^2>R_X\})
\notag\\
&\overset{\mathrm{(d)}}{\leq}(\alpha-\delta)+\delta=\alpha.
\label{EQ_false_alarm_union}
\end{align}
In \eqref{EQ_false_alarm_union}, (a) uses total probability and
$\eta/n=F_{2n,2{\hyperref[EQ_calibration]{\nu}}}^{-1}(1-\alpha+\delta)$.
Step (b) uses the defining quantile of $R_X$.
Step (c) applies the two acceptance conditions in \eqref{EQ_decision},
and (d) uses the union bound. This proves Proposition~\ref{PROP_FA}.
\end{proof}
\endgroup

\section{Preparation Observations and Disclosed Information}
\label{APP_joint_information}
We identify the observations used in the security and estimation-error
bounds. The proofs disclose additional information to Trudy, so their
conditional Gaussian calculations do not require the full distribution
of the channels given her actual observations.

\subsection{What Alice's AN Conceals}
To distinguish the forwarded preparation and fresh AN in
Trudy's reception of \eqref{EQ_response}, define
the useful response channel ${\symref{protected}{\mathbf a}}\in\mathbb C^q$ and
the AN channel ${\symref{protected}{\mathbf J}}\in\mathbb C^{q\times d}$ as
\symdef{protected}
\begin{equation}
{\symref{protected}{\mathbf a}}:={\hyperref[SUBSEC_channels]{\mathbf G_A}}{\hyperref[EQ_spatial]{\mathbf f_A}},\qquad
{\symref{protected}{\mathbf J}}:={\hyperref[SUBSEC_channels]{\mathbf G_A}}{\hyperref[EQ_spatial]{\mathbf B}}.
\label{EQ_intercept_channels}
\end{equation}
Let ${\hyperref[EQ_protected_intercepts]{\mathbf Y_{B,T,0}}},{\hyperref[EQ_protected_intercepts]{\mathbf Y_{A,T,0}}}\in\mathbb C^{q\times m}$
be Trudy's receptions of Bob's preparation and Alice's response.
Their receiver-noise blocks ${\hyperref[EQ_protected_intercepts]{\mathbf Z_{B,T,0}}},{\hyperref[EQ_protected_intercepts]{\mathbf Z_{A,T,0}}}$
have independent $\CN(0,{\hyperref[SUBSEC_channels]{\sigma_T^2}})$ entries and are independent
of legitimate receiver noise and AN.
Let ${\hyperref[EQ_protected_intercepts]{\boldsymbol\zeta_0}}\in\mathbb C^{q\times m}$ denote the total
noise in Trudy's response reception. Then
\begin{align}
{\hyperref[EQ_protected_intercepts]{\mathbf Y_{B,T,0}}}&={\hyperref[SUBSEC_channels]{\mathbf G_B}}{\hyperref[EQ_private_probe]{\mathbf X_0}}+{\hyperref[EQ_protected_intercepts]{\mathbf Z_{B,T,0}}},\notag\\
{\hyperref[EQ_protected_intercepts]{\mathbf Y_{A,T,0}}}&={\hyperref[SUBSEC_channels]{\mathbf G_A}}{\hyperref[EQ_response]{\mathbf X_A}}+{\hyperref[EQ_protected_intercepts]{\mathbf Z_{A,T,0}}}\notag\\
&\overset{\mathrm{(a)}}{=}{\hyperref[EQ_response]{\kappa}}{\symref{protected}{\mathbf a}}{\hyperref[SUBSEC_spatial]{\mathbf h}}{\hyperref[EQ_private_probe]{\mathbf X_0}}
+{\hyperref[EQ_protected_intercepts]{\boldsymbol\zeta_0}},\notag\\
{\hyperref[EQ_protected_intercepts]{\boldsymbol\zeta_0}}&={\hyperref[EQ_response]{\kappa}}{\symref{protected}{\mathbf a}}{\hyperref[SUBSEC_spatial]{\mathbf u_A}}^H{\symref{block}{\mathbf Z_A}}
+{\symref{protected}{\mathbf J}}{\hyperref[EQ_response]{\mathbf A_A}}+{\hyperref[EQ_protected_intercepts]{\mathbf Z_{A,T,0}}}.
\label{EQ_protected_intercepts}
\end{align}
In \eqref{EQ_protected_intercepts}, (a) substitutes
\eqref{EQ_response}, uses \eqref{EQ_intercept_channels}, and
collects the forwarded Alice noise, transmitted AN, and Trudy noise.
The two receptions contain the same ${\hyperref[EQ_private_probe]{\mathbf X_0}}$ but independent
receiver noises.

\symdef{RA}
To quantify this noise, let ${\symref{RA}{\mathbf R_A}}\in\mathbb C^{q\times q}$ be the covariance
of a column ${\symref{RA}{\boldsymbol\zeta_{0,j}}}$ of ${\hyperref[EQ_protected_intercepts]{\boldsymbol\zeta_0}}$
given ${\symref{condition}{\mathcal C_0}}$, for $1\leq j\leq m$. We obtain
\begin{align}
{\symref{RA}{\mathbf R_A}}&:=\operatorname{Cov}({\symref{RA}{\boldsymbol\zeta_{0,j}}}\mid{\symref{condition}{\mathcal C_0}})
\notag\\
&\overset{\mathrm{(a)}}{=}{\hyperref[EQ_response]{\kappa}}^2{\hyperref[SUBSEC_channels]{\sigma^2}}{\symref{protected}{\mathbf a}}{\symref{protected}{\mathbf a}}^H
+\frac{\beta_AP_A}{d}{\symref{protected}{\mathbf J}}{\symref{protected}{\mathbf J}}^H+{\hyperref[SUBSEC_channels]{\sigma_T^2}}\Id_q.
\label{EQ_intercept_noise}
\end{align}
In \eqref{EQ_intercept_noise}, (a) adds the covariances of the
independent terms in \eqref{EQ_protected_intercepts}, using
$\|{\hyperref[SUBSEC_spatial]{\mathbf u_A}}\|=1$, the AN variance in \eqref{EQ_response},
and the receiver-noise variances in Section~\ref{SUBSEC_channels}.
The columns are conditionally independent proper Gaussian vectors,
and ${\symref{RA}{\mathbf R_A}}\succ0$ because ${\hyperref[SUBSEC_channels]{\sigma_T^2}}>0$.

\subsection{Information Used in the Bounds}
To specify the disclosures, first collect Trudy's actual observations.
\symdef{directions}
Let $\mathcal D:=\{\mathbf f_A,\mathbf B,\mathbf u_B,\mathbf u_A,\mathbf f_B\}$
be the public directions in Section~\ref{SUBSEC_spatial}.
For $U\in\{A,B\}$, let $\mathbf T_U\in\mathbb C^{q\times p}$ be
Trudy's initial-training reception, with receiver noise
$\mathbf N_U\in\mathbb C^{q\times p}$ having independent
$\CN(0,\sigma_T^2)$ entries. The training model is
\begin{equation}
\mathbf T_U=\mathbf G_U\mathbf P_{\rm tr}+\mathbf N_U.
\label{EQ_initial_channel_observation}
\end{equation}
The two preparation receptions are defined in
\eqref{EQ_protected_intercepts}.
\symdef{currentinfo}
Let $\mathbf Y_{B,T,1}\in\mathbb C^{q\times n}$ be Trudy's current
challenge reception and let $\mathbf Z_{B,T,1}$ be its receiver noise,
with fresh independent $\CN(0,\sigma_T^2)$ entries. Then
\begin{equation}
\mathbf Y_{B,T,1}=\mathbf G_B\mathbf X+\mathbf Z_{B,T,1}.
\label{EQ_current_challenge_observation}
\end{equation}
Her information when choosing a response is
\begin{equation}
\mathcal O_T:=\{\mathbf T_A,\mathbf T_B,\mathcal D,
\mathbf Y_{B,T,0},\mathbf Y_{A,T,0},\mathbf Y_{B,T,1}\}.
\label{EQ_actual_information}
\end{equation}
\symdef{Oplus}
The information disclosed for the forward-estimation bound is
\begin{equation}
\mathcal O_T^+:=\{\mathcal O_T,\mathbf G_A,\mathbf G_B,\mathbf X_0\}.
\label{EQ_disclosed_information}
\end{equation}
The preparation receptions share $\mathbf X_0$, and Alice's
preparation noise contributes to both her intercepted response and
Bob's calibration record. Consequently, the bounds do not assume
that Bob's calibration is independent of $\mathcal O_T$.
Theorem~\ref{THM_echo_bound} discloses Bob's calibration and the
channels, while Proposition~\ref{PROP_information} discloses
$\mathbf G_A,\mathbf G_B,\mathbf X_0$ and keeps his calibration private.
These disclosures give the conditional Gaussian models used in
Appendices~\ref{APP_security} and \ref{APP_information}.

The last $d$ forward coefficients form
${\hyperref[SUBSEC_spatial]{\mathbf h_\perp}}\in\mathbb C^{1\times d}$ in Section~\ref{SUBSEC_spatial}.
We lower-bound Trudy's estimation error for these coefficients
by disclosing additional information, which cannot increase her
minimum estimation error.
\begin{proposition}[Forward information under additional disclosure]
\label{PROP_information}
Let ${\symref{Oplus}{\mathcal O_T^+}}$ contain Trudy's observations and the disclosed
${\hyperref[SUBSEC_channels]{\mathbf G_A}},{\hyperref[SUBSEC_channels]{\mathbf G_B}},{\hyperref[EQ_private_probe]{\mathbf X_0}}$, as specified in
\eqref{EQ_disclosed_information}. Keep ${\hyperref[SUBSEC_spatial]{\mathbf h}}$ and
$({\hyperref[EQ_calibration]{\widehat{\mathbf k}}},{\hyperref[EQ_calibration]{\widehat v}})$ private.
\symdef{qh}
Conditional on ${\symref{Oplus}{\mathcal O_T^+}}$, each coefficient of
${\hyperref[SUBSEC_spatial]{\mathbf h_\perp}}$ has variance
\begin{equation}
q_h=\left(S^{-1}+{\hyperref[EQ_response]{\kappa}}^2E_0{\symref{protected}{\mathbf a}}^H
{\symref{RA}{\mathbf R_A}}^{-1}{\symref{protected}{\mathbf a}}\right)^{-1}.
\label{EQ_forward_information}
\end{equation}
For fixed ${\symref{protected}{\mathbf a}},{\symref{protected}{\mathbf J}}$ and energy budgets, $q_h$ is
nondecreasing in ${\hyperref[SEC_protocol]{\beta_A}}$.
Any estimate of ${\hyperref[SUBSEC_spatial]{\mathbf h_\perp}}$ based on Trudy's actual
information $\mathcal O_T$ in \eqref{EQ_actual_information} has mean-square error
per coefficient at least $\mathbb E_{{\symref{protected}{\mathbf a}},{\symref{protected}{\mathbf J}}}[q_h]$.
\end{proposition}
Thus Alice's AN can increase the remaining uncertainty even when
the preparation matrix and interception channels are disclosed.
Appendix~\ref{APP_information} proves the conditional variance, its
monotonicity, and the lower bound on the mean-square error.

\section{Challenge and Response Replacement after Interception}
\label{APP_security}
To prove Theorem~\ref{THM_echo_bound}, we bound Bob's acceptance
probability for response replacement, challenge replacement, and both,
averaging over the channel and
preparation distributions.

\subsection{Proof of Theorem~\ref{THM_echo_bound}}
\begin{proof}
We upper-bound Bob's acceptance probability under response
replacement, including replacement of both blocks, and then
under challenge replacement alone. Using Trudy's remaining
uncertainty about Bob's current AN, we obtain
\eqref{EQ_upper_exact} and derive the scalar bound
\eqref{EQ_upper_scalar}.
\emph{Step 1: upper-bound the acceptance probability under response replacement.}
\symdef{Jinfo}
Let ${\symref{Jinfo}{\mathcal J}}$ collect the physical channels, all initial-training
and preparation observations, the private matrix
${\hyperref[EQ_private_probe]{\mathbf X_0}}$, Bob's calibration record, and Trudy's private
randomization. It excludes current AN and receiver noises. Disclosing ${\symref{Jinfo}{\mathcal J}}$ cannot decrease Trudy's
maximum acceptance probability.
To bound that probability, first use the two conditions for
$I_B=1$ in \eqref{EQ_decision}. With $c_X=1+R_X/E_0$ from
Theorem~\ref{THM_echo_bound},
\begin{align}
I_B=1&\ \Longrightarrow\ \|{\hyperref[EQ_challenge]{\mathbf X}}\|_F^2\leq R_X,\quad T\leq{\hyperref[EQ_decision]{\eta}},
\notag\\
{\hyperref[EQ_residual_covariance]{\mathbf V_X}}&\overset{\mathrm{(a)}}{\preceq}c_X\Id_n,
\qquad {\hyperref[EQ_cancellation]{\mathbf r}}{\hyperref[EQ_residual_covariance]{\mathbf V_X}}^{-1}{\hyperref[EQ_cancellation]{\mathbf r}}^H\geq\|{\hyperref[EQ_cancellation]{\mathbf r}}\|^2/c_X,
\notag\\
\{I_B=1\}&\overset{\mathrm{(b)}}{\subseteq}
\{\|{\hyperref[EQ_cancellation]{\mathbf r}}\|^2\leq{\hyperref[EQ_decision]{\eta}} c_X{\hyperref[EQ_calibration]{\widehat v}}\}.
\label{EQ_acceptance_norm_bound}
\end{align}
In \eqref{EQ_acceptance_norm_bound}, (a) uses
${\hyperref[EQ_residual_covariance]{\mathbf V_X}}=\Id_n+{\hyperref[EQ_challenge]{\mathbf X}}^H{\hyperref[EQ_challenge]{\mathbf X}}/E_0$ in \eqref{EQ_residual_covariance}
and the energy condition. Step (b) combines the resulting quadratic-form
inequality with $T\leq{\hyperref[EQ_decision]{\eta}}$ in \eqref{EQ_decision}.
We therefore need the distribution of ${\hyperref[EQ_cancellation]{\mathbf r}}$ given Trudy's information.

\symdef{forge}
Allow Trudy to choose any received signal row
${\symref{forge}{\mathbf w}}\in\mathbb C^{1\times n}$ before Bob's receiver noise.
For response replacement, \eqref{EQ_cancellation} becomes
${\hyperref[EQ_cancellation]{\mathbf r}}={\symref{forge}{\mathbf w}}-{\hyperref[EQ_calibration]{\widehat{\mathbf k}}}{\hyperref[EQ_challenge]{\mathbf X}}+{\symref{forge}{\mathbf z_B}}$,
where ${\symref{forge}{\mathbf z_B}}\in\mathbb C^{1\times n}$ has independent
$\CN(0,{\hyperref[SUBSEC_channels]{\sigma^2}})$ entries. Thus the residual uncertainty comes
from the part of ${\hyperref[EQ_challenge]{\mathbf X}}$ that Trudy cannot infer.
\symdef{YT}
Her current reception is ${\symref{YT}{\mathbf Y_T}}={\hyperref[SUBSEC_channels]{\mathbf G_B}}{\hyperref[EQ_challenge]{\mathbf X}}+{\symref{YT}{\mathbf Z_T}}
\in\mathbb C^{q\times n}$, also denoted ${\hyperref[EQ_current_challenge_observation]{\mathbf Y_{B,T,1}}}$ in
\eqref{EQ_current_challenge_observation}. The entries of ${\symref{YT}{\mathbf Z_T}}$ are
independent $\CN(0,{\hyperref[SUBSEC_channels]{\sigma_T^2}})$. Subtracting the public component gives
\begin{equation}
{\symref{YT}{\mathbf Y_T}}-{\hyperref[SUBSEC_channels]{\mathbf G_B}}{\hyperref[EQ_challenge]{\mathbf C}}={\hyperref[SUBSEC_channels]{\mathbf G_B}}{\hyperref[EQ_challenge]{\mathbf A_B}}+{\symref{YT}{\mathbf Z_T}}.
\label{EQ_current_intercept}
\end{equation}
For $1\leq j\leq n$, let $\mathbf a_j\in\mathbb C^p$ be column $j$ of ${\hyperref[EQ_challenge]{\mathbf A_B}}$, and let $\mathbf y_j,\mathbf z_j\in\mathbb C^q$
be the corresponding columns of ${\symref{YT}{\mathbf Y_T}}-{\hyperref[SUBSEC_channels]{\mathbf G_B}}{\hyperref[EQ_challenge]{\mathbf C}}$
and ${\symref{YT}{\mathbf Z_T}}$. Given ${\symref{Jinfo}{\mathcal J}}$, the equation
$\mathbf y_j={\hyperref[SUBSEC_channels]{\mathbf G_B}}\mathbf a_j+\mathbf z_j$ has independent
$\mathbf a_j\sim\CN(\mathbf0,b\Id_p)$ and
$\mathbf z_j\sim\CN(\mathbf0,{\hyperref[SUBSEC_channels]{\sigma_T^2}}\Id_q)$ by \eqref{EQ_challenge}.
Let $\boldsymbol\mu\in\mathbb C^p$ and ${\hyperref[EQ_Trudy_covariance]{\mathbf Q_T}}\in\mathbb C^{p\times p}$
denote the conditional mean and covariance of $\mathbf a_j$. Then
\begin{align}
f(\mathbf a_j\mid\mathbf y_j,{\symref{Jinfo}{\mathcal J}})
&\overset{\mathrm{(a)}}{\propto}
f(\mathbf a_j\mid{\symref{Jinfo}{\mathcal J}})f(\mathbf y_j\mid\mathbf a_j,{\symref{Jinfo}{\mathcal J}})\notag\\
&\overset{\mathrm{(b)}}{\propto}
\exp[-\mathbf a_j^H\mathbf a_j/b-\|\mathbf y_j-{\hyperref[SUBSEC_channels]{\mathbf G_B}}\mathbf a_j\|^2/{\hyperref[SUBSEC_channels]{\sigma_T^2}}]
\notag\\
&\overset{\mathrm{(c)}}{\propto}
\exp[-\mathbf a_j^H(b^{-1}\Id_p+\sigma_T^{-2}{\hyperref[SUBSEC_channels]{\mathbf G_B}}^H{\hyperref[SUBSEC_channels]{\mathbf G_B}})\mathbf a_j
\notag\\
&\qquad+2\Re\{\mathbf a_j^H\sigma_T^{-2}{\hyperref[SUBSEC_channels]{\mathbf G_B}}^H\mathbf y_j\}]\notag\\
&\overset{\mathrm{(d)}}{\propto}
\exp[-(\mathbf a_j-\boldsymbol\mu)^H{\hyperref[EQ_Trudy_covariance]{\mathbf Q_T}}^{-1}(\mathbf a_j-\boldsymbol\mu)],
\label{EQ_Trudy_conditional_density}\\
{\hyperref[EQ_Trudy_covariance]{\mathbf Q_T}}&=(b^{-1}\Id_p+\sigma_T^{-2}{\hyperref[SUBSEC_channels]{\mathbf G_B}}^H{\hyperref[SUBSEC_channels]{\mathbf G_B}})^{-1},
\label{EQ_Trudy_covariance}\\
\boldsymbol\mu&={\hyperref[EQ_Trudy_covariance]{\mathbf Q_T}}\sigma_T^{-2}{\hyperref[SUBSEC_channels]{\mathbf G_B}}^H\mathbf y_j.\notag
\end{align}
In \eqref{EQ_Trudy_conditional_density}--\eqref{EQ_Trudy_covariance},
(a) applies Bayes' rule, (b) substitutes the Gaussian densities,
(c) expands the exponent and drops terms independent of $\mathbf a_j$,
and (d) completes the square. Hence
$\mathbf a_j\mid\mathbf y_j,{\symref{Jinfo}{\mathcal J}}\sim\CN(\boldsymbol\mu,{\hyperref[EQ_Trudy_covariance]{\mathbf Q_T}})$.
Given ${\symref{Jinfo}{\mathcal J}}$, different columns use independent AN and noises.
Observing the other columns of ${\symref{YT}{\mathbf Y_T}}$ therefore leaves this
column's conditional covariance equal to ${\hyperref[EQ_Trudy_covariance]{\mathbf Q_T}}$.

\symdef{mean}
Let ${\symref{mean}{\widehat{\mathbf X}_T}}:=\mathbb E_{{\hyperref[EQ_challenge]{\mathbf X}}\mid{\symref{Jinfo}{\mathcal J}},{\symref{YT}{\mathbf Y_T}}}
[{\hyperref[EQ_challenge]{\mathbf X}}]\in\mathbb C^{p\times n}$ be Trudy's conditional challenge mean,
obtained column by column from \eqref{EQ_Trudy_covariance}.
The conditional residual variance per symbol is
\begin{equation}
q_0:=\widehat{\mathbf k}\mathbf Q_T\widehat{\mathbf k}^H+\sigma^2>0.
\label{EQ_attack_residual_variance}
\end{equation}
For every $\mathbf w$ chosen from $\mathcal J,\mathbf Y_T$,
the conditional residual distribution and acceptance probability satisfy
\begin{align}
\mathbf r^H\mid\mathcal J,\mathbf Y_T,\mathbf w
&\overset{\mathrm{(a)}}{\sim}
\CN((\mathbf w-\widehat{\mathbf k}\widehat{\mathbf X}_T)^H,q_0\Id_n),
\label{EQ_attack_ncx}\\
\Pr\{I_B=1\mid\mathcal J,\mathbf Y_T,\mathbf w\}
&\overset{\mathrm{(b)}}{\leq}
\Pr\{\|\mathbf r\|^2\leq\eta c_X\widehat v
\mid\mathcal J,\mathbf Y_T,\mathbf w\}\notag\\
&\overset{\mathrm{(c)}}{\leq}G_n(\eta c_X\widehat v/q_0).
\label{EQ_ball_proof}
\end{align}
In \eqref{EQ_attack_ncx}--\eqref{EQ_ball_proof}, (a) substitutes
\eqref{EQ_Trudy_covariance} into
$\mathbf r=\mathbf w-\widehat{\mathbf k}\mathbf X+\mathbf z_B$.
Step (b) uses \eqref{EQ_acceptance_norm_bound}.
In (c), Anderson's inequality \cite{Anderson1955} states that
a centered Gaussian maximizes the probability of a centered ball
over all mean shifts. The centered squared norm divided by $q_0$
has distribution $\Gamma(n,1)$, whose CDF is
\eqref{EQ_appendix_gamma_cdf}.
The bound holds for every ${\symref{forge}{\mathbf w}}$, including when both blocks
are replaced. It therefore bounds every feasible response attack,
without asserting optimality against Bob's actual test.

\emph{Step 2: upper-bound the acceptance probability under challenge replacement alone.}
Let $\pi\in\mathfrak A$ be any strategy that replaces the challenge alone. To upper-bound this probability, give Trudy the entire ${{\symref{YT}{\mathbf Y_T}}}$ before she chooses
${{\hyperref[SUBSEC_spatial]{\mathbf X_{T\to A}}}}$. This relaxes the timing restriction in
Section~\ref{SUBSEC_spatial} and cannot reduce her maximum acceptance probability. Conditional on ${{\symref{Jinfo}{\mathcal J}}},{{\symref{YT}{\mathbf Y_T}}}$,
\symdef{wA}
the forged challenge contributes the following row to Bob's
combined received signal:
${{\symref{wA}{\mathbf w_A}}}:={{\hyperref[EQ_response]{\kappa}}} {g}{{\hyperref[SUBSEC_spatial]{\mathbf u_A}}}^H{{\hyperref[SUBSEC_channels]{\mathbf K_A}}}
{{\hyperref[SUBSEC_spatial]{\mathbf X_{T\to A}}}}\in{\mathbb C}^{1\times {n}}$.
To derive Bob's received row, let
${{\symref{delivery}{\mathbf Y_{A,1}}}},{{\symref{delivery}{\mathbf Z_{A,1}}}}\in{\mathbb C}^{{{\hyperref[SEC_system_model]{N_R}}}\times {n}}$
be Alice's current received block and receiver noise. The channel delivers
\symdef{delivery}
\begin{equation}
{{\symref{delivery}{\mathbf Y_{A,1}}}}=\begin{cases}
{{\hyperref[EQ_TDD_channels]{\mathbf H_A}}}{{\hyperref[EQ_challenge]{\mathbf X}}}+{{\symref{delivery}{\mathbf Z_{A,1}}}},&\text{no }{T}\to A,\\
{{\hyperref[SUBSEC_channels]{\mathbf K_A}}}{{\hyperref[SUBSEC_spatial]{\mathbf X_{T\to A}}}}+{{\symref{delivery}{\mathbf Z_{A,1}}}},&{T}\to A.
\end{cases}
\label{EQ_challenge_delivery}
\end{equation}
Alice applies the authentication response in Section~\ref{SUBSEC_response}
to ${{\symref{delivery}{\mathbf Y_{A,1}}}}$.
\symdef{e1}
Its noise ${\symref{e1}{\mathbf e_1}}$ contains the three terms in
\eqref{EQ_echo_proof} with $L=n$. Conditional on ${\symref{Jinfo}{\mathcal J}}$, it is independent
of ${\hyperref[EQ_challenge]{\mathbf X}}$ and the interception noise in
\eqref{EQ_current_intercept}, with covariance ${\hyperref[EQ_response_noise_covariance]{v_0}}\Id_n$
for ${\symref{e1}{\mathbf e_1}}^H$.
Applying Section~\ref{SUBSEC_response} to \eqref{EQ_challenge_delivery} gives
\begin{align}
{{\hyperref[EQ_cancellation]{\mathbf y_1}}}&\overset{\mathrm{(a)}}{=}{{\symref{wA}{\mathbf w_A}}}+{{\symref{e1}{\mathbf e_1}}},\notag\\
{{\hyperref[EQ_cancellation]{\mathbf r}}}^H\mid{{\symref{Jinfo}{\mathcal J}}},{{\symref{YT}{\mathbf Y_T}}},{{\symref{wA}{\mathbf w_A}}}
&\overset{\mathrm{(b)}}{\sim}
{\CN}\!\bigl(({{\symref{wA}{\mathbf w_A}}}-{{\hyperref[EQ_calibration]{\widehat{\mathbf k}}}}{{\symref{mean}{\widehat{\mathbf X}_T}}})^H,\notag\\[-2pt]
&\qquad({{\hyperref[EQ_attack_residual_variance]{q_0}}}+{{\hyperref[EQ_response_noise_covariance]{v_0}}}-{{\hyperref[SUBSEC_channels]{\sigma^2}}}){\Id}_{n}\bigr).
\label{EQ_challenge_residual}
\end{align}
In (a) of \eqref{EQ_challenge_residual}, substitute
the forged input from \eqref{EQ_challenge_delivery} into the response
in Section~\ref{SUBSEC_response}, using
$g$ from \eqref{EQ_return_gain}.
Step (b) uses the residual definition \eqref{EQ_cancellation},
the challenge covariance \eqref{EQ_Trudy_covariance}, and
the independent current noise ${\symref{e1}{\mathbf e_1}}$ in \eqref{EQ_response_noise_covariance}.
The conditional residual variance per symbol is
${\hyperref[EQ_calibration]{\widehat{\mathbf k}}}{\hyperref[EQ_Trudy_covariance]{\mathbf Q_T}}{\hyperref[EQ_calibration]{\widehat{\mathbf k}}}^H+{\hyperref[EQ_response_noise_covariance]{v_0}}
={\hyperref[EQ_attack_residual_variance]{q_0}}+{\hyperref[EQ_response_noise_covariance]{v_0}}-{\hyperref[SUBSEC_channels]{\sigma^2}}$.
Bounding Bob's acceptance probability by the probability of
the ball in \eqref{EQ_ball_proof} gives
\begin{align}
\Pr_{\pi}\{{{\hyperref[EQ_objective]{I_B}}}=1\mid{{\symref{Jinfo}{\mathcal J}}},{{\symref{YT}{\mathbf Y_T}}}\}
&\overset{\mathrm{(a)}}{\leq}
{{\hyperref[EQ_appendix_gamma_cdf]{G_n}}}\!\left(\frac{{{\hyperref[EQ_decision]{\eta}}} {{\hyperref[THM_echo_bound]{c_X}}}{{\hyperref[EQ_calibration]{\widehat v}}}}{{{\hyperref[EQ_attack_residual_variance]{q_0}}}+{{\hyperref[EQ_response_noise_covariance]{v_0}}}-{{\hyperref[SUBSEC_channels]{\sigma^2}}}}\right)\notag\\
&\overset{\mathrm{(b)}}{\leq}{{\hyperref[EQ_appendix_gamma_cdf]{G_n}}}({{\hyperref[EQ_decision]{\eta}}} {{\hyperref[THM_echo_bound]{c_X}}}{{\hyperref[EQ_calibration]{\widehat v}}}/{{\hyperref[EQ_attack_residual_variance]{q_0}}}).
\label{EQ_challenge_bound}
\end{align}
In (a) of \eqref{EQ_challenge_bound}, apply the event containment in
\eqref{EQ_acceptance_norm_bound} and maximize over ${{\symref{wA}{\mathbf w_A}}}$. In (b),
${{\hyperref[EQ_response_noise_covariance]{v_0}}}\geq{{\hyperref[SUBSEC_channels]{\sigma^2}}}$ by \eqref{EQ_response_noise_covariance} and ${{\hyperref[EQ_appendix_gamma_cdf]{G_n}}}$ is increasing.
Equations~\eqref{EQ_ball_proof} and \eqref{EQ_challenge_bound} thus
give the same conditional upper bound on Bob's acceptance
probability under either or both replacements. This also covers
a strategy that selects the replacement direction using Trudy's
available information.
Averaging the common conditional bound and taking the supremum
over admissible strategies gives the intermediate bound $B_2$:
\begin{equation}
\begin{aligned}
P_2^\star&\leq B_2,\\
B_2&:=\mathbb E_{\widehat{\mathbf k},\widehat v,\mathbf G_B}
\!\left[G_n\!\left(
\frac{\eta c_X\widehat v}
{\widehat{\mathbf k}\mathbf Q_T\widehat{\mathbf k}^H+\sigma^2}
\right)\right].
\end{aligned}
\label{EQ_upper_exact}
\end{equation}

\emph{Step 3: reduce $B_2$ to scalar probabilities.}
The denominator in \eqref{EQ_upper_exact} contains
${\hyperref[EQ_calibration]{\widehat{\mathbf k}}}{\hyperref[EQ_Trudy_covariance]{\mathbf Q_T}}{\hyperref[EQ_calibration]{\widehat{\mathbf k}}}^H$.
We first lower-bound this quadratic form to reduce $B_2$ to a
probability involving the estimate norm. We then bound that
probability while retaining the dependence of the TDD channels.
\symdef{strength}
Let $Q:=\|{\hyperref[EQ_calibration]{\widehat{\mathbf k}}}\|^2/v_0$ be the normalized estimate energy,
and let $Z_{\hyperref[EQ_calibration]{\nu}}:={\hyperref[EQ_calibration]{\nu}}{\hyperref[EQ_calibration]{\widehat v}}/v_0$ be the normalized preparation
residual energy from \eqref{EQ_regression_laws}.
\symdef{direction}
Define the unit column ${\symref{direction}{\mathbf v}}:={\hyperref[EQ_calibration]{\widehat{\mathbf k}}}^H/\|{\hyperref[EQ_calibration]{\widehat{\mathbf k}}}\|$
and the normalized interception-channel energy
$Z_q:=\|{\hyperref[SUBSEC_channels]{\mathbf G_B}}{\symref{direction}{\mathbf v}}\|^2/S_T$. The estimate is nonzero almost surely.
Then
\begin{align}
{\hyperref[EQ_calibration]{\widehat{\mathbf k}}}{\hyperref[EQ_Trudy_covariance]{\mathbf Q_T}}{\hyperref[EQ_calibration]{\widehat{\mathbf k}}}^H
&\overset{\mathrm{(a)}}{\geq}
\frac{\|{\hyperref[EQ_calibration]{\widehat{\mathbf k}}}\|^2}{b^{-1}+S_TZ_q/{\hyperref[SUBSEC_channels]{\sigma_T^2}}},\notag\\
Z_q\mid{\hyperref[EQ_calibration]{\widehat{\mathbf k}}},{\hyperref[EQ_calibration]{\widehat v}}
&\overset{\mathrm{(b)}}{\sim}\Gamma(q,1),
\label{EQ_directional_precision}\\
B_2&\overset{\mathrm{(c)}}{\leq}
\mathbb E_{Q,Z_{\hyperref[EQ_calibration]{\nu}},Z_q}\!\left[
G_n\!\left(\frac{{\hyperref[EQ_decision]{\eta}} c_XZ_{\hyperref[EQ_calibration]{\nu}}(b^{-1}+S_TZ_q/{\hyperref[SUBSEC_channels]{\sigma_T^2}})}{{\hyperref[EQ_calibration]{\nu}} Q}\right)
\right].
\label{EQ_scalar_acceptance_reduction}
\end{align}
In \eqref{EQ_directional_precision}--\eqref{EQ_scalar_acceptance_reduction},
(a) uses $({\symref{direction}{\mathbf v}}^H{\hyperref[EQ_Trudy_covariance]{\mathbf Q_T}}{\symref{direction}{\mathbf v}})
({\symref{direction}{\mathbf v}}^H{\hyperref[EQ_Trudy_covariance]{\mathbf Q_T}}^{-1}{\symref{direction}{\mathbf v}})\geq1$ and \eqref{EQ_Trudy_covariance}.
In (b), ${\hyperref[SUBSEC_channels]{\mathbf G_B}}$ is independent of legitimate calibration and has
independent $\CN(0,S_T)$ entries by Section~\ref{SUBSEC_channels}.
Its projection onto any fixed unit ${\symref{direction}{\mathbf v}}$ therefore gives the
same Gamma distribution, so $Z_q$ is independent of $Q,Z_{\hyperref[EQ_calibration]{\nu}}$.
Step (c) substitutes (a) into \eqref{EQ_upper_exact}, drops the positive
${\hyperref[SUBSEC_channels]{\sigma^2}}$ from its denominator, and uses ${\hyperref[EQ_calibration]{\widehat v}}/v_0=Z_{\hyperref[EQ_calibration]{\nu}}/{\hyperref[EQ_calibration]{\nu}}$.
By \eqref{EQ_regression_laws}, $Z_{\hyperref[EQ_calibration]{\nu}}\sim\Gamma({\hyperref[EQ_calibration]{\nu}},1)$ is independent
of $Q$ after averaging over the channels as well.

To use the estimate-energy CDF $F_Q(x):=\Pr\{Q\leq x\}$ for $x\geq0$,
introduce an independent $Z_n\sim\Gamma(n,1)$, so that
$G_n(y)=\Pr\{Z_n\leq y\}$ for $y\geq0$.
Equation~\eqref{EQ_scalar_acceptance_reduction} becomes
\begin{align}
B_2&\overset{\mathrm{(a)}}{\leq}
\Pr\!\left\{QZ_n\leq\frac{{\hyperref[EQ_decision]{\eta}} c_XZ_{\hyperref[EQ_calibration]{\nu}}}{{\hyperref[EQ_calibration]{\nu}}}
(b^{-1}+S_TZ_q/{\hyperref[SUBSEC_channels]{\sigma_T^2}})\right\}\notag\\
&\overset{\mathrm{(b)}}{=}
\mathbb E_{Z_{\hyperref[EQ_calibration]{\nu}},Z_n,Z_q}\!\left[
F_Q\!\left(
\frac{{\hyperref[EQ_decision]{\eta}} c_XZ_{\hyperref[EQ_calibration]{\nu}}(b^{-1}+S_TZ_q/{\hyperref[SUBSEC_channels]{\sigma_T^2}})}{{\hyperref[EQ_calibration]{\nu}} Z_n}
\right)\right].
\label{EQ_estimate_energy_event}
\end{align}
In \eqref{EQ_estimate_energy_event}, (a) writes the Gamma CDF as an
event probability. Step (b) conditions on the three Gamma variables
and uses their independence from $Q$ to evaluate the conditional
probability by $F_Q$ at the displayed threshold.

We therefore need $\Pr\{Q\leq x\}$ for $x\geq0$.
TDD correlates ${\hyperref[SUBSEC_spatial]{\mathbf h}}$ with return training, so first condition
on the joint physical-channel and training information ${\symref{condition}{\mathcal C_0}}$
from \eqref{EQ_response_noise_covariance}. Let
$H_f:=\|{\hyperref[SUBSEC_spatial]{\mathbf h}}\|^2/S$ be the normalized forward energy.
The equivalent channel ${\hyperref[EQ_echo]{\mathbf k}}={\hyperref[EQ_response]{\kappa}} g{\hyperref[SUBSEC_spatial]{\mathbf h}}$ in \eqref{EQ_echo}
has normalized squared norm $\|{\hyperref[EQ_echo]{\mathbf k}}\|^2/v_0={\hyperref[EQ_physical_ratio]{\mathcal A}} H_f$, where
\begin{equation}
{\hyperref[EQ_physical_ratio]{\mathcal A}}:=\frac{{\hyperref[EQ_response]{\kappa}}^2|g|^2S}{v_0},
\qquad 0\leq{\hyperref[EQ_physical_ratio]{\mathcal A}}<S/{\hyperref[SUBSEC_channels]{\sigma^2}}.
\label{EQ_physical_ratio}
\end{equation}
The conditional Gaussian estimate in \eqref{EQ_regression_laws} gives
\begin{equation}
2E_0Q\mid{\symref{condition}{\mathcal C_0}}\sim\chi'^2_{2p}(2E_0{\hyperref[EQ_physical_ratio]{\mathcal A}} H_f).
\label{EQ_estimate_strength}
\end{equation}
Choose $0<{\hyperref[EQ_upper_scalar]{\theta}}<S/{\hyperref[SUBSEC_channels]{\sigma^2}}$ and write
$F_{{\hyperref[EQ_physical_ratio]{\mathcal A}}}({\hyperref[EQ_upper_scalar]{\theta}}):=\Pr\{{\hyperref[EQ_physical_ratio]{\mathcal A}}<{\hyperref[EQ_upper_scalar]{\theta}}\}$.
Using $\Phi_{k,\lambda}$ defined before \eqref{EQ_guard_law},
split according to ${\hyperref[EQ_physical_ratio]{\mathcal A}}$:
\begin{align}
\Pr\{Q\leq x\}
&=\Pr\{Q\leq x,{\hyperref[EQ_physical_ratio]{\mathcal A}}<{\hyperref[EQ_upper_scalar]{\theta}}\}
+\Pr\{Q\leq x,{\hyperref[EQ_physical_ratio]{\mathcal A}}\geq{\hyperref[EQ_upper_scalar]{\theta}}\}\notag\\
&\overset{\mathrm{(a)}}{\leq}F_{{\hyperref[EQ_physical_ratio]{\mathcal A}}}({\hyperref[EQ_upper_scalar]{\theta}})
\notag\\
&\quad+\mathbb E_{{\hyperref[EQ_physical_ratio]{\mathcal A}},H_f}\!\left[
\ind\{{\hyperref[EQ_physical_ratio]{\mathcal A}}\geq{\hyperref[EQ_upper_scalar]{\theta}}\}
\Phi_{2p,2E_0{\hyperref[EQ_physical_ratio]{\mathcal A}} H_f}(2E_0x)\right]\notag\\
&\overset{\mathrm{(b)}}{\leq}F_{{\hyperref[EQ_physical_ratio]{\mathcal A}}}({\hyperref[EQ_upper_scalar]{\theta}})
\notag\\
&\quad+\mathbb E_{{\hyperref[EQ_physical_ratio]{\mathcal A}},H_f}\!\left[
\ind\{{\hyperref[EQ_physical_ratio]{\mathcal A}}\geq{\hyperref[EQ_upper_scalar]{\theta}}\}
\Phi_{2p,2E_0{\hyperref[EQ_upper_scalar]{\theta}} H_f}(2E_0x)\right]\notag\\
&\overset{\mathrm{(c)}}{\leq}F_{{\hyperref[EQ_physical_ratio]{\mathcal A}}}({\hyperref[EQ_upper_scalar]{\theta}})
+\mathbb E_{H_f}\!\left[\Phi_{2p,2E_0{\hyperref[EQ_upper_scalar]{\theta}} H_f}(2E_0x)\right].
\label{EQ_cutoff_domination}
\end{align}
In \eqref{EQ_cutoff_domination}, (a) bounds the first event by
$F_{{\hyperref[EQ_physical_ratio]{\mathcal A}}}({\hyperref[EQ_upper_scalar]{\theta}})$ and uses \eqref{EQ_estimate_strength} for the second.
Step (b) decreases the noncentrality by replacing ${\hyperref[EQ_physical_ratio]{\mathcal A}}\geq{\hyperref[EQ_upper_scalar]{\theta}}$
with ${\hyperref[EQ_upper_scalar]{\theta}}$, which increases the CDF. Step (c) removes the indicator.
This leaves an expectation over the marginal $H_f$, without assuming
that $H_f$ and ${\hyperref[EQ_physical_ratio]{\mathcal A}}$ are independent.

To evaluate the last expectation in \eqref{EQ_cutoff_domination}, let
$\mathbf z_{\rm aux}\in\mathbb C^{1\times p}$ have independent $\CN(0,1)$ entries and be independent of ${\hyperref[SUBSEC_spatial]{\mathbf h}}$,
and define the auxiliary row
$\mathbf v_{\hyperref[EQ_upper_scalar]{\theta}}:=\sqrt{{\hyperref[EQ_upper_scalar]{\theta}}/S}{\hyperref[SUBSEC_spatial]{\mathbf h}}+E_0^{-1/2}\mathbf z_{\rm aux}$.
Its two terms are independent zero-mean Gaussian rows with covariances
${\hyperref[EQ_upper_scalar]{\theta}}\Id_p$ and $E_0^{-1}\Id_p$ by \eqref{EQ_effective_laws}. Hence
\begin{align}
\mathbb E_{H_f}[\Phi_{2p,2E_0{\hyperref[EQ_upper_scalar]{\theta}} H_f}(2E_0x)]
&\overset{\mathrm{(a)}}{=}\Pr\{\|\mathbf v_{\hyperref[EQ_upper_scalar]{\theta}}\|^2\leq x\}\notag\\
&\overset{\mathrm{(b)}}{=}G_p\!\left(\frac{x}{{\hyperref[EQ_upper_scalar]{\theta}}+E_0^{-1}}\right).
\label{EQ_cutoff_gaussian_average}
\end{align}
In \eqref{EQ_cutoff_gaussian_average}, (a) conditions on ${\hyperref[SUBSEC_spatial]{\mathbf h}}$,
for which $2E_0\|\mathbf v_{\hyperref[EQ_upper_scalar]{\theta}}\|^2$ has the displayed noncentral
chi-square distribution. In (b), the summed covariance gives
$\|\mathbf v_{\hyperref[EQ_upper_scalar]{\theta}}\|^2/({\hyperref[EQ_upper_scalar]{\theta}}+E_0^{-1})\sim\Gamma(p,1)$.
Substitution into \eqref{EQ_cutoff_domination} supplies the CDF bound
needed in \eqref{EQ_estimate_energy_event}.

\symdef{args}
The argument of $G_p$ obtained from \eqref{EQ_estimate_energy_event}
and \eqref{EQ_cutoff_gaussian_average} is affine in $Z_q$.
Write it as $d_0+e_0Z_q$, with common random factor $D>0$
and positive coefficients $d_0,e_0$ defined by
\begin{equation}
D:=\frac{{\hyperref[EQ_decision]{\eta}} c_XZ_{\hyperref[EQ_calibration]{\nu}}}{{\hyperref[EQ_calibration]{\nu}} Z_n({\hyperref[EQ_upper_scalar]{\theta}}+E_0^{-1})},\qquad
d_0:=D/b,\qquad e_0:=DS_T/{\hyperref[SUBSEC_channels]{\sigma_T^2}}.
\label{EQ_bound_arguments}
\end{equation}
\symdef{affine}
Let $Z_p\sim\Gamma(p,1)$ be independent of $Z_n,Z_{\hyperref[EQ_calibration]{\nu}},Z_q$ and,
for $x,y>0$, let ${\symref{affine}{\mathcal B_{p,q}}}(x,y):=\Pr\{Z_p\leq x+yZ_q\}$.
Then
\begin{align}
B_2&\overset{\mathrm{(a)}}{\leq}
F_{{\hyperref[EQ_physical_ratio]{\mathcal A}}}({\hyperref[EQ_upper_scalar]{\theta}})+\Pr\{Z_p\leq d_0+e_0Z_q\},\notag\\
{\hyperref[EQ_upper_scalar]{\overline P_2}}&\overset{\mathrm{(b)}}{=}
\min\{1,F_{{\hyperref[EQ_physical_ratio]{\mathcal A}}}({\hyperref[EQ_upper_scalar]{\theta}})+
\mathbb E_{Z_n,Z_{\hyperref[EQ_calibration]{\nu}}}[{\symref{affine}{\mathcal B_{p,q}}}(d_0,e_0)]\}.
\label{EQ_upper_scalar_proof}
\end{align}
In \eqref{EQ_upper_scalar_proof}, (a) applies
\eqref{EQ_cutoff_domination}--\eqref{EQ_cutoff_gaussian_average}
inside \eqref{EQ_estimate_energy_event} and uses \eqref{EQ_bound_arguments}.
Step (b) conditions on $Z_n,Z_{\hyperref[EQ_calibration]{\nu}}$ and caps the probability bound at one.
This proves \eqref{EQ_upper_scalar} while preserving the TDD dependence.

For numerical evaluation, calculate the probability
${\symref{affine}{\mathcal B_{p,q}}}(x,y)$ that now appears in \eqref{EQ_upper_scalar_proof}.
For integer $p,q$, with indices $0\leq k<p$ and $0\leq j\leq k$,
\begin{align}
{\symref{affine}{\mathcal B_{p,q}}}(x,y)
&=\mathbb E_{Z_q}[G_p(x+yZ_q)]\notag\\
&\overset{\mathrm{(a)}}{=}1-\mathbb E_{Z_q}\!\left[
e^{-(x+yZ_q)}\sum_{k=0}^{p-1}\frac{(x+yZ_q)^k}{k!}\right]\notag\\
&\overset{\mathrm{(b)}}{=}1-e^{-x}\sum_{k=0}^{p-1}\sum_{j=0}^k
\frac{x^{k-j}y^j\Gamma(q+j)}{(k-j)!j!\Gamma(q)(1+y)^{q+j}}.
\label{EQ_affine_gamma}
\end{align}
In \eqref{EQ_affine_gamma}, (a) integrates \eqref{EQ_appendix_gamma_cdf}
by parts for integer $p$ to write its complementary probability as a finite sum.
Step (b) expands the power and evaluates
$\mathbb E_{Z_q}[Z_q^je^{-yZ_q}]=\Gamma(q+j)/[\Gamma(q)(1+y)^{q+j}]$
using the Gamma density. Every deterministic cutoff ${\hyperref[EQ_upper_scalar]{\theta}}$ is valid.
Appendix~\ref{APP_numerical} evaluates $F_{{\hyperref[EQ_physical_ratio]{\mathcal A}}}({\hyperref[EQ_upper_scalar]{\theta}})$.
Together with \eqref{EQ_ball_proof} and \eqref{EQ_challenge_bound},
this completes the proof of Theorem~\ref{THM_echo_bound}.
\end{proof}

\section{One-Way Missed-Detection Probabilities}
\label{APP_oneway}
This appendix constructs a feasible one-way attack and compares its
acceptance probability with the two-way upper bound.

\subsection{Matched Noisy Preparation}
\label{APP_oneway_noisy}
We calculate the acceptance probability of the feasible attack in
Section~\ref{SUBSEC_oneway}. This gives a lower bound on the worst-case
one-way missed detection probability for comparison with ${\hyperref[EQ_upper_scalar]{\overline P_2}}$.
\begin{proof}[Derivation of the one-way missed detection probability]
\symdef{errors}
Let $E_1:={\hyperref[SUBSEC_oneway]{\widehat g_1}}-{\hyperref[SUBSEC_oneway]{\widehat g_0}}$ be the difference between the
gain Bob estimates from Trudy's forged pilot and the gain he estimated
from Alice's preparation. Let $t_1>0$ be the squared threshold in \eqref{EQ_one_threshold}.
The required probability is $P_{\mathrm{MD},1}({\hyperref[SUBSEC_oneway]{\rho}})=\Pr\{|E_1|^2\leq t_1\}$.
We determine the threshold $t_1$, the conditional distribution of $E_1$,
and then average over the random channels and preparation variance.

\emph{Step 1: identify the threshold and preparation uncertainty.}
\symdef{oneconstants}
\symdef{t1}
With ${\hyperref[EQ_one_squared_threshold]{\nu_1}}:=2m-1$ complex preparation residual degrees of freedom,
the squared acceptance threshold from \eqref{EQ_one_threshold} is
\begin{equation}
t_1:={\hyperref[SUBSEC_oneway]{\widehat v_1}}\left(\frac1{2n}+\frac1{2m}\right)
F_{2,2{\hyperref[EQ_one_squared_threshold]{\nu_1}}}^{-1}(1-{\hyperref[EQ_decision]{\alpha}}).
\label{EQ_one_squared_threshold}
\end{equation}
Its randomness comes from the preparation variance estimate
${\hyperref[SUBSEC_oneway]{\widehat v_1}}$. To determine its distribution, let
\symdef{Lambda}
${\symref{Lambda}{\Lambda}}:=\|{\hyperref[EQ_projection_proof2]{\boldsymbol\ell}}\|^2$ be the AN leakage energy and let
\symdef{vbeta}
$v_{\rm th}$ and $v_{\hyperref[SUBSEC_oneway]{\beta}}$ be the thermal and total noise variances
after normalization by the useful pilot. Section~\ref{SUBSEC_oneway} gives
\begin{equation}
v_{\rm th}:=\frac{{\hyperref[SUBSEC_channels]{\sigma^2}}}{(1-{\hyperref[SUBSEC_oneway]{\beta}})P_1},\qquad
v_{\hyperref[SUBSEC_oneway]{\beta}}:=v_{\rm th}+\frac{{\hyperref[SUBSEC_oneway]{\beta}}{\symref{Lambda}{\Lambda}}}{d(1-{\hyperref[SUBSEC_oneway]{\beta}})}.
\label{EQ_one_noise}
\end{equation}
\symdef{Z1}
Let $Z_1:={\hyperref[EQ_one_squared_threshold]{\nu_1}}{\hyperref[SUBSEC_oneway]{\widehat v_1}}/v_{\hyperref[SUBSEC_oneway]{\beta}}$ be the preparation residual
energy normalized by its noise variance. The sample mean and the
orthogonal residuals are independent Gaussian projections, as in
\eqref{EQ_regression_laws}. Conditional on the channels and initial training,
\begin{align}
Z_1&\overset{\mathrm{(a)}}{\sim}\Gamma({\hyperref[EQ_one_squared_threshold]{\nu_1}},1),\qquad
{\hyperref[SUBSEC_oneway]{\widehat v_1}}=v_{\hyperref[SUBSEC_oneway]{\beta}} Z_1/{\hyperref[EQ_one_squared_threshold]{\nu_1}},\notag\\
\left.\frac{|{\hyperref[SUBSEC_oneway]{\widehat g_1}}-{\hyperref[SUBSEC_oneway]{\widehat g_0}}|^2}
{{\hyperref[SUBSEC_oneway]{\widehat v_1}}(1/(2n)+1/(2m))}\right|{\hyperref[EQ_objective]{\mathcal H_0}}
&\overset{\mathrm{(b)}}{\sim}F_{2,2{\hyperref[EQ_one_squared_threshold]{\nu_1}}}=F_{2,4m-2}.
\label{EQ_one_FA_law}
\end{align}
In \eqref{EQ_one_FA_law}, (a) sums the ${\hyperref[EQ_one_squared_threshold]{\nu_1}}$ independent residual
squared magnitudes divided by $v_{\hyperref[SUBSEC_oneway]{\beta}}$.
In (b), the honest mean difference has variance
$v_{\hyperref[SUBSEC_oneway]{\beta}}(1/(2n)+1/(2m))$, so its normalized squared magnitude is
$\Gamma(1,1)$, independently of $Z_1$.
Their normalized ratio gives the F-distribution and hence false-alarm
probability ${\hyperref[EQ_decision]{\alpha}}$ for \eqref{EQ_one_squared_threshold}.

\emph{Step 2: determine the gain received from Trudy's pilot.}
Trudy chooses a public constant $a>0$ before the channel realization
as a guess for Bob's private preparation estimate ${\hyperref[SUBSEC_oneway]{\widehat g_0}}$.
To make the received gain have conditional mean $a$, she uses her
channel estimate to choose the transmit beam.
\symdef{hT}
Let ${\symref{hT}{\mathbf h_T}}^{\mathsf T}:={\hyperref[EQ_spatial]{\mathbf u_B}}^H{\hyperref[SUBSEC_channels]{\mathbf K_B}}\in\mathbb C^{1\times q}$
be her channel after Bob's receive combining, and let
${\hyperref[EQ_attack_error_fraction]{\widehat{\mathbf h}_T}}\in\mathbb C^q$ be its minimum mean-square error
estimate from the common reverse-link training.
\symdef{CSI}
To determine the received-gain error, let
$\widetilde{\mathbf h}_T\in\mathbb C^q$ be the pilot-normalized training
observation, with independent noise
$\mathbf n_T\sim\CN(\mathbf0,\sigma_T^2\Id_q/E_{\rm seed})$ and prior
$\mathbf h_T\sim\CN(\mathbf0,S_T\Id_q)$.
Let $\epsilon_T\in(0,1)$ be the estimation-error variance divided by $S_T$,
and let $Z_T>0$ be the estimate energy divided by its per-entry variance.
Then
\begin{align}
\widetilde{\mathbf h}_T
&\overset{\mathrm{(a)}}{=}\mathbf h_T+\mathbf n_T,
\label{EQ_attack_training_observation}\\
\epsilon_T&\overset{\mathrm{(b)}}{=}
\frac{\sigma_T^2}{S_TE_{\rm seed}+\sigma_T^2},\qquad
\widehat{\mathbf h}_T=(1-\epsilon_T)\widetilde{\mathbf h}_T,
\label{EQ_attack_error_fraction}\\
\mathbf h_T-\widehat{\mathbf h}_T\mid\widetilde{\mathbf h}_T
&\sim\CN(\mathbf0,S_T\epsilon_T\Id_q),\notag\\
\operatorname{Cov}(\widehat{\mathbf h}_T)
&\overset{\mathrm{(c)}}{=}(1-\epsilon_T)^2
(S_T+\sigma_T^2/E_{\rm seed})\Id_q\notag\\
&=S_T(1-\epsilon_T)\Id_q,\notag\\
\widehat{\mathbf h}_T&\overset{\mathrm{(d)}}{\sim}
\CN(\mathbf0,S_T(1-\epsilon_T)\Id_q),
\label{EQ_attack_estimate_law}\\
Z_T&:=\frac{\|\widehat{\mathbf h}_T\|^2}{S_T(1-\epsilon_T)}
\overset{\mathrm{(e)}}{\sim}\Gamma(q,1).
\label{EQ_attack_CSI}
\end{align}
In \eqref{EQ_attack_training_observation}--\eqref{EQ_attack_CSI},
(a) applies the pilot normalization in \eqref{EQ_training_pilots}
to \eqref{EQ_initial_channel_observation} and projects onto the unit
receive direction $\mathbf u_B$.
Step (b) applies the Gaussian conditioning calculation in
\eqref{EQ_Trudy_conditional_density} with prior covariance $S_T\Id_q$,
observation matrix $\Id_q$, and noise covariance
$\sigma_T^2\Id_q/E_{\rm seed}$.
Step (c) takes the covariance of the linear estimate in
\eqref{EQ_attack_error_fraction}, and (d) also uses its zero mean
and Gaussianity. Step (e) sums the $q$ normalized squared Gaussian
magnitudes. The Gaussian error has zero cross-covariance with the estimate
by \eqref{EQ_attack_error_fraction}, so they are independent.
Moreover, Trudy's channel is independent of the legitimate channels,
and the displayed distributions hold for every fixed unit $\mathbf u_B$.
Thus $Z_T$ is independent of $s,\Lambda$.
\symdef{beam}
She sends the public pilot with scale $\sqrt{(1-{\hyperref[SUBSEC_oneway]{\beta}})P_1}$ and beam
\begin{equation}
{\symref{beam}{\mathbf x_T}}(a):=a{\hyperref[EQ_attack_error_fraction]{\widehat{\mathbf h}_T}}^*/\|{\hyperref[EQ_attack_error_fraction]{\widehat{\mathbf h}_T}}\|^2
\in\mathbb C^q.
\label{EQ_feasible_beam}
\end{equation}
Let ${\hyperref[EQ_one_received_gain_error]{\delta_T}}:={\symref{hT}{\mathbf h_T}}^{\mathsf T}{\symref{beam}{\mathbf x_T}}(a)-a$ be the
received-gain error caused by Trudy's imperfect channel estimate. Then
\begin{align}
{\hyperref[EQ_one_received_gain_error]{\delta_T}}&\overset{\mathrm{(a)}}{=}
({\symref{hT}{\mathbf h_T}}-{\hyperref[EQ_attack_error_fraction]{\widehat{\mathbf h}_T}})^{\mathsf T}{\symref{beam}{\mathbf x_T}}(a),\notag\\
\mathbb E_{{\hyperref[EQ_one_received_gain_error]{\delta_T}}\mid{\hyperref[EQ_attack_error_fraction]{\widehat{\mathbf h}_T}}}[{\hyperref[EQ_one_received_gain_error]{\delta_T}}]&=0,\notag\\
\operatorname{Var}({\hyperref[EQ_one_received_gain_error]{\delta_T}}\mid{\hyperref[EQ_attack_error_fraction]{\widehat{\mathbf h}_T}})
&\overset{\mathrm{(b)}}{=}\frac{a^2S_T{\hyperref[EQ_attack_error_fraction]{\epsilon_T}}}{\|{\hyperref[EQ_attack_error_fraction]{\widehat{\mathbf h}_T}}\|^2}
=\frac{a^2{\hyperref[EQ_attack_error_fraction]{\epsilon_T}}}{(1-{\hyperref[EQ_attack_error_fraction]{\epsilon_T}})Z_T}.
\label{EQ_one_received_gain_error}
\end{align}
In \eqref{EQ_one_received_gain_error}, (a) uses
${\hyperref[EQ_attack_error_fraction]{\widehat{\mathbf h}_T}}^{\mathsf T}{\symref{beam}{\mathbf x_T}}(a)=a$ from \eqref{EQ_feasible_beam}.
Step (b) projects the independent Gaussian estimation error onto
that beam and substitutes \eqref{EQ_attack_CSI}.
The beam is finite almost surely and is admissible under the
unrestricted attack power in Section~\ref{SUBSEC_spatial}.
For $q\geq2$, its average transmit energy is also finite.

\emph{Step 3: derive the distribution of the gain difference $E_1$.}
To evaluate $\Pr\{|E_1|^2\leq t_1\}$, separate the errors in the two
gain estimates. Let $\overline e_0,\overline z_1\in\mathbb C$ be Bob's
normalized noise averages over the $2m$ preparation symbols and
$2n$ forged symbols, respectively. The forged block contains no
Alice signal or AN, whereas preparation retains AN leakage.
Given $(s,\Lambda,Z_T)$, the four errors
$\delta_T,g-s,\overline e_0,\overline z_1$ are independent and proper
Gaussian. This follows from \eqref{EQ_return_gain_conditional_density},
the independent attack-channel error in \eqref{EQ_attack_training_observation}--%
\eqref{EQ_attack_CSI}, and fresh receiver noises.
\symdef{V1}
Let $V_1>0$ denote the conditional variance of $E_1$. Then
\begin{align}
\widehat g_0&=g+\overline e_0,\qquad
\widehat g_1=a+\delta_T+\overline z_1,\notag\\
E_1&\overset{\mathrm{(a)}}{=}(a-s)+\delta_T-(g-s)-\overline e_0+\overline z_1,
\label{EQ_one_error_decomposition}\\
V_1&:=\operatorname{Var}(E_1\mid s,\Lambda,Z_T)\notag\\
&\overset{\mathrm{(b)}}{=}
\tau+\frac{v_\beta}{2m}+\frac{v_{\rm th}}{2n}
+\frac{a^2\epsilon_T}{(1-\epsilon_T)Z_T},
\label{EQ_one_attack_variance}\\
E_1\mid s,\Lambda,Z_T&\overset{\mathrm{(c)}}{\sim}\CN(a-s,V_1),\notag\\
\left.\frac{2|E_1|^2}{V_1}\right|s,\Lambda,Z_T
&\overset{\mathrm{(d)}}{\sim}\chi'^2_2(2(a-s)^2/V_1).
\label{EQ_one_law_proof}
\end{align}
In \eqref{EQ_one_error_decomposition}--\eqref{EQ_one_law_proof},
(a) subtracts the two gain estimates.
Step (b) adds the four independent error variances: $\tau$ from
\eqref{EQ_effective_laws}, $v_\beta/(2m)$ and $v_{\rm th}/(2n)$ from
averaging the independent noise samples in \eqref{EQ_one_noise}, and
the attack-channel variance in \eqref{EQ_one_received_gain_error}.
The fixed term $a-s$ contributes no conditional variance.
Step (c) uses the zero error means and Gaussianity.
Step (d) divides the real and imaginary parts by $\sqrt{V_1/2}$ and
sums their squares. The preparation mean and residual independence
in \eqref{EQ_one_FA_law} also makes $Z_1$ independent of $E_1$
given $(s,\Lambda,Z_T)$.

\emph{Step 4: average Bob's acceptance probability.}
Using the CDF $\Phi_{k,\lambda}$ defined before \eqref{EQ_guard_law},
condition on the variables that determine $t_1,V_1$:
\begin{align}
P_{\mathrm{MD},1}({\hyperref[SUBSEC_oneway]{\rho}})&=\Pr\{|E_1|^2\leq t_1\}\notag\\
&\overset{\mathrm{(a)}}{=}\mathbb E_{s,{\symref{Lambda}{\Lambda}},Z_T,Z_1}
[\Pr\{|E_1|^2\leq t_1\mid s,{\symref{Lambda}{\Lambda}},Z_T,Z_1\}]\notag\\
&\overset{\mathrm{(b)}}{=}\mathbb E_{s,{\symref{Lambda}{\Lambda}},Z_T,Z_1}
\left[\Phi_{2,2(a-s)^2/V_1}(2t_1/V_1)\right].
\label{EQ_one_lower}
\end{align}
In \eqref{EQ_one_lower}, (a) uses total probability and (b) substitutes
\eqref{EQ_one_law_proof} and the threshold \eqref{EQ_one_squared_threshold}.
Finally, attack feasibility and Theorem~\ref{THM_echo_bound} give
\begin{align}
P_{\mathrm{MD},1}({\hyperref[SUBSEC_oneway]{\rho}})&\overset{\mathrm{(a)}}{\leq}P_{\mathrm{MD},1}^{\star}({\hyperref[SUBSEC_oneway]{\rho}}),\notag\\
P_2^\star({\hyperref[SUBSEC_oneway]{\rho}})&\overset{\mathrm{(b)}}{\leq}{\hyperref[EQ_upper_scalar]{\overline P_2}}({\hyperref[SUBSEC_oneway]{\rho}}).
\label{EQ_comparison_proof}
\end{align}
In \eqref{EQ_comparison_proof}, (a) includes \eqref{EQ_feasible_beam}
among the admissible attacks, and (b) uses \eqref{EQ_upper_scalar}.
Consequently, the sufficient comparison condition is
\begin{equation}
\overline P_2(\rho)<P_{\mathrm{MD},1}(\rho)
\quad\Longrightarrow\quad
P_2^\star(\rho)<P_{\mathrm{MD},1}^{\star}(\rho).
\label{EQ_comparison}
\end{equation}
\end{proof}

\subsection{Proof of Proposition~\ref{PROP_two_way_comparison}}
\label{APP_comparison_proof}
\begingroup
\allowdisplaybreaks[4]

We compare quadratic and linear bounds to solve the sufficient
inequality in \eqref{EQ_comparison}.
\begin{proof}
Assume $n>p$, $1\leq {\hyperref[SEC_system_model]{N_R}}<p$, and $q\geq1$.
Vary only ${\hyperref[SUBSEC_channels]{S_T}}={\hyperref[SUBSEC_oneway]{\rho}} {\hyperref[EQ_TDD_channels]{S}}$, keeping other model and design parameters
fixed. Derived quantities such as ${\hyperref[EQ_attack_error_fraction]{\epsilon_T}}$ vary accordingly.
For the fixed $R>0$ in Proposition~\ref{PROP_two_way_comparison},
we establish
\begin{equation}
\begin{aligned}
{{\hyperref[EQ_upper_scalar]{\overline P_2}}}({\hyperref[SUBSEC_oneway]{\rho}})&\leq {\hyperref[EQ_explicit_quadratic_coefficients]{A_0}}+{\hyperref[EQ_explicit_quadratic_coefficients]{A_1}}{\hyperref[SUBSEC_oneway]{\rho}}+{\hyperref[EQ_explicit_quadratic_coefficients]{A_2}}{\hyperref[SUBSEC_oneway]{\rho}}^2,\\
{{\hyperref[SUBSEC_oneway]{P_{\mathrm{MD},1}}}}({\hyperref[SUBSEC_oneway]{\rho}})&\geq c\rho,
\qquad 0<{\hyperref[SUBSEC_oneway]{\rho}}\leq R.
\end{aligned}
\label{EQ_explicit_envelopes_proof}
\end{equation}

\emph{Step 1: bound the probability $F_{{\hyperref[EQ_physical_ratio]{\mathcal A}}}({\hyperref[EQ_upper_scalar]{\theta}})$.}
The two-way bound \eqref{EQ_upper_scalar} has two contributions:
$F_{{\hyperref[EQ_physical_ratio]{\mathcal A}}}({\hyperref[EQ_upper_scalar]{\theta}})$ and an expectation of $G_p$.
We first bound $F_{{\hyperref[EQ_physical_ratio]{\mathcal A}}}({\hyperref[EQ_upper_scalar]{\theta}})$ by a constant independent of ${\hyperref[SUBSEC_oneway]{\rho}}$.
For $0<{\hyperref[EQ_upper_scalar]{\theta}}<S/{\hyperref[SUBSEC_channels]{\sigma^2}}$, let
$c_{\hyperref[EQ_upper_scalar]{\theta}}:={\hyperref[EQ_upper_scalar]{\theta}}/[{\hyperref[EQ_response]{\kappa}}^2(S-{\hyperref[EQ_upper_scalar]{\theta}}{\hyperref[SUBSEC_channels]{\sigma^2}})]$.
The definition of ${\hyperref[EQ_physical_ratio]{\mathcal A}}$ in \eqref{EQ_physical_ratio}, with $v_0$ from \eqref{EQ_response_noise_covariance}, gives
$\{{\hyperref[EQ_physical_ratio]{\mathcal A}}<{\hyperref[EQ_upper_scalar]{\theta}}\}=\{|g|^2<c_{\hyperref[EQ_upper_scalar]{\theta}}({\hyperref[SUBSEC_channels]{\sigma^2}}+\beta_AP_A{\symref{Lambda}{\Lambda}}/d)\}$.
Choose $t>0$ with $tc_{\hyperref[EQ_upper_scalar]{\theta}}\beta_AP_A{\hyperref[EQ_training]{\tau}}/d<1$. To separate the
gain and leakage averages, bound the event indicator by an exponential:
\begin{align}
F_{{\hyperref[EQ_physical_ratio]{\mathcal A}}}({\hyperref[EQ_upper_scalar]{\theta}})
&=\Pr\{|g|^2<c_{\hyperref[EQ_upper_scalar]{\theta}}({\hyperref[SUBSEC_channels]{\sigma^2}}+\beta_AP_A{\symref{Lambda}{\Lambda}}/d)\}\notag\\
&\overset{\mathrm{(a)}}{=}\mathbb E_{g,{\symref{Lambda}{\Lambda}}}
[\ind\{|g|^2<c_{\hyperref[EQ_upper_scalar]{\theta}}({\hyperref[SUBSEC_channels]{\sigma^2}}+\beta_AP_A{\symref{Lambda}{\Lambda}}/d)\}]\notag\\
&\overset{\mathrm{(b)}}{\leq}\mathbb E_{g,{\symref{Lambda}{\Lambda}}}
[e^{-t|g|^2+tc_{\hyperref[EQ_upper_scalar]{\theta}}({\hyperref[SUBSEC_channels]{\sigma^2}}+\beta_AP_A{\symref{Lambda}{\Lambda}}/d)}]
\label{EQ_exponential_indicator_bound}\\
&\overset{\mathrm{(c)}}{=}
\frac{e^{tc_{\hyperref[EQ_upper_scalar]{\theta}}{\hyperref[SUBSEC_channels]{\sigma^2}}}}{1+t{\hyperref[EQ_training]{\tau}}}
\mathbb E_s[e^{-ts^2/(1+t{\hyperref[EQ_training]{\tau}})}]
\mathbb E_{\symref{Lambda}{\Lambda}}[e^{tc_{\hyperref[EQ_upper_scalar]{\theta}}\beta_AP_A{\symref{Lambda}{\Lambda}}/d}].
\label{EQ_amplification_exponential_reduction}
\end{align}
In \eqref{EQ_exponential_indicator_bound}--\eqref{EQ_amplification_exponential_reduction},
(a) writes the probability as an indicator expectation.
In (b), the exponential exceeds one on the indicated event and is
positive elsewhere. Step (c) uses the independence of ${\symref{Lambda}{\Lambda}}$ and
$(s,g)$ from \eqref{EQ_return_gain_conditional_density} and
$\mathbb E_{g\mid s}[e^{-t|g|^2}]=(1+t{\hyperref[EQ_training]{\tau}})^{-1}e^{-ts^2/(1+t{\hyperref[EQ_training]{\tau}})}$
from \eqref{EQ_effective_laws}.

We retain the expectation over $s$ in
\eqref{EQ_amplification_exponential_reduction} and evaluate the
leakage expectation. Since
$\Lambda/\tau\sim\Gamma(d,1)$ as derived after \eqref{EQ_return_gain_conditional_density},
\begin{align}
\mathbb E_\Lambda[e^{tc_\theta\beta_AP_A\Lambda/d}]
&\overset{\mathrm{(a)}}{=}\frac1{\Gamma(d)}\int_0^\infty
v^{d-1}e^{-(1-tc_\theta\beta_AP_A\tau/d)v}\,\mathrm dv\notag\\
&\overset{\mathrm{(b)}}{=}(1-tc_\theta\beta_AP_A\tau/d)^{-d}.
\label{EQ_leakage_exponential_expectation}
\end{align}
In \eqref{EQ_leakage_exponential_expectation}, (a) uses the density
of $v=\Lambda/\tau$, and (b) evaluates the Gamma integral.
The restriction on $t$ makes it finite.
\symdef{epsilonA}
Substitution into \eqref{EQ_amplification_exponential_reduction} gives
\begin{equation}
F_{\mathcal A}(\theta)\leq
\frac{e^{tc_\theta\sigma^2}\mathbb E_s[e^{-ts^2/(1+t\tau)}]}
{(1+t\tau)(1-tc_\theta\beta_AP_A\tau/d)^d}
=\varepsilon_A.
\label{EQ_explicit_amplification_tail}
\end{equation}
The coefficient $\varepsilon_A$ is independent of $\rho$ because the legitimate-channel and
training distributions are fixed. Appendix~\ref{APP_numerical}
evaluates the expectation for the numerical setting.

\emph{Step 2: bound the second term in ${\hyperref[EQ_upper_scalar]{\overline P_2}}({\hyperref[SUBSEC_oneway]{\rho}})$ by a polynomial.}
Next bound the expectation of ${\hyperref[EQ_appendix_gamma_cdf]{G_p}}$ in
\eqref{EQ_upper_scalar}. Its argument, after setting
$S_T=\rho S$, is
$\eta c_XZ_\nu(b^{-1}+\rho SZ_q/\sigma_T^2)/
[\nu Z_n(\theta+E_0^{-1})]$.
To obtain powers of $\rho$, first bound $G_p(x)$ by a power of
its nonnegative argument $x$:
\begin{align}
{\hyperref[EQ_appendix_gamma_cdf]{G_p}}(x)
&=\frac1{\Gamma(p)}\int_0^x v^{p-1}e^{-v}\,\mathrm dv\notag\\
&\overset{\mathrm{(a)}}{\leq}
\frac1{\Gamma(p)}\int_0^x v^{p-1}\,\mathrm dv
\overset{\mathrm{(b)}}{=}\frac{x^p}{\Gamma(p+1)}.
\label{EQ_explicit_gamma_cdf_bound}
\end{align}
In \eqref{EQ_explicit_gamma_cdf_bound}, (a) uses $e^{-v}\leq1$
for $v\geq0$, and (b) integrates and uses $\Gamma(p+1)=p\Gamma(p)$.
Substitution into \eqref{EQ_upper_scalar} raises $Z_\nu/Z_n$ to
the power $p$, so the required expectation is
\begin{align}
\mathbb E_{{\hyperref[EQ_regression_laws]{Z_\nu}},{\hyperref[EQ_estimate_energy_event]{Z_n}}}[({\hyperref[EQ_regression_laws]{Z_\nu}}/{\hyperref[EQ_estimate_energy_event]{Z_n}})^p]
&\overset{\mathrm{(a)}}{=}\mathbb E_{{\hyperref[EQ_regression_laws]{Z_\nu}}}[{\hyperref[EQ_regression_laws]{Z_\nu}}^p]\mathbb E_{{\hyperref[EQ_estimate_energy_event]{Z_n}}}[{\hyperref[EQ_estimate_energy_event]{Z_n}}^{-p}]
\notag\\
&\overset{\mathrm{(b)}}{=}
\frac{\Gamma({\hyperref[EQ_calibration]{\nu}}+p)\Gamma(n-p)}{\Gamma({\hyperref[EQ_calibration]{\nu}})\Gamma(n)}.
\label{EQ_explicit_gamma_ratio_moment}
\end{align}
In \eqref{EQ_explicit_gamma_ratio_moment}, (a) uses the independence
in \eqref{EQ_upper_scalar}, and (b) integrates the powers against
the Gamma densities. The inverse moment is finite because $n>p$.

To express the bound as a polynomial in $\rho$, define the
positive factor $M_\theta$, coefficients $w_j$ for $0\leq j\leq p$,
and quadratic coefficients $A_0,A_1,A_2$ by
\begin{align}
M_\theta&:=\left[\frac{\eta c_X}{\nu(\theta+E_0^{-1})}\right]^p
\frac{\Gamma(\nu+p)\Gamma(n-p)}
{\Gamma(\nu)\Gamma(n)\Gamma(p+1)},
\label{EQ_explicit_moment_factor}\\
w_j&:=M_\theta\binom{p}{j}b^{-(p-j)}
\left(\frac{S}{\sigma_T^2}\right)^j
\frac{\Gamma(q+j)}{\Gamma(q)},
\label{EQ_explicit_polynomial_coefficients}\\
A_0&:=\varepsilon_A+w_0,\qquad A_1:=w_1,\qquad
A_2:=\sum_{j=2}^{p}w_jR^{j-2}.
\label{EQ_explicit_quadratic_coefficients}
\end{align}
Combining \eqref{EQ_explicit_amplification_tail},
\eqref{EQ_explicit_gamma_cdf_bound}, and
\eqref{EQ_explicit_gamma_ratio_moment} gives
\begin{align}
{{\hyperref[EQ_upper_scalar]{\overline P_2}}}({\hyperref[SUBSEC_oneway]{\rho}})
&\overset{\mathrm{(a)}}{\leq}{\symref{epsilonA}{\varepsilon_A}}+
\mathbb E_{{\hyperref[EQ_regression_laws]{Z_\nu}},{\hyperref[EQ_estimate_energy_event]{Z_n}},{\hyperref[EQ_directional_precision]{Z_q}}}\!\biggl[
{\hyperref[EQ_appendix_gamma_cdf]{G_p}}\!\Bigl(\frac{{\hyperref[EQ_decision]{\eta}} c_XZ_\nu(b^{-1}+{\hyperref[SUBSEC_oneway]{\rho}} SZ_q/{\hyperref[SUBSEC_channels]{\sigma_T^2}})}
{{\hyperref[EQ_calibration]{\nu}} {\hyperref[EQ_estimate_energy_event]{Z_n}}({\hyperref[EQ_upper_scalar]{\theta}}+{\hyperref[EQ_private_probe]{E_0}}^{-1})}\Bigr)\biggr]\notag\\
&\overset{\mathrm{(b)}}{\leq}{\symref{epsilonA}{\varepsilon_A}}+
M_{\hyperref[EQ_upper_scalar]{\theta}}\mathbb E_{{\hyperref[EQ_directional_precision]{Z_q}}}[(b^{-1}+{\hyperref[SUBSEC_oneway]{\rho}} SZ_q/{\hyperref[SUBSEC_channels]{\sigma_T^2}})^p]
\notag\\
&\overset{\mathrm{(c)}}{=}{\symref{epsilonA}{\varepsilon_A}}+\sum_{j=0}^pw_j\rho^j
\notag\\
&\overset{\mathrm{(d)}}{\leq}{\hyperref[EQ_explicit_quadratic_coefficients]{A_0}}+{\hyperref[EQ_explicit_quadratic_coefficients]{A_1}}{\hyperref[SUBSEC_oneway]{\rho}}+{\hyperref[EQ_explicit_quadratic_coefficients]{A_2}}{\hyperref[SUBSEC_oneway]{\rho}}^2.
\label{EQ_explicit_two_way_polynomial}
\end{align}
In \eqref{EQ_explicit_two_way_polynomial}, (a) uses
\eqref{EQ_upper_scalar} and \eqref{EQ_explicit_amplification_tail},
omitting the minimum with one.
Step (b) uses \eqref{EQ_explicit_gamma_cdf_bound} and
\eqref{EQ_explicit_gamma_ratio_moment}, giving $M_{\hyperref[EQ_upper_scalar]{\theta}}$ in
\eqref{EQ_explicit_moment_factor}.
Step (c) expands the power and uses
$\mathbb E_{{\hyperref[EQ_directional_precision]{Z_q}}}[{\hyperref[EQ_directional_precision]{Z_q}}^j]=\Gamma(q+j)/\Gamma(q)$, giving
\eqref{EQ_explicit_polynomial_coefficients}.
Step (d) uses ${\hyperref[SUBSEC_oneway]{\rho}}^j\leq R^{j-2}{\hyperref[SUBSEC_oneway]{\rho}}^2$ for $j\geq2$ and
\eqref{EQ_explicit_quadratic_coefficients}.

\emph{Step 3: lower-bound ${\hyperref[SUBSEC_oneway]{P_{\mathrm{MD},1}}}({\hyperref[SUBSEC_oneway]{\rho}})$ by $c\rho$.}
Starting from \eqref{EQ_one_lower}, we retain a positive-probability
set of channel and preparation realizations on which a fixed disk
is accepted. A uniform lower bound on the conditional Gaussian
density over this disk then lower-bounds $P_{\mathrm{MD},1}(\rho)$.
To control the disk and density, choose
fixed gain tolerance $h>0$, leakage limit ${\hyperref[EQ_explicit_one_way_event]{\lambda_+}}>0$,
estimate-energy limits $0<{\hyperref[EQ_explicit_one_way_event]{z_-}}<{\hyperref[EQ_explicit_one_way_event]{z_+}}$, and a positive constant $z_v>0$. To lower-bound $P_{\mathrm{MD},1}$ in \eqref{EQ_one_lower},
we count outcomes in $\mathcal E$ whose squared gain difference
$|E_1|^2$ is at most a fixed threshold $t_->0$. If $t_-\leq t_1$
on $\mathcal E$, their probability satisfies
$\Pr\{\mathcal E,\ |E_1|^2\leq t_-\}\leq P_{\mathrm{MD},1}.
$
We ensure $t_1\geq t_-$ by including $Z_1\geq z_v$ in
$\mathcal E$, using \eqref{EQ_one_FA_law} and
\eqref{EQ_one_squared_threshold}.
\symdef{event}
Define the event
\begin{equation}
{{\symref{event}{\mathcal E}}}:=
\{|s-a|\leq h,\ {\symref{Lambda}{\Lambda}}\leq\lambda_+,\ {\hyperref[EQ_explicit_one_way_event]{z_-}}\leq {\hyperref[EQ_attack_CSI]{Z_T}}\leq {\hyperref[EQ_explicit_one_way_event]{z_+}},\ {\symref{Z1}{Z_1}}\geq {\hyperref[EQ_explicit_one_way_event]{z_v}}\}.
\label{EQ_explicit_one_way_event}
\end{equation}
\symdef{pis}\symdef{pie}
Let $\pi_s$ and $\pi_{\mathcal E}$ denote the gain-interval and event
probabilities. By the independent distributions in \eqref{EQ_one_lower},
\begin{align}
\pi_s&:=\Pr\{|s-a|\leq h\},
\label{EQ_explicit_gain_interval}\\
\pi_{\mathcal E}&:=\Pr\{\mathcal E\}\notag\\
&\overset{\mathrm{(a)}}{=}\pi_sG_d(\lambda_+/\tau)
[G_q(z_+)-G_q(z_-)]\notag\\
&\quad\times[1-G_{\nu_1}(z_v)]>0.
\label{EQ_explicit_event_probability}
\end{align}
In \eqref{EQ_explicit_event_probability}, (a) multiplies the four
independent event probabilities in \eqref{EQ_explicit_one_way_event}.
The Gaussian training matrix has positive density everywhere, so its
largest singular value assigns positive probability to every open
interval in $(0,\infty)$. Thus $a,h>0$ imply $\pi_s>0$.
The other three probabilities in \eqref{EQ_explicit_event_probability}
are positive because $\lambda_+>0$, $0<z_-<z_+$, and
$0<z_v<\infty$.

To lower-bound the density of $E_1$ in \eqref{EQ_one_error_decomposition}, whose conditional
distribution is given in \eqref{EQ_one_law_proof}, we need a fixed acceptance region $\{E_1\in\mathbb C:|E_1|^2\leq t_-\}$.
\symdef{Cbeta}
The fixed threshold $t_-$ and positive variance constants $C_+,D_0$ are
\begin{equation}
\begin{aligned}
t_-&:=v_{\rm th}\frac{z_v}{\nu_1}
\left(\frac1{2n}+\frac1{2m}\right)F_{2,2\nu_1}^{-1}(1-\alpha),\\
C_+&:=\tau+v_{\rm th}\left(\frac1{2m}+\frac1{2n}\right)
+\frac{\beta\lambda_+}{2md(1-\beta)},\\
D_0&:=\frac{a^2\sigma_T^2}{SE_{\rm seed}}.
\end{aligned}
\label{EQ_explicit_main_one_way_constants}
\end{equation}
Define the noise-variance upper bound ${\symref{Cbeta}{v_{\beta,+}}}$ and the
combined variance $C_\beta$ of the legitimate return-gain
error $g-s$ and Bob's preparation and current noise averages. On ${\symref{event}{\mathcal E}}$,
\begin{align}
{\symref{Cbeta}{v_{\beta,+}}}&:={\hyperref[EQ_one_noise]{v_{\rm th}}}+\frac{{\hyperref[SUBSEC_oneway]{\beta}}{{\hyperref[EQ_explicit_one_way_event]{\lambda_+}}}}{d(1-{\hyperref[SUBSEC_oneway]{\beta}})},\,\,{\symref{Cbeta}{C_\beta}}:={\hyperref[EQ_training]{\tau}}+\frac{{\symref{vbeta}{v_\beta}}}{2m}+\frac{{\hyperref[EQ_one_noise]{v_{\rm th}}}}{2n},
\label{EQ_explicit_one_way_constants}\\
{\symref{t1}{t_1}}&\overset{\mathrm{(a)}}{=}
\frac{{\symref{vbeta}{v_\beta}} {\symref{Z1}{Z_1}}}{{\hyperref[EQ_one_squared_threshold]{\nu_1}}}\left(\frac1{2n}+\frac1{2m}\right)
F_{2,2{\hyperref[EQ_one_squared_threshold]{\nu_1}}}^{-1}(1-{\hyperref[EQ_decision]{\alpha}})\geq {\hyperref[EQ_explicit_main_one_way_constants]{t_-}},\notag\\
{\symref{V1}{V_1}}({\hyperref[SUBSEC_oneway]{\rho}})&\overset{\mathrm{(b)}}{=}{\symref{Cbeta}{C_\beta}}+\frac{{\hyperref[EQ_explicit_main_one_way_constants]{D_0}}}{{\hyperref[SUBSEC_oneway]{\rho}} {\hyperref[EQ_attack_CSI]{Z_T}}},\notag\\
\frac{{\hyperref[EQ_explicit_main_one_way_constants]{D_0}}}{{\hyperref[SUBSEC_oneway]{\rho}} {\hyperref[EQ_explicit_one_way_event]{z_+}}}&\overset{\mathrm{(c)}}{\leq}{\symref{V1}{V_1}}({\hyperref[SUBSEC_oneway]{\rho}})
\leq {\hyperref[EQ_explicit_main_one_way_constants]{C_+}}+\frac{{\hyperref[EQ_explicit_main_one_way_constants]{D_0}}}{{\hyperref[SUBSEC_oneway]{\rho}} {\hyperref[EQ_explicit_one_way_event]{z_-}}}.
\label{EQ_explicit_variance_bounds}
\end{align}
In \eqref{EQ_explicit_one_way_constants}--\eqref{EQ_explicit_variance_bounds}, (a) uses
\eqref{EQ_one_squared_threshold}, ${\symref{vbeta}{v_\beta}}\geq {\hyperref[EQ_one_noise]{v_{\rm th}}}$,
and ${\symref{Z1}{Z_1}}\geq {\hyperref[EQ_explicit_one_way_event]{z_v}}$.
Step (b) substitutes ${\hyperref[EQ_attack_error_fraction]{\epsilon_T}}/(1-{\hyperref[EQ_attack_error_fraction]{\epsilon_T}})
={\hyperref[SUBSEC_channels]{\sigma_T^2}}/({\hyperref[SUBSEC_oneway]{\rho}} SE_{\rm seed})$ from \eqref{EQ_attack_error_fraction}
into \eqref{EQ_one_attack_variance}.
Step (c) uses ${\symref{Cbeta}{C_\beta}}\leq {\hyperref[EQ_explicit_main_one_way_constants]{C_+}}$ and ${\hyperref[EQ_explicit_one_way_event]{z_-}}\leq {\hyperref[EQ_attack_CSI]{Z_T}}\leq {\hyperref[EQ_explicit_one_way_event]{z_+}}$.

\symdef{densityrho}
Let ${\symref{densityrho}{f_\rho}}(z)$ be the density function of ${\symref{errors}{E_1}}$ conditional on
$s,{\symref{Lambda}{\Lambda}},{\hyperref[EQ_attack_CSI]{Z_T}},{\symref{Z1}{Z_1}}$, with $z\in\mathbb C$.
For $|z|^2\leq {\hyperref[EQ_explicit_main_one_way_constants]{t_-}}$, ${\symref{event}{\mathcal E}}$, and $0<{\hyperref[SUBSEC_oneway]{\rho}}\leq R$,
\begin{align}
{{\symref{densityrho}{f_\rho}}}(z)
&\overset{\mathrm{(a)}}{=}\frac{e^{-|z-(a-s)|^2/{\symref{V1}{V_1}}({\hyperref[SUBSEC_oneway]{\rho}})}}{\pi {\symref{V1}{V_1}}({\hyperref[SUBSEC_oneway]{\rho}})}
\notag\\
&\overset{\mathrm{(b)}}{\geq}\frac{e^{-(h+\sqrt{{\hyperref[EQ_explicit_main_one_way_constants]{t_-}}})^2/{\symref{V1}{V_1}}({\hyperref[SUBSEC_oneway]{\rho}})}}{\pi {\symref{V1}{V_1}}({\hyperref[SUBSEC_oneway]{\rho}})}
\notag\\
&\overset{\mathrm{(c)}}{\geq}
\frac{{\hyperref[SUBSEC_oneway]{\rho}} {\hyperref[EQ_explicit_one_way_event]{z_-}}}{\pi({\hyperref[EQ_explicit_main_one_way_constants]{D_0}}+C_+Rz_-)}
\exp\!\left[-\frac{Rz_+(h+\sqrt{{\hyperref[EQ_explicit_main_one_way_constants]{t_-}}})^2}{{\hyperref[EQ_explicit_main_one_way_constants]{D_0}}}\right].
\label{EQ_explicit_density_lower}
\end{align}
In \eqref{EQ_explicit_density_lower}, (a) uses
\eqref{EQ_one_law_proof}, and (b) uses
$|z-(a-s)|\leq\sqrt{{\hyperref[EQ_explicit_main_one_way_constants]{t_-}}}+h$.
Step (c) applies the upper variance bound in
\eqref{EQ_explicit_variance_bounds} to the prefactor and the
lower variance bound to the exponent, then uses ${\hyperref[SUBSEC_oneway]{\rho}}\leq R$.

Integrating this density with the notation $\mathrm d^2z$ for area
integration on $\mathbb C$, we obtain
\begin{align}
{\hyperref[SUBSEC_oneway]{P_{\mathrm{MD},1}^{\star}}}(\rho)
&\overset{\mathrm{(a)}}{\geq}
{\hyperref[SUBSEC_oneway]{P_{\mathrm{MD},1}}}(\rho)\notag\\
&\overset{\mathrm{(b)}}{\geq}
\mathbb E_{s,{\symref{Lambda}{\Lambda}},
{\hyperref[EQ_attack_CSI]{Z_T}},{\symref{Z1}{Z_1}}}
\!\left[
{\symref{notation}{\ind}}_{\symref{event}{\mathcal E}}
\int_{|z|^2\leq t_-}
{\symref{densityrho}{f_\rho}}(z)\,\mathrm d^2z
\right]\notag\\
&\overset{\mathrm{(c)}}{\geq}
\frac{\rho z_-}{\pi(D_0+C_+Rz_-)}
e^{-Rz_+(h+\sqrt{t_-})^2/D_0}\notag\\
&\quad\times
\mathbb E_{s,\Lambda,Z_T,Z_1}\!\left[
\ind_{\mathcal E}
\int_{|z|^2\leq t_-}1\,\mathrm d^2z
\right]\notag\\
&\overset{\mathrm{(d)}}{=}
\frac{\rho t_-z_-}{D_0+C_+Rz_-}
e^{-Rz_+(h+\sqrt{t_-})^2/D_0}\notag\\
&\quad\times\mathbb E_{s,\Lambda,Z_T,Z_1}[\ind_{\mathcal E}]
\notag\\
&\overset{\mathrm{(e)}}{=}
\frac{{\symref{pie}{\pi_{\mathcal E}}}\rho t_-z_-}
{D_0+C_+Rz_-}
e^{-Rz_+(h+\sqrt{t_-})^2/D_0}
\overset{\mathrm{(f)}}{=:}c\rho.
\label{EQ_explicit_one_way_linear}
\end{align}
In \eqref{EQ_explicit_one_way_linear}, (a) uses attack feasibility.
Step (b) restricts \eqref{EQ_one_lower} to $\mathcal E$ and
$|z|^2\leq t_-$, which implies acceptance by
\eqref{EQ_explicit_variance_bounds}.
Step (c) substitutes the uniform density lower bound in
\eqref{EQ_explicit_density_lower}.
Step (d) uses
$\int_{|z|^2\leq t_-}1\,\mathrm d^2z=\pi t_-$.
Step (e) uses
$\mathbb E_{s,\Lambda,Z_T,Z_1}[\ind_{\mathcal E}]
=\Pr\{\mathcal E\}=\pi_{\mathcal E}$
from \eqref{EQ_explicit_event_probability}.
Step (f) defines the positive coefficient $c$, which is independent of $\rho$.

\emph{Step 4: solve the sufficient inequality for ${\hyperref[SUBSEC_oneway]{\rho}}$.}
The bounds in \eqref{EQ_explicit_envelopes} give a strict
comparison when ${\hyperref[EQ_explicit_quadratic_coefficients]{A_0}}+({\hyperref[EQ_explicit_quadratic_coefficients]{A_1}}-c){\hyperref[SUBSEC_oneway]{\rho}}+{\hyperref[EQ_explicit_quadratic_coefficients]{A_2}}{\hyperref[SUBSEC_oneway]{\rho}}^2<0$.
\symdef{discriminant}
The condition $c>{\hyperref[EQ_explicit_quadratic_coefficients]{A_1}}+2\sqrt{{{\hyperref[EQ_explicit_quadratic_coefficients]{A_0}}}{\hyperref[EQ_explicit_quadratic_coefficients]{A_2}}}$ makes its discriminant
${\symref{discriminant}{\Delta_\rho}}:=(c-{\hyperref[EQ_explicit_quadratic_coefficients]{A_1}})^2-4{{\hyperref[EQ_explicit_quadratic_coefficients]{A_0}}}{\hyperref[EQ_explicit_quadratic_coefficients]{A_2}}$ positive and its roots
${\hyperref[EQ_explicit_roots]{\rho_-}},{\hyperref[EQ_explicit_roots]{\rho_+}}$ in \eqref{EQ_explicit_roots} distinct and positive.
For ${\hyperref[EQ_explicit_roots]{\rho_-}}<{\hyperref[SUBSEC_oneway]{\rho}}<{\hyperref[EQ_explicit_roots]{\rho_+}}$ and ${\hyperref[SUBSEC_oneway]{\rho}}\leq R$,
\begin{align}
{\hyperref[EQ_upper_scalar]{\overline P_2}}({\hyperref[SUBSEC_oneway]{\rho}})-{\hyperref[SUBSEC_oneway]{P_{\mathrm{MD},1}}}({\hyperref[SUBSEC_oneway]{\rho}})
&\overset{\mathrm{(a)}}{\leq}{\hyperref[EQ_explicit_quadratic_coefficients]{A_0}}+({\hyperref[EQ_explicit_quadratic_coefficients]{A_1}}-c){\hyperref[SUBSEC_oneway]{\rho}}+{\hyperref[EQ_explicit_quadratic_coefficients]{A_2}}{\hyperref[SUBSEC_oneway]{\rho}}^2\notag\\
&\overset{\mathrm{(b)}}{=}{\hyperref[EQ_explicit_quadratic_coefficients]{A_2}}({\hyperref[SUBSEC_oneway]{\rho}}-{\hyperref[EQ_explicit_roots]{\rho_-}})({\hyperref[SUBSEC_oneway]{\rho}}-{\hyperref[EQ_explicit_roots]{\rho_+}})
\overset{\mathrm{(c)}}{<}0.
\label{EQ_explicit_general_condition}
\end{align}
In \eqref{EQ_explicit_general_condition}, (a) combines
\eqref{EQ_explicit_two_way_polynomial} and \eqref{EQ_explicit_one_way_linear},
(b) factors using \eqref{EQ_explicit_roots}, and (c) uses the
opposite signs of the two factors.
The interval is nonempty exactly when $R>{\hyperref[EQ_explicit_roots]{\rho_-}}$.
Equation~\eqref{EQ_comparison_proof} then gives
${\hyperref[EQ_objective]{P_2^\star}}<{\hyperref[SUBSEC_oneway]{P_{\mathrm{MD},1}^{\star}}}$, proving the proposition.
\end{proof}
\endgroup

\subsection{Perfect Current Channel and Noise Knowledge}
\label{APP_perfect_benchmark}
The two-way advantage over \eqref{EQ_one_threshold} might partly
result from errors in the one-way Bob's preparation estimates
$\widehat g_0$ and $\widehat v_1$. To check whether the advantage persists without these errors,
we give Bob in the one-way scheme the actual $g$ and
${\symref{vbeta}{v_\beta}}$.
\begin{proof}
Alice retains the estimated AN basis, so ${\symref{vbeta}{v_\beta}}$ includes its leakage.
Let ${\symref{perfect}{t_{1,\mathrm{perf}}}}>0$ be the squared acceptance threshold \eqref{EQ_one_squared_threshold},
and let ${\symref{perfect}{V_{1,\mathrm{perf}}}}$ be the attacked gain-difference variance \eqref{EQ_one_attack_variance}
conditional on $(s,{\symref{Lambda}{\Lambda}},{\hyperref[EQ_attack_CSI]{Z_T}})$.
We choose $t_{1,\mathrm{perf}}$ to give false-alarm probability
$\alpha$ and calculate the conditional variance
$V_{1,\mathrm{perf}}$ of the gain difference $\widehat g_1-g$:
\symdef{perfect}
\begin{align}
{\hyperref[EQ_decision]{\alpha}}&\overset{\mathrm{(a)}}{=}
\Pr_{{\hyperref[EQ_objective]{\mathcal H_0}}}\{|{\hyperref[SUBSEC_oneway]{\widehat g_1}}-g|^2>{\symref{perfect}{t_{1,\mathrm{perf}}}}
\mid g,{\symref{vbeta}{v_\beta}}\}\notag\\
&\overset{\mathrm{(b)}}{=}e^{-2nt_{1,\mathrm{perf}}/{\symref{vbeta}{v_\beta}}},\notag\\
{\symref{perfect}{V_{1,\mathrm{perf}}}}&\overset{\mathrm{(c)}}{=}{\symref{V1}{V_1}}-\frac{{\symref{vbeta}{v_\beta}}}{2m}
={\hyperref[EQ_training]{\tau}}+\frac{{\hyperref[EQ_one_noise]{v_{\rm th}}}}{2n}
+\frac{a^2{\hyperref[EQ_attack_error_fraction]{\epsilon_T}}}{(1-{\hyperref[EQ_attack_error_fraction]{\epsilon_T}}){\hyperref[EQ_attack_CSI]{Z_T}}}.
\label{EQ_perfect_benchmark}
\end{align}
In \eqref{EQ_perfect_benchmark}, (a) sets the false-alarm
probability to $\alpha$, and (b) uses
$\widehat g_1-g\sim\CN(0,v_\beta/(2n))$ under $\mathcal H_0$
from \eqref{EQ_one_noise}. Taking logarithms in (b) gives
$t_{1,\mathrm{perf}}$.
Step (c) replaces $\widehat g_0=g+\overline e_0$ by $g$
in \eqref{EQ_one_error_decomposition}, subtracting
$v_\beta/(2m)$ from $V_1$ in \eqref{EQ_one_attack_variance}.
The variance $\tau$ remains because $g-s$ is still uncertain
to Trudy.
Equation~\eqref{EQ_one_lower} evaluates the same feasible attack
using $t_{1,\mathrm{perf}}$ and $V_{1,\mathrm{perf}}$, without
averaging over $Z_1$. Its acceptance probability lower-bounds
worst-case missed detection for this benchmark.
\end{proof}

\section{Forward-Channel Uncertainty with Alice's AN}
\label{APP_information}
To lower-bound Trudy's minimum mean-square error in estimating
$\mathbf h_\perp$ from her actual information $\mathcal O_T$,
we additionally reveal $\mathbf G_A,\mathbf G_B$ and preparation matrix $\mathbf X_0$,
forming $\mathcal O_T^+$ in \eqref{EQ_disclosed_information}.
Because Trudy can ignore this additional information, her
minimum error using $\mathcal O_T^+$ cannot exceed that using
$\mathcal O_T$, and therefore provides the desired lower bound.
\subsection{Proof of Proposition~\ref{PROP_information}}
\begin{proof}
To prove the estimation-error bound in
Proposition~\ref{PROP_information}, we first derive the
conditional variance $q_h$ of each entry $h_{\perp,j}$,
$1\leq j\leq d$, of $\mathbf h_\perp$ defined in
Section~\ref{SUBSEC_spatial}, given $\mathcal O_T^+$.
The conditional-mean estimator attains conditional mean-square
error $q_h$, so averaging gives
$\mathbb E_{\mathbf a,\mathbf J}[q_h]$.
Separately, we prove that $q_h$ is nondecreasing in $\beta_A$,
as also claimed in the proposition.

\emph{Step 1: obtain the observations of
$\hyperref[SUBSEC_spatial]{\mathbf h_\perp}$.}
Our target is the mean-square error for the last $d$ entries
of $\mathbf h$, which form $\mathbf h_\perp$ in
Section~\ref{SUBSEC_spatial}.
We give Trudy $\mathbf G_A,\mathbf G_B,\mathbf X_0$, forming
$\mathcal O_T^+$ in \eqref{EQ_disclosed_information}.
She can ignore this disclosure, so her minimum mean-square
error with $\mathcal O_T^+$ lower-bounds that with
$\mathcal O_T$ in \eqref{EQ_actual_information}.

Let $h_{\perp,j}\in\mathbb C$ be entry $j$ of $\mathbf h_\perp$,
where $1\leq j\leq d$.
To separate these coefficients in the intercepted response
$\mathbf Y_{A,T,0}$ in \eqref{EQ_protected_intercepts}, let
$\mathbf X_{0,\perp}\in\mathbb C^{d\times m}$ contain the last
$d$ rows of $\mathbf X_0$.
Let $\mathbf M_\perp,\mathbf N_\perp\in\mathbb C^{q\times d}$
be the projected observation and noise, with columns
$\mathbf m_j,\mathbf n_j\in\mathbb C^q$.
The row orthogonality in \eqref{EQ_private_probe} gives
\begin{align}
\mathbf M_\perp
&:=\mathbf Y_{A,T,0}\mathbf X_{0,\perp}^H/E_0\notag\\
&\overset{\mathrm{(a)}}{=}
\kappa\mathbf a\mathbf h\mathbf X_0\mathbf X_{0,\perp}^H/E_0
+\boldsymbol\zeta_0\mathbf X_{0,\perp}^H/E_0\notag\\
&\overset{\mathrm{(b)}}{=}
\kappa\mathbf a\mathbf h_\perp+\mathbf N_\perp,
\quad
\mathbf N_\perp:=\boldsymbol\zeta_0\mathbf X_{0,\perp}^H/E_0,
\notag\\
\mathbf n_j\mid\mathbf a,\mathbf J,\mathbf X_0
&\overset{\mathrm{(c)}}{\sim}
\CN(\mathbf0,\mathbf R_A/E_0).
\label{EQ_projected_forward_observation}
\end{align}
In \eqref{EQ_projected_forward_observation}, (a) substitutes
\eqref{EQ_protected_intercepts}.
Step (b) uses $\mathbf X_0\mathbf X_{0,\perp}^H/E_0
=[\mathbf0_{d\times N_R},\Id_d]^{\mathsf T}$, selecting
the last $d$ entries of $\mathbf h$.
Step (c) uses \eqref{EQ_intercept_noise} and
$\mathbf X_{0,\perp}\mathbf X_{0,\perp}^H=E_0\Id_d$.

To identify the observations needed to estimate $h_{\perp,j}$,
we observe the Markov chain
\begin{equation}
h_{\perp,j}
\longrightarrow
(\mathbf m_j,\mathbf a,\mathbf J,\mathbf X_0)
\longrightarrow
\mathcal O_T^+.
\label{EQ_forward_information_markov}
\end{equation}
Conditional on $\mathbf a,\mathbf J,\mathbf X_0$,
the Gaussian noise projections in
\eqref{EQ_projected_forward_observation} are independent.
With $\mathbf u_A$ fixed, $h_{\perp,j}$ uses only row
$N_R+j$ of $\mathbf H$ in \eqref{EQ_TDD_channels}.
This row is independent of the rows determining the other
forward coefficients and $\mathbf H_B$, and of the return-training
noise, giving the claimed independence.
Given the disclosed channels and $\mathbf X_0$,
$\mathbf T_A,\mathbf T_B,\mathbf Y_{B,T,0},\mathbf Y_{B,T,1}$
in \eqref{EQ_actual_information} involve independent signals
and noises, which explains why the chain holds.
By \eqref{EQ_forward_information_markov}, we therefore
calculate the conditional distribution of $h_{\perp,j}$
given $\mathcal O_T^+$ using only
$(\mathbf m_j,\mathbf a,\mathbf J,\mathbf X_0)$.
Then we have
\begin{equation}
\begin{aligned}
&f_{h_{\perp,j}\mid\mathcal O_T^+}
 (z\mid\mathcal O_T^+)=
f_{h_{\perp,j}\mid\mathbf m_j,\mathbf a,\mathbf J,\mathbf X_0}
 (z\mid\mathbf m_j,\mathbf a,\mathbf J,\mathbf X_0).
\end{aligned}
\label{EQ_forward_information_reduction_density}
\end{equation}
Therefore, the conditional distribution required for
Proposition~\ref{PROP_information} can be calculated using
the single-column observation
$\mathbf m_j=\kappa\mathbf a h_{\perp,j}+\mathbf n_j$
in \eqref{EQ_projected_forward_observation}.

\emph{Step 2: calculate the variance conditional on
$\mathcal O_T^+$.}
To calculate the estimation error after observing $\mathbf m_j$,
we first determine the conditional distribution of $h_{\perp,j}$.
By \eqref{EQ_effective_laws} and Step~1, each $h_{\perp,j}$
has prior distribution $\CN(0,S)$ and observation
$\mathbf m_j=\kappa\mathbf a h_{\perp,j}+\mathbf n_j$.
Let $\mu_j\in\mathbb C$ and $q_h>0$ be its conditional mean
and variance given $\mathcal O_T^+$.
For $z\in\mathbb C$, the conditional density of $h_{\perp,j}$
given $\mathcal O_T^+$ satisfies
\begin{align}
f_{h_{\perp,j}\mid\mathcal O_T^+}(z\mid\mathcal O_T^+)
&\overset{\mathrm{(a)}}{\propto}
\exp\bigl[-|z|^2/S\notag\\
&\qquad-E_0(\mathbf m_j-\kappa\mathbf az)^H
\mathbf R_A^{-1}(\mathbf m_j-\kappa\mathbf az)\bigr]\notag\\
&\overset{\mathrm{(b)}}{\propto}
\exp[-|z-\mu_j|^2/q_h],\label{EQ_forward_posterior_square}\\
\operatorname{Cov}(\mathbf h_\perp^H\mid\mathcal O_T^+)
&\overset{\mathrm{(c)}}{=}q_h\Id_d.
\label{EQ_forward_information_proof}
\end{align}
In \eqref{EQ_forward_posterior_square}--%
\eqref{EQ_forward_information_proof}, (a) uses the
Markov chain in \eqref{EQ_forward_information_markov} and
Bayes' rule to combine
$h_{\perp,j}\sim\CN(0,S)$ from \eqref{EQ_effective_laws}
with the observation
$\mathbf m_j=\kappa\mathbf a h_{\perp,j}+\mathbf n_j$
and noise distribution
$\mathbf n_j\mid\mathbf a,\mathbf J,\mathbf X_0
\sim\CN(\mathbf0,\mathbf R_A/E_0)$
in \eqref{EQ_projected_forward_observation}.
Step (b) completes the square in $z$, using $q_h^{-1}
=S^{-1}+\kappa^2E_0\mathbf a^H\mathbf R_A^{-1}\mathbf a$ and
$\mu_j=q_h\kappa E_0\mathbf a^H\mathbf R_A^{-1}\mathbf m_j.$
For (c), the coefficients $h_{\perp,j}$ are independent
by \eqref{EQ_TDD_channels}, and the conditional noise columns
$\mathbf n_j$ are independent by
\eqref{EQ_projected_forward_observation}.
Since each $\mathbf m_j$ involves only $h_{\perp,j}$,
the conditional density factors over $j$.
Each coefficient has conditional variance $q_h$ from
\eqref{EQ_forward_posterior_square}, giving covariance
$q_h\Id_d$ and proving \eqref{EQ_forward_information}.

\emph{Step 3: obtain the mean-square-error lower bound.}
Let $\widehat h_j=\widehat h_j(\mathcal O_T)\in\mathbb C$ be
any estimate of $h_{\perp,j}$ based on Trudy's actual information.
Both $\widehat h_j$ and the conditional mean
$\mu_j=\mathbb E_{h_{\perp,j}\mid\mathcal O_T^+}[h_{\perp,j}]$
are fixed given $\mathcal O_T^+$, which contains $\mathcal O_T$.
We separate the conditional variance from the squared difference
between these estimates, then average:
\begin{align}
\mathbb E_{h_{\perp,j}\mid\mathcal O_T^+}
 [|h_{\perp,j}-\widehat h_j|^2]&\overset{\mathrm{(a)}}{=}
\mathbb E_{h_{\perp,j}\mid\mathcal O_T^+}
 [|h_{\perp,j}-\mu_j|^2]\notag\\
&\quad+|\mu_j-\widehat h_j|^2\notag\\
&\overset{\mathrm{(b)}}{=}
q_h+|\mu_j-\widehat h_j|^2\geq q_h,
\notag\\
\mathbb E_{h_{\perp,j},\mathcal O_T}
 [|h_{\perp,j}-\widehat h_j|^2]
&\overset{\mathrm{(c)}}{=}
\mathbb E_{\mathcal O_T^+}\!\left[
 \mathbb E_{h_{\perp,j}\mid\mathcal O_T^+}
 [|h_{\perp,j}-\widehat h_j|^2]\right]\notag\\
&\overset{\mathrm{(d)}}{\geq}
\mathbb E_{\mathcal O_T^+}[q_h]
\overset{\mathrm{(e)}}{=}\mathbb E_{\mathbf a,\mathbf J}[q_h].
\label{EQ_information_lower_bound}
\end{align}
In \eqref{EQ_information_lower_bound}, (a) expands the squared
error about $\mu_j$. The cross term vanishes because
$\mathbb E_{h_{\perp,j}\mid\mathcal O_T^+}
[h_{\perp,j}-\mu_j]=0$.
Step (b) uses \eqref{EQ_forward_information_proof}.
Thus the conditional-mean estimator attains conditional
mean-square error $q_h$ when $\mathcal O_T^+$ is available.
Step (c) applies total expectation, and (d) uses the preceding
inequality. Step (e) uses the fact that \eqref{EQ_forward_information} depends on
the observations only through $\mathbf a,\mathbf J$, whose
joint distribution therefore determines the average.
The bound holds for every $j$, so it also bounds the mean-square
error per coefficient of any estimate of $\mathbf h_\perp$.

\emph{Step 4: determine how Alice's AN changes $q_h$.}
We next show that $q_h$ in \eqref{EQ_forward_information}
is nondecreasing in $\beta_A$, with $\mathbf a,\mathbf J$
and all other parameters fixed.
Hold $\mathbf a,\mathbf J$ and all other model and energy
parameters fixed while varying $\beta_A\in[0,1)$.
Both $\kappa$ in \eqref{EQ_response} and $\mathbf R_A$ in
\eqref{EQ_intercept_noise} depend on $\beta_A$.
To compare the noise for the same signal coefficient $\mathbf a$,
divide \eqref{EQ_projected_forward_observation} by $\kappa>0$.
The noise covariance becomes
$\widetilde{\mathbf R}_A(\beta_A)/E_0$, where the positive
definite matrix
$\widetilde{\mathbf R}_A(\beta_A):=\mathbf R_A/\kappa^2
\in\mathbb C^{q\times q}$ satisfies
\begin{align}
\widetilde{\mathbf R}_A(\beta_A)
&\!\overset{\mathrm{(a)}}{=}\!\sigma^2\mathbf a\mathbf a^H
\!\!+\!\frac{\beta_A(SP_B\!+\!\sigma^2)}{d(1\!-\!\beta_A)}
 \mathbf J\mathbf J^H\notag\\
&\quad+\frac{\sigma_T^2(SP_B+\sigma^2)}
 {(1\!-\!\beta_A)P_A}\Id_q,\notag\\
q_h(\beta_A)
&\overset{\mathrm{(b)}}{=}
\left[S^{-1}+E_0\mathbf a^H
\widetilde{\mathbf R}_A(\beta_A)^{-1}\mathbf a\right]^{-1}.
\label{EQ_forward_information_reduction}
\end{align}
In \eqref{EQ_forward_information_reduction}, (a) uses
$\kappa^2$ from \eqref{EQ_response} and $\mathbf R_A$ from
\eqref{EQ_intercept_noise}.
Step (b) uses
$\mathbf R_A=\kappa^2\widetilde{\mathbf R}_A$
into \eqref{EQ_forward_information}.
The coefficients of $\mathbf J\mathbf J^H$ and $\Id_q$
increase with $\beta_A$, so for two AN fractions
$0\leq\beta_1\leq\beta_2<1$,
\begin{align}
\widetilde{\mathbf R}_A(\beta_2)
&\overset{\mathrm{(a)}}{\succeq}
\widetilde{\mathbf R}_A(\beta_1),\notag\\
\mathbf a^H\widetilde{\mathbf R}_A(\beta_2)^{-1}\mathbf a
&\overset{\mathrm{(b)}}{\leq}
\mathbf a^H\widetilde{\mathbf R}_A(\beta_1)^{-1}\mathbf a,
\notag\\
q_h(\beta_2)&\overset{\mathrm{(c)}}{\geq}q_h(\beta_1).
\label{EQ_forward_information_monotonicity}
\end{align}
In \eqref{EQ_forward_information_monotonicity}, (a) follows
from \eqref{EQ_forward_information_reduction} because
$\mathbf J\mathbf J^H$ and $\Id_q$ are positive semidefinite.
Step (b) uses the reversed order of inverses of positive definite
matrices, and (c) substitutes into the expression for $q_h$ in
\eqref{EQ_forward_information_reduction}.
Thus Alice's AN cannot decrease the conditional variance $q_h$
under $\mathcal O_T^+$.
Together with \eqref{EQ_forward_information_proof} and
\eqref{EQ_information_lower_bound}, this proves
Proposition~\ref{PROP_information}.
\end{proof}

\section{Computation of the Probability Bounds}
\label{APP_numerical}
We calculate the two-way upper bound $\overline P_2$ in
\eqref{EQ_upper_scalar} and the one-way attack acceptance
probability $P_{\mathrm{MD},1}$ in \eqref{EQ_one_lower}.
We then verify that \eqref{EQ_comparison} holds throughout
the interval in \eqref{EQ_explicit_example_range}.

\emph{Evaluation of the probabilities.}
The two-way bound \eqref{EQ_upper_scalar} requires the actual
probability $F_{{\hyperref[EQ_physical_ratio]{\mathcal A}}}({\hyperref[EQ_upper_scalar]{\theta}})$, in addition to the exponential
upper bound used for the sufficient comparison condition.
To evaluate ${\hyperref[EQ_cutoff_domination]{F_{\mathcal A}}}({\hyperref[EQ_upper_scalar]{\theta}})$, convert its event into a
threshold on $|g|^2$, whose conditional distribution is known.
\symdef{rtheta}
For $0<{\hyperref[EQ_upper_scalar]{\theta}}<{\hyperref[EQ_TDD_channels]{S}}/{\hyperref[SUBSEC_channels]{\sigma^2}}$, define the threshold ${\symref{rtheta}{r_\theta}}({\symref{Lambda}{\Lambda}}):=
\frac{{\hyperref[EQ_upper_scalar]{\theta}}(\beta_AP_A\Lambda/d+{\hyperref[SUBSEC_channels]{\sigma^2}})}{{\hyperref[EQ_response]{\kappa}}^2({\hyperref[EQ_TDD_channels]{S}}-{\hyperref[EQ_upper_scalar]{\theta}}{{\hyperref[SUBSEC_channels]{\sigma^2}}})}.$
Then 
\begin{align}
{\hyperref[EQ_cutoff_domination]{F_{\mathcal A}}}({\hyperref[EQ_upper_scalar]{\theta}})
&\overset{\mathrm{(a)}}{=}\Pr\{|g|^2<{\symref{rtheta}{r_\theta}}({\symref{Lambda}{\Lambda}})\}\notag\\
&\overset{\mathrm{(b)}}{=}\mathbb E_{s,{\symref{Lambda}{\Lambda}}}
[\Pr\{|g|^2<{\symref{rtheta}{r_\theta}}({\symref{Lambda}{\Lambda}})\mid s,{\symref{Lambda}{\Lambda}}\}]\notag\\
&\overset{\mathrm{(c)}}{=}\mathbb E_{s,{\symref{Lambda}{\Lambda}}}
\!\left[\Phi_{2,2s^2/{\hyperref[EQ_training]{\tau}}}\!\left(\frac{2{\symref{rtheta}{r_\theta}}({\symref{Lambda}{\Lambda}})}{{\hyperref[EQ_training]{\tau}}}\right)\right].
\label{EQ_TDD_ratio_cdf}
\end{align}
In \eqref{EQ_TDD_ratio_cdf},
(a) rearranges \eqref{EQ_physical_ratio},
(b) conditions on $s,{\symref{Lambda}{\Lambda}}$, and (c) uses
$2|g|^2/{\hyperref[EQ_training]{\tau}}\mid s\sim\chi'^2_2(2s^2/{\hyperref[EQ_training]{\tau}})$ and independence
of ${\symref{Lambda}{\Lambda}}/{\hyperref[EQ_training]{\tau}}\sim\Gamma(d,1)$ and $(s,g)$ from \eqref{EQ_return_gain_conditional_density}.

To calculate $\pi_s$ in \eqref{EQ_explicit_gain_interval}, we need
$\Pr\{s\leq r_s\}$ for a gain threshold $r_s\geq0$.
The expectations in \eqref{EQ_TDD_ratio_cdf},
\eqref{EQ_one_lower}, and \eqref{EQ_explicit_amplification_tail}
also involve $s$. Its distribution is therefore needed
to perform the averaging over $s$.

\symdef{gaincdf}
For $N_R=2$ as in Section~\ref{SEC_numerical}, let
$\lambda_1>\lambda_2>0$ be the ordered eigenvalues of
$\widehat{\mathbf H}_B\widehat{\mathbf H}_B^H/(S-\tau)$.
The SVD specified before \eqref{EQ_spatial} also gives
$\widehat{\mathbf H}_B^H\mathbf u_B=s\mathbf f_A$. Hence,
\begin{align}
\frac{\widehat{\mathbf H}_B\widehat{\mathbf H}_B^H}
     {S-\tau}\mathbf u_B
&\overset{\mathrm{(a)}}{=}
\frac{s\widehat{\mathbf H}_B\mathbf f_A}{S-\tau}
=\frac{s^2}{S-\tau}\mathbf u_B,\notag\\
s^2&\overset{\mathrm{(b)}}{=}(S-\tau)\lambda_1.
\label{EQ_gain_eigenvalue_relation}
\end{align}
In \eqref{EQ_gain_eigenvalue_relation}, (a) uses
$\widehat{\mathbf H}_B^H\mathbf u_B=s\mathbf f_A$ and
$\widehat{\mathbf H}_B\mathbf f_A=s\mathbf u_B$ from
\eqref{EQ_spatial}.
Step (b) identifies $s^2/(S-\tau)$ as the largest
eigenvalue $\lambda_1$, because $s$ is the largest
singular value.

By \eqref{EQ_training},
$\widehat{\mathbf H}_B/\sqrt{S-\tau}$
has independent $\CN(0,1)$ entries. Its Gram matrix therefore has
the complex Wishart eigenvalue density \cite{Dumitriu2002},
which for two receivers is
\begin{equation}
f_{\rm eig}(u,v)=
\frac{e^{-u-v}u^{p-2}v^{p-2}(u-v)^2}{\Gamma(p-1)\Gamma(p)},
\qquad 0<v<u.
\label{EQ_gain_eigenvalue_density}
\end{equation}
Let $F_{p,2}$ and $f_p$ be the CDF and density of $\lambda_1$.
Set $x=r_s^2/(S-\tau)$. Using $u,v\geq0$,
\begin{align}
\Pr\{s\leq r_s\}
&\overset{\mathrm{(a)}}{=}\Pr\{\lambda_1\leq x\}=F_{p,2}(x)
\label{EQ_gain_cdf_normalization}\\
&\overset{\mathrm{(b)}}{=}
\int_0^x\!\int_0^u f_{\rm eig}(u,v)\,\mathrm dv\,\mathrm du\notag\\
&\overset{\mathrm{(c)}}{=}
\frac{\int_0^x\!\int_0^x e^{-u-v}u^{p-2}v^{p-2}(u-v)^2
\,\mathrm du\,\mathrm dv}{2\Gamma(p-1)\Gamma(p)}\notag\\
&\overset{\mathrm{(d)}}{=}
\frac{\Gamma(p+1)}{\Gamma(p)}G_{p-1}(x)G_{p+1}(x)\notag\\
&\quad-\frac{\Gamma(p)}{\Gamma(p-1)}G_p(x)^2\notag\\
&\overset{\mathrm{(e)}}{=}pG_{p-1}(x)G_{p+1}(x)-(p-1)G_p(x)^2.
\label{EQ_explicit_gain_two_receivers}
\end{align}
In \eqref{EQ_gain_cdf_normalization}--\eqref{EQ_explicit_gain_two_receivers},
(a) uses $s^2=(S-\tau)\lambda_1$ from \eqref{EQ_gain_eigenvalue_relation}.
Step (b) integrates \eqref{EQ_gain_eigenvalue_density} over
$0<v<u\leq x$. Step (c) uses the symmetry of
$e^{-u-v}u^{p-2}v^{p-2}(u-v)^2$ under exchanging $u$ and $v$.
Its integrals over $0<v<u<x$ and $0<u<v<x$ are equal,
so the integral over $0<v<u<x$ is half the integral
over $[0,x]^2$. Step (d) expands $(u-v)^2=u^2-2uv+v^2$
and uses \eqref{EQ_appendix_gamma_cdf} to evaluate each scalar integral.
Step (e) uses $\Gamma(p+1)=p\Gamma(p)$ and
$\Gamma(p)=(p-1)\Gamma(p-1)$.
To obtain the density $f_p$ of the largest eigenvalue
$\lambda_1$, integrate the joint density in
\eqref{EQ_gain_eigenvalue_density} over $\lambda_2$:
\begin{align}
f_p(\lambda_1)
&{=}
\int_0^{\lambda_1}
f_{\rm eig}(\lambda_1,\lambda_2)\,\mathrm d\lambda_2
\notag\\
&\overset{\mathrm{(a)}}{=}
\frac{e^{-\lambda_1}\lambda_1^{p-2}}
     {\Gamma(p-1)\Gamma(p)}\notag\\
&\quad\times\int_0^{\lambda_1}
e^{-\lambda_2}\lambda_2^{p-2}
(\lambda_1-\lambda_2)^2\,\mathrm d\lambda_2,
\, \lambda_1>0.
\label{EQ_singular_density}
\end{align}
In \eqref{EQ_singular_density}, (a) substitutes
\eqref{EQ_gain_eigenvalue_density}.
We now evaluate the expectation over $s$ in
\eqref{EQ_explicit_amplification_tail} and the probability
$\pi_s$ in \eqref{EQ_explicit_gain_interval}:
\begin{align}
&\mathbb E_s[e^{-ts^2/(1+t\tau)}]\notag\\
&\overset{\mathrm{(a)}}{=}
\int_0^\infty
e^{-t(S-\tau)\lambda_1/(1+t\tau)}
f_p(\lambda_1)\,\mathrm d\lambda_1,
\notag\\
\pi_s
&\overset{\mathrm{(b)}}{=}
F_{p,2}\!\left(\frac{(a+h)^2}{S-\tau}\right)\notag\\
&\quad-F_{p,2}\!\left(
\frac{\max\{0,a-h\}^2}{S-\tau}
\right).
\label{EQ_gain_numerical_averages}
\end{align}
In \eqref{EQ_gain_numerical_averages}, (a) uses
$s^2=(S-\tau)\lambda_1$ from \eqref{EQ_gain_eigenvalue_relation},
together with the density in \eqref{EQ_singular_density}.
Step (b) applies \eqref{EQ_gain_cdf_normalization} to
the two endpoints of $|s-a|\leq h$ in
\eqref{EQ_explicit_gain_interval}.

\subsection{Numerical Integration Checks}
\label{APP_numerical_checks}
For the two-way calculation, integration over $s$ uses
\eqref{EQ_singular_density}, and integration over $\Lambda/\tau$
uses a 32-node Gaussian quadrature rule. The independent Gamma
variables $Z_n,Z_\nu$ in \eqref{EQ_upper_scalar_proof} use
40-node rules. Increasing these orders to 48 and 64, respectively,
changes the plotted optimized upper bounds by less than
$1.3\times10^{-13}$.
For the one-way calculation in \eqref{EQ_one_lower}, increasing
the principal order from 96 to 144 and the auxiliary orders from
16 to 24 changes the values at
$(p=q,\rho)=(6,0.034),(8,0.060),(10,0.081)$ by less than
$7\times10^{-18}$. These changes assess numerical convergence
and are not rigorous numerical error bounds.

\subsection{Parameters for the Example Interval}
\label{APP_example_parameters}
For the six-antenna example in Section~\ref{SEC_numerical},
use $t=9$, $h=0.8$, $z_-=4$, $z_+=12$,
$z_v=0.9\nu_1$, and $\lambda_+=3\tau d$, together with the
stated $\theta=40$ and $R=0.005$.
These choices satisfy $tc_\theta\beta_AP_A\tau/d<1$.
Substitution into \eqref{EQ_explicit_amplification_tail},
\eqref{EQ_explicit_moment_factor}--\eqref{EQ_explicit_quadratic_coefficients},
and \eqref{EQ_explicit_one_way_linear} gives the coefficient
bounds reported in Section~\ref{SEC_numerical}.
The endpoint calculation there establishes \eqref{EQ_comparison}
throughout \eqref{EQ_explicit_example_range}.


\begin{thebibliography}{99}
\bibitem{Xiao2008}
L.~Xiao, L.~J. Greenstein, N.~B. Mandayam, and W.~Trappe,
``Using the physical layer for wireless authentication in time-variant
channels,'' \emph{IEEE Trans. Wireless Commun.}, vol.~7, no.~7,
pp.~2571--2579, Jul. 2008.
\bibitem{Baracca2012}
P.~Baracca, N.~Laurenti, and S.~Tomasin, ``Physical layer authentication
over MIMO fading wiretap channels,'' \emph{IEEE Trans. Wireless Commun.},
vol.~11, no.~7, pp.~2564--2573, 2012, doi: 10.1109/TWC.2012.051512.111481.
\bibitem{Senigagliesi2021}
L.~Senigagliesi, M.~Baldi, and E.~Gambi, ``Comparison of statistical
and machine learning techniques for physical layer authentication,''
\emph{IEEE Trans. Inf. Forensics Security}, vol.~16, pp.~1506--1521,
2021, doi: 10.1109/TIFS.2020.3033454.
\bibitem{SpatialPLASurvey2026}
Y.~Chen, B.~He, J.~Luo, S.~Wang, Y.~Ma, H.~Xu, L.~Yao, J.~Lei,
A.~Nallanathan, and K.-K.~Wong, ``A survey of physical-layer authentication
enhanced by emerging spatial domain technologies,'' arXiv:2606.28807, 2026.
\bibitem{Srinivasan2024}
M.~Srinivasan, L.~Senigagliesi, H.~Chen, A.~Chorti, M.~Baldi, and
H.~Wymeersch, ``AoA-based physical layer authentication in analog arrays
under impersonation attacks,'' in \emph{Proc. IEEE SPAWC}, 2024.
\bibitem{Pham2026}
T.~M.~Pham, L.~Senigagliesi, M.~Baldi, R.~F.~Schaefer, G.~P.~Fettweis,
and A.~Chorti, ``Leveraging angle of arrival estimation against impersonation
attacks in physical layer authentication,'' \emph{IEEE Trans. Inf. Forensics
Security}, vol.~21, pp.~3226--3239, 2026, doi: 10.1109/TIFS.2026.3675885.
\bibitem{NearField2026}
H.~El~Hassani, L.~Senigagliesi, and A.~Chorti, ``Near-field physical-layer
authentication under impersonation attacks,'' arXiv:2609.04879, 2026.
\bibitem{Senigagliesi2025}
L.~Senigagliesi, A.~V.~Guglielmi, M.~Baldi, and S.~Tomasin, ``Security
analysis of RIS-assisted physical-layer authentication over multipath
channels,'' in \emph{Proc. IEEE WIFS}, 2025, pp.~31--36,
doi: 10.1109/WIFS66636.2025.00014.
\bibitem{Jorswieck2015}
E.~Jorswieck, S.~Tomasin, and A.~Sezgin, ``Broadcasting into the uncertainty:
Authentication and confidentiality by physical-layer processing,''
\emph{Proc. IEEE}, vol.~103, no.~10, pp.~1702--1724, 2015,
doi: 10.1109/JPROC.2015.2469602.
\bibitem{Shan2013}
D.~Shan, K.~Zeng, W.~Xiang, P.~Richardson, and Y.~Dong,
``{PHY-CRAM}: Physical layer challenge-response authentication mechanism
for wireless networks,'' \emph{IEEE J. Sel. Areas Commun.}, vol.~31,
no.~9, pp.~1817--1827, Sep. 2013.
\bibitem{Crosara2026}
L.~Crosara, S.~Tomasin, and M.~Mirmohseni, ``Hybrid channel- and coding-based
challenge-response physical-layer authentication with IRS,'' in
\emph{Proc. IEEE ICC}, 2026, doi: 10.1109/ICC59461.2026.11588191.
\bibitem{Zhang2026}
Y.~Zhang, R.~J\"antti, Y.~Zheng, Y.~Dang, M.~Kaveh, and Z.~Han,
``Physical layer challenge-response authentication between ambient
backscatter devices,'' \emph{IEEE Trans. Inf. Forensics Security},
vol.~21, pp.~1512--1527, 2026, doi: 10.1109/TIFS.2026.3655191.
\bibitem{Guglielmi2026}
A.~V. Guglielmi, L.~Crosara, and S.~Tomasin,
``Design of physical layer challenge response authentication with {RIS}
and multiple-antenna devices,'' \emph{IEEE Trans. Inf. Forensics Security},
vol.~21, pp.~3677--3688, 2026.
\bibitem{Rowhani2025}
S.~B.~Rowhani, A.~V.~Guglielmi, D.~Scazzoli, M.~Magarini, and S.~Tomasin,
``Challenge-response PLA with amplify and forward relay: Design and
experimental validation,'' in \emph{Proc. EUSIPCO}, 2025.
\bibitem{GoelNegi2008}
S.~Goel and R.~Negi, ``Guaranteeing secrecy using artificial
noise,'' \emph{IEEE Trans. Wireless Commun.}, vol.~7, no.~6,
pp.~2180--2189, June 2008.
\bibitem{Lin2013}
P.-H.~Lin, S.-H.~Lai, S.-C.~Lin, and H.-J.~Su, ``On secrecy rate of
the generalized artificial-noise assisted secure beamforming for wiretap
channels,'' \emph{IEEE J. Sel. Areas Commun.}, vol.~31, no.~9,
pp.~1728--1740, 2013, doi: 10.1109/JSAC.2013.130917.
\bibitem{Huang2013}
C.-W. Huang, T.-H. Chang, X. Zhou, and Y.-W. P. Hong,
``Two-way training for discriminatory channel estimation in wireless
MIMO systems,'' 2013, arXiv:1301.4625,
doi: 10.1109/TSP.2013.2245124.
\bibitem{AN_TIFS2021}
J.~B.~Perazzone, P.~L.~Yu, B.~M.~Sadler, and R.~S.~Blum,
``Artificial noise-aided MIMO physical layer authentication with imperfect
CSI,'' \emph{IEEE Trans. Inf. Forensics Security}, vol.~16,
pp.~2173--2185, 2021, doi: 10.1109/TIFS.2021.3050599.
\bibitem{Maurer1993}
U.~M.~Maurer, ``Secret key agreement by public discussion from common
information,'' \emph{IEEE Trans. Inf. Theory}, vol.~39, no.~3,
pp.~733--742, 1993, doi: 10.1109/18.256484.
\bibitem{Akhlaghpasand2018}
H.~Akhlaghpasand, S.~M. Razavizadeh, E.~Bj\"ornson, and T.~T. Do,
``Jamming detection in massive {MIMO} systems,''
\emph{IEEE Wireless Commun. Lett.}, vol.~7, no.~2,
pp.~242--245, Apr. 2018, doi: 10.1109/LWC.2017.2769650.
\bibitem{Anderson1955}
T.~W. Anderson, ``The integral of a symmetric unimodal function over
a symmetric convex set and some probability inequalities,''
\emph{Proc. Amer. Math. Soc.}, vol.~6, no.~2, pp.~170--176, 1955,
doi: 10.1090/S0002-9939-1955-0069229-1.
\bibitem{Dumitriu2002}
I.~Dumitriu and A.~Edelman, ``Matrix models for beta ensembles,''
\emph{J. Math. Phys.}, vol.~43, no.~11, pp.~5830--5847, Nov. 2002,
doi: 10.1063/1.1507823.
\end{thebibliography}
\end{document}